\documentclass[fleqn,usenatbib]{mnras}

\usepackage{newtxtext,newtxmath}  
\usepackage[T1]{fontenc}
\usepackage{ae,aecompl}
\usepackage{gensymb}

\usepackage{graphicx}	
\usepackage{amsmath}  	
\usepackage{amssymb}	
\usepackage{multirow}

\newcommand{\kepler}{\textit{Kepler}}
\newcommand{\Gaia}{\textit{Gaia}}

\newcommand{\teff}{$\mathrm{T_{eff}}$}

\newcommand{\thalf}{$t_{1/2}$}
\newcommand{\NGTS}{NGTS}
\newcommand{\tess}{\textit{TESS}}

\newcommand{\gaia}{\textit{Gaia}}

\newcommand{\sysrem}{{\scshape sysrem}}
\newcommand{\mad}{$MAD$}

\newcommand{\starnumber}{\textcolor{black}{339}}
\newcommand{\fieldnumber}{24}

\newcommand{\flarenumber}{\textcolor{black}{610}}

\newcommand{\lxlbol}{$L_{x}/L_{bol}$}

\newcommand{\U}{$\mathrm{\textit{U}_{LSR}}$}
\newcommand{\V}{$\mathrm{\textit{V}_{LSR}}$}
\newcommand{\W}{$\mathrm{\textit{W}_{LSR}}$}
\newcommand{\vtan}{$V_{tan}$}
\newcommand{\halpha}{H$\alpha$}

\newcommand{\minflareenergy}{$1.5\times10^{31}$ erg}
\newcommand{\maxflareenergy}{$3.4\times10^{36}$ erg}
\newcommand{\firstnight}{2015 September 21}
\newcommand{\lastnight}{2017 Jan 31}

\newcommand{\Msun}{$M_{\odot}$}

\newcommand{\rewrite}[1]{\textcolor{black}{#1}}
\newcommand{\edit}[1]{\textcolor{black}{#1}}

\title[Stellar flares detected with NGTS]{Stellar flares detected with the Next Generation Transit Survey}

\author[J. A. G. Jackman et al.]{James A. G. Jackman,$^{1,2,3}$\thanks{E-mail: jamesjackman@asu.edu}
Peter J. Wheatley,$^{1,2}$\thanks{E-mail: P.J.Wheatley@warwick.ac.uk}
Jack~S.~Acton,$^{4}$\newauthor
David~R.~Anderson,$^{1,2}$
Daniel Bayliss,$^{1,2}$
Joshua~T.~Briegal,$^{5}$
Matthew~R.~Burleigh,$^{4}$\newauthor
Sarah~L.~Casewell,$^{4}$
Boris~T.~G\"ansicke,$^{1,2}$
Samuel Gill,$^{1}$
Edward~Gillen,$^{6,5}$\thanks{Winton Fellow}\newauthor
Michael~R.~Goad,$^{4}$
Maximilian~N.~G{\"u}nther,$^{7}$\thanks{Juan Carlos Torres Fellow}
Beth~A.~Henderson,$^{4}$\newauthor
Simon~T.~Hodgkin,$^{8}$
James~S.~Jenkins,$^{9, 10}$
Chloe Pugh,$^{1}$
Didier~Queloz,$^{5}$\newauthor
Liam~Raynard,$^{4}$
Rosanna~H.~Tilbrook,$^{4}$
Christopher A. Watson,$^{11}$
Richard G. West,$^{1}$
\\
$^{1}$Dept. of Physics, University of Warwick, Gibbet Hill Road, Coventry CV4 7AL, UK\\
$^{2}$Centre for Exoplanets and Habitability, University of Warwick, Gibbet Hill Road, Coventry CV4 7AL, UK\\
$^{3}$School of Earth and Space Exploration, Arizona State University, Tempe, AZ 85287, USA\\
$^{4}$School of Physics and Astronomy, University of Leicester, Leicester LE1 7RH, UK\\ 
$^{5}$Astrophysics Group, Cavendish Laboratory, J.J. Thomson Avenue, Cambridge CB3 0HE, UK\\
$^{6}$Astronomy Unit, Queen Mary University of London, Mile End Road, London E1 4NS, UK\\
$^{7}$Department of Physics, and Kavli Institute for Astrophysics and Space Research, Massachusetts Institute of Technology,\\ Cambridge, MA 02139, USA\\ $^{8}$Institute of Astronomy, Madingley Road, Cambridge, CB3 0HA, UK\\
$^{9}$Departamento de Astronomia, Universidad de Chile, Casilla 36-D, Santiago, Chile\\
$^{10}$Centro de Astrof\'isica y Tecnolog\'ias Afines (CATA), Casilla 36-D, Santiago, Chile.\\
$^{11}$Astrophysics Research Centre, Queen's University of Belfast, 1 University Road, Belfast BT7 1NN, UK \\}

\date{Accepted XXX. Received YYY; in original form ZZZ}

\pubyear{2021}

\begin{document}
\label{firstpage}
\pagerange{\pageref{firstpage}--\pageref{lastpage}}
\maketitle

\begin{abstract}
We present the results of a search for stellar flares in the first data release from the Next Generation Transit Survey (NGTS). We have found \flarenumber\ flares from \starnumber\ stars, with spectral types between F8 and M6, the majority of which belong to the Galactic thin disc. We have used the 13 second cadence NGTS lightcurves to measure flare properties such as the flare amplitude, duration and bolometric energy. We have measured the average flare occurrence rates of K and early to mid M stars and present a generalised method to measure these rates while accounting for changing detection sensitivities. We find that field age K and early M stars show similar flare behaviour, while fully convective M stars exhibit increased white-light flaring activity, which we attribute to their increased spin down time. We have also studied the average flare rates of pre-main sequence K and M stars, showing they exhibit increased flare activity relative to their main sequence counterparts. 

\end{abstract}

\begin{keywords}
stars: flare -- stars: starspots -- stars: rotation
\end{keywords}

\section{Introduction}
The study of stellar flares in recent years has benefited greatly from the introduction of wide-field long-duration surveys. The first observations of these stellar phenomena were either from targeted observations \citep[e.g.][]{Lacy76,Alekseev94,Andrews96}, or were by chance  \citep[e.g. F and G star superflares][]{Schaefer_2000}. Targeted and wide-field studies have revealed that stars down to and including the L spectral type \citep[e.g.][]{Schmidt07,Schmidt16,Gizis17b} could produce flares with \edit{radiated} energies hundreds to thousands of times that produced by the Sun \citep[$10^{32}$ erg for the 1859 Carrington event;][]{Carrington_1859, Carrington_Energy}. These flares occur on nearly all main-sequence stars with an outer convective envelope, with the trigger believed to be due to magnetic reconnection \citep[][]{Shibata1999, Benz10}. These reconnection events cause the acceleration of charged particles into the chromosphere, resulting in plasma evaporation and the release of energy from radio to X-ray wavelengths \citep[][]{Benz10, Benz17}. When observed in visible wavelengths, these flares are termed ``white-light flares'' \citep[e.g.][]{Paudel18}. 
The observable behaviour of flares is known to change with spectral type, with K and M stars showing more frequent detectable high-energy flares than their higher mass counterparts \citep[][]{Balona15}. \edit{This is due in part to the increased contrast between the flare and quiescent spectrum for lower mass stars and in part due to the decreased photospheric luminosity of these stars relative to their solar-type counterparts. Both of these factors result in flares of a given energy appearing larger in a given filter and more easily detectable on low-mass stars.} Stars with more active magnetic fields have also shown greater flaring activity than their inactive counterparts \citep[][]{Hilton11}. 

Due to their likelihood to flare at least once in a given night, early ground-based observations of white-light flares typically focused on nearby magnetically active M dwarfs such as AD Leo \citep[e.g.][]{Oskanian69, Pettersen84}. Such campaigns have been used to study statistical distributions of flare properties \citep[e.g. the flare energy occurrence rate][]{Lacy76,Gershberg83} on these single stars, along with the temporal morphology of flares \citep[e.g.][]{Rodono74}. In addition to these studies, targeted spectroscopic campaigns have been used to study the spectral signature of flares \citep[e.g.][]{Kowalski13}. However, as these works are typically limited to nearby active M stars it is difficult to scale up the highly detailed observations from these studies to large sample sizes.

Wide-field exoplanet surveys have helped to partially overcome this limitation, by providing simultaneous observations of thousands of stars for long periods of time. Most notable for white-light flare studies was the \kepler\ space telescope, which provided continuous observations of over 150,000 target stars for four years in its primary mission \citep[][]{Borucki2010}. 
Stars were observed in either long or short cadence modes of 30 minutes and 1 minute respectively. 
Results from \kepler\ include the detection of ``superflares'' on solar-type stars \citep[][]{Maehara2012,Shibayama13} which before \kepler\ were rarely observed events \citep[][]{Schaefer_1989,Schaefer_2000}, studies of quasi-periodic pulsations \citep[][]{Balona15,Pugh2016} and improved understanding of flare occurrence rates \citep[e.g.][]{Hawley14}. This work is now being carried forward with the Transiting Exoplanet Survey Satellite  \citep[\tess;][]{Ricker15}, in particular with the two-minute short-cadence mode during the primary mission and more recently, the 20-second fast-cadence mode. The two minute cadence \tess\ observations have already been used to find flares down to late-M stars \citep[e.g.][]{Guenther20,Medina20} and to study the relation between flare occurrence and starspot phase \citep[][]{Doyle19, Feinstein20}. However, it should be noted that the short cadence mode observations of both \kepler\ and \tess\ represent only a small fraction of the total data collected \citep[e.g. approximately 0.3 per cent for the \kepler\ Q1 data;][]{Gilliland2011}. The vast majority of observations for both missions are instead at the 30-minute long-cadence mode, of pre-selected targets for \kepler\ and full-frame images for \tess. As the typical timescale of stellar flares is on the order of minutes to hours \citep[e.g.][]{Hawley14}, longer cadence modes can have the effect of smearing out their appearance in lightcurves and can result in inaccurate measurements of flare parameters \citep[][]{Yang18}. Along with this, substructure can be removed and short duration flares are missed entirely. Consequently, for precise measurements of flare parameters on a large sample of stars, high-cadence observations with a wide field of view are required.

Such observations have become possible in recent years with ground-based optical surveys such as the Next Generation Transit Survey (\NGTS). NGTS is a ground-based transiting exoplanet survey located at the ESO Paranal Observatory in Chile. \NGTS\ is formed of 12, 20\,cm f/2.8, optical telescopes operating with an exposure time of 10 seconds and total field of view of $\approx$ 96 square degrees \citep[][]{Wheatley18}. \NGTS\ is designed to be sensitive to the bright K and M stars (\textit{I}$\leq$16), so each telescope is fitted with a custom Andor iKon-L 936 camera, with a back-illuminated deep-depletion CCD. The bandpass is between 520--890\,nm. Unlike the \kepler\ and \tess\ short cadence modes, \NGTS\ observes without a set target list, meaning that high cadence lightcurves can be generated for all stars in our field of view. 

Results to date from searching for stellar flares with NGTS include 
G star superflares \citep[][]{Jackman18}, the detection of quasi-periodic pulsations from a flare on a pre-main sequence M star \citep[][]{JackmanQPP} and the 
first detection of a white-light flare from an L2.5 dwarf \citep[][]{Jackman19}. These are single flare events and it is clear that the wide field of view full-frame images of NGTS can also lend themselves to larger scale studies of flare parameters.

In this work we present a survey of stellar flares detected in the first data release of \NGTS\ (\NGTS\ DR1\footnote{\url{http://eso.org/rm/api/v1/public/releaseDescriptions/122}}). We describe our detection method and our method for measuring stellar and flare parameters, then proceed to present our results. We discuss the properties of the measured flare parameters, and compare with
previous results from \kepler\ and from the ground. We also discuss how our observations complement those from previous surveys. 

\section{Observations} \label{sec:FLARE_DETECT}
The \fieldnumber\ fields of data used in \NGTS\ DR1 span observation dates between a earliest date of \firstnight\ and a latest date of \lastnight. A full list of the fields used and information about them can be found in Tab.\,\ref{tab:field_info}. Throughout this work we use the standard \NGTS\ apertures which are generated for all stars within our field of view with \textit{I}<16, as described by
\citet{Wheatley18}.

\begin{table*}
	\centering
	\begin{tabular}[width=\textwidth]{|c|c|c|c|c|}
    \hline
    Field & First Night & Last Night & Nights &$\mathrm{N_{Sources}}$\tabularnewline\hline
	NG0522-2518 & 2015 Sep 21 & 2016 May 03 & 149 & 7435 \tabularnewline
	NG0531-0826 & 2015 Sep 23 & 2016 Apr 20 & 138 & 10178 \tabularnewline
	NG0549-3345 & 2016 Aug 06 & 2017 Jan 31 & 126 & 9074 \tabularnewline
	NG0612-2518 & 2015 Sep 21 & 2016 May 14 & 157 & 13452 \tabularnewline
	NG0618-6441 & 2015 Sep 21 & 2016 May 24 & 159 & 11689 \tabularnewline
	NG1135-2518 & 2015 Nov 26 & 2016 Aug 03 & 126 & 6407 \tabularnewline
	NG1213-3633 & 2015 Nov 28 & 2016 Aug 04 & 116 & 10625 \tabularnewline
	NG1315-2807 & 2016 Jan 05 & 2016 Aug 30 & 105 & 8176 \tabularnewline
	NG1340-3345 & 2016 Apr 18 & 2016 Aug 31 & 67 & 11433 \tabularnewline
	NG1416-2518 & 2016 Jan 05 & 2016 Sep 11 & 100 & 8589 \tabularnewline
	NG1421+0000 & 2016 Jan 05 & 2016 Aug 31 & 100 & 4279 \tabularnewline
	NG1428-2518 & 2016 Jan 05 & 2016 Sep 15 & 109 & 8840 \tabularnewline
	NG1444+0537 & 2016 Jan 13 & 2016 Sep 03 & 99 & 4181 \tabularnewline
	NG2025-1941 & 2016 Apr 20 & 2016 Nov 28 & 89 & 12862 \tabularnewline
	NG2028-2518 & 2016 Apr 20 & 2016 Nov 30 & 121 & 11496 \tabularnewline
	NG2047-0248 & 2016 Apr 20 & 2016 Nov 28 & 122 & 12565 \tabularnewline
	NG2058-0248 & 2016 Apr 20 & 2016 Dec 01 & 126 & 10491 \tabularnewline
	NG2126-1652 & 2016 Apr 20 & 2016 Dec 12 & 126 & 6766 \tabularnewline
	NG2132+0248 & 2016 Apr 20 & 2016 Dec 02 & 119 & 8537 \tabularnewline
	NG2142+0826 & 2016 Apr 20 & 2016 Dec 02 & 121 & 9077 \tabularnewline
	NG2145-3345 & 2016 Apr 20 & 2016 Sep 02 & 60 & 5615 \tabularnewline
	NG2150-3922 & 2016 Apr 20 & 2016 Dec 21 & 145 & 5905 \tabularnewline
	NG2331-3922 & 2016 May 03 & 2017 Jan 10 & 129 & 3433 \tabularnewline
	NG2346-3633 & 2016 May 03 & 2017 Jan 10 & 150 & 3495 \tabularnewline
	\hline
	\end{tabular}
    \caption{NGTS fields used in this work, along with the number of nights and sources observed per field.}\label{tab:field_info} 
\end{table*}

\subsection{Characterising Stars} \label{sec:stellar_characterisation}
The NGTS pipeline is designed to cross match all input catalogues with a range of catalogues to provide broadband photometric information for all observed sources. This pipeline currently performs matching with \gaia\ DR2 \citep{Gaia2016,GaiaDR2}, 2MASS \citep{2MASS_2006}, APASS \citep{APASS_14}, UCAC4 \citep{UCAC4_13}, RAVE \citep{RAVE17}, GALEX \citep{GALEX_05} and ALLWISE \citep{ALLWISE2014}. Along with this, distance information is obtained from \Gaia\ DR2 using the catalogue of \citet{BailerJones18}. 

To obtain the stellar parameters for each source, flaring or non-flaring, we used the revised \tess\ Input Catalog \citep[TIC;][]{Stassun19}. The TIC is a collection of cross matched catalogue photometry and stellar parameters for sources across the whole sky. Importantly for our analysis the TIC provides effective temperatures and radii (using \Gaia\ DR2 parallaxes) for effectively all sources in our input catalogue. For each NGTS field we matched the results of our cross-matching with the TIC using the \Gaia\ DR2 source ID, or the 2MASS source ID where \Gaia\ DR2 did not provide a match. For this work we have used the stellar parameters provided in the TIC. We found that many radii values in the TIC did not have an associated uncertainty. In these cases we assumed a 10 per cent uncertainty on the quoted radius. We associated stars with a spectral type using their TIC effective temperature and the effective temperature-spectral type tables from \citet{Pecaut13} \footnote{\url{http://www.pas.rochester.edu/~emamajek/EEM_dwarf_UBVIJHK_colors_Teff.txt}}.

\begin{figure}
	\includegraphics[trim={1cm 0 0 0}, width=\columnwidth]{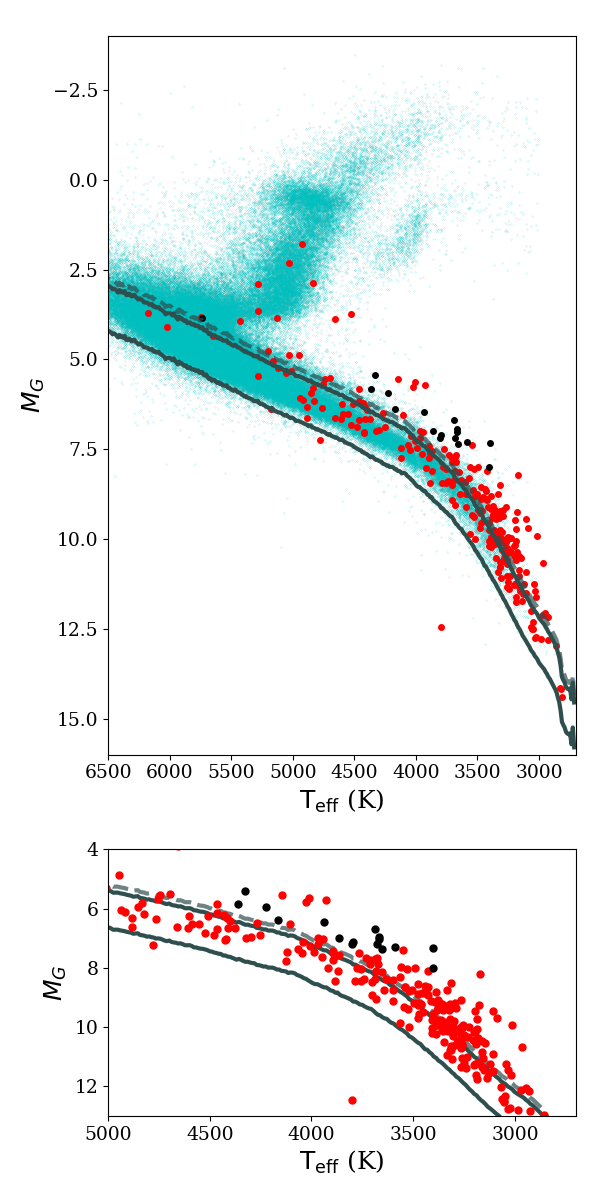}
    \caption{Hertzsprung-Russell diagram of the stars observed with \NGTS\ and crossmatched with \Gaia\ DR2 and the TIC v8 catalogue, using the TIC v8 effective temperatures. All stars from the NGTS fields are shown in cyan, while the flaring stars are depicted with red circles. The black circles are sources associated with Orion, using the best fitting SEDs from \citet{Jackman20}. The dark grey lines indicate the upper and lower main sequence limits specified in Sect.\,\ref{sec:stellar_characterisation} and the dashed line is the pre-main sequence cut off. The bottom plot shows a zoomed section with only the flare stars and main sequence limits displayed.}
    \label{fig:HR_diagram}
\end{figure}

At times in our analysis it was useful to separate main sequence and non-main sequence stars. To do this we used a Hertzsprung-Russell diagram, as shown in Fig.\,\ref{fig:HR_diagram}. We calculated a median absolute magnitude curve using all astrometrically and photometrically ``clean'' \Gaia\ sources \citep[i.e. those passing the filters of][]{Arenou18} within \rewrite{100} pc \citep[e.g.][]{Gaia_HR}. This curve was measured as a function of the \rewrite{TIC v8 effective temperature.} We then defined empirical upper and lower limits in absolute \gaia\ G magnitude as $\Delta M_{G}$=+0.65 and $\Delta M_{G}$=-0.55 from the median curve. We used $\Delta M_{G}$=-0.55 as a lower limit to remove the majority of equal mass binaries and pre-main sequence stars and ensure our main sequence sample is dominated by field age stars. It has previously been noted that binary systems pervade throughout the entirety of the HR diagram \citep[e.g.][]{Gaia_variable_18,Winters19}, making it difficult to choose a limit which will remove them without removing large numbers of legitimate isolated main sequence stars. To analyse pre-main sequence stars alone we use a separate limit of $\Delta M_{G}$=-0.753. This is something we discuss further in Sect.\,\ref{sec:discussion}.
The $\Delta M_{G}$=+0.65 upper limit is chosen empirically in order to remove cool subdwarfs while keeping most main sequence stars. Our chosen limits and how they compare to the sample of stars observed with NGTS can be seen in Fig.\,\ref{fig:HR_diagram}.

From this analysis we identified 35612 K0V-K8V stars and 4135 M0-M5 stars that are consistent with our definition of being isolated and on the main sequence. Of the K stars, 27694 are K0V-K4V and 7918 are K5V-K8V. Of the M stars, 3757 are M0V-M2V and 378 are M3V-M5V.

\subsection{Flare Detection} \label{sec:FLARE DETECT}
To search for stellar flares in each \NGTS\ field, we started with the raw lightcurves. We detrended these lightcurves using a custom version of the \NGTS\ \sysrem\ process \citep[][]{Wheatley18}. During this process we applied an extra filter to remove images with excess variance (e.g. clouds across the image). 

This filter was defined empirically using the airmass, $X$, and $\sigma_{excess}$ values calculated by the NGTS pipeline and \sysrem. $\sigma_{excess}$ is a \edit{measure of the per-image variance introduced by external factors (e.g. clouds) and is expressed in units of magnitude \citep[][]{Collier06}}. $\sigma_{excess}$ used by the \sysrem\ pipeline to down-weight images where there are lots of outlying points across multiple stars. We removed all observations with $\sigma_{excess}>$0.02 for airmasses below or equal to 1.2. For airmasses above 1.2, we removed all points with $\sigma_{excess}>$0.008 + 0.01$X$. This filter limited the effects of clouds without penalising images at higher airmasses which tend to have inherently higher variance, along with not removing outlying data points due to large flares. 

When searching for flares in the detrended lightcurve for each star, we split the lightcurve into individual nights. The majority of reported stellar flare durations are on the order of minutes to hours \citep[e.g.][]{Hawley14,Maehara16}, with only the largest exceeding this. This means almost all of the observed flares in the NGTS lightcurves should be confined within single nights. Splitting the data into individual nights also reduces the effect of longer-timescale variations.

For each night we performed a two step flare detection method.  This is the same method as was used by \citet{Jackman18,JackmanQPP,Jackman20}. For clarity, we outline the method here as well. Firstly, we look for discontinuities within the night, by searching for regions where there are at least 3 consecutive points above 6 median absolute deviations (\mad) from the median of the night. \edit{The \mad\ is a measure of the variance in a sample and is calculated as the median of the absolute deviations from the sample median.} We use \mad\ over the standard deviation as it is a robust parameter, so is less likely to be altered by the presence of single point outliers, such as cosmic rays. Regions with at least 3 consecutive points above 6 \mad\ are flagged as flare candidates. As a second independent step, we compare the median of the night against the median of the entire lightcurve for that object. If the median of the night is at least five \mad\ above the median lightcurve flux, then it is flagged. We do this as this is indicative of a long duration and high amplitude flare which may not be fully observed within one night \citep[such as the flare discussed in][that dominated the entire night of observation]{JackmanQPP}. Once this automated flagging procedure was complete, we inspected each flagged night visually and removed any false positives. Such false positives can be due to astrophysical events (e.g. short-period variable stars), human-made (e.g. satellites passing through our aperture) or systematics (e.g. overlapping apertures).

Events that passed our checks and were classified as flares were also inspected for signs of complexity. Flares in previous works have been classified as ``simple'' and ``complex''. Simple flares typically follow the classical fast rise, exponential decay shape \citep[e.g.][]{Moffett74}. Complex flares deviate from this template, showing multiple peaks \citep[e.g.][]{Davenport2014}, oscillations \citep[e.g.][]{Anfinogentov13} or other substructure which might complicate their analysis. \edit{Complex flares may also arise from the superposition of two simultaneous independent flare events.} Each flare lightcurve was visually inspected for signs of multiple peaks and oscillations, which if present resulted in a complex flag being applied to that event.

\subsubsection{Centroiding}
We have performed a centroiding analysis for each of our flare candidates. This was done in order to confirm the true source of each flaring event. If a star has a flaring companion nearby, we would expect the centroid position to shift relative to the centre of the aperture for the duration of the flare. By using this shift we can identify cases where the detected flare is not coming from the expected source and where possible assign the correct \gaia\ source to the flaring star \citep[e.g.][]{Gunther17}. 

Flux from nearby stars within an aperture will dilute the observed lightcurves. This will result in reduced measured flare amplitudes, hindering their detection. The reduced flare amplitudes will also result in decreased measured flare energies. Close stars may also affect the quality of the stellar characterisation discussed in Sect.\,\ref{sec:stellar_characterisation}. Accurate stellar temperatures and radii are essential for calculating flare energies. Therefore, for each NGTS target, we searched for neighbouring \gaia\ DR2 or 2MASS sources within 15\arcsec, the radius of an NGTS aperture. Stars with nearby neighbours which could cause dilution are flagged during our analysis and only used in sections which do not require accurate stellar characterisation.

\subsection{Flare Amplitude}
\label{sec:flare_amplitude_method}
We calculated our flare amplitudes using $\frac{\Delta F}{F_{q}}$. $\Delta F$ is the change in observed flux due to the flare, while $F_{q}$ is the quiescent flux. An amplitude of 1 indicates that the flare is emitting as much flux in the NGTS filter as the quiescent star. We measured the quiescent flux for each flare event by taking the median of the 20 minutes preceding the flare. If the flare dominated the entire night, we used the median value of the previous night. We confirmed the calculated quiescent flux and amplitude for each flare through visual inspection. For events where the quiescent flux calculation was incorrect (e.g due to including another flare), we manually corrected the values.

\subsection{Flare Energy} \label{sec:flare_energy}
To calculate our flare energies we have followed the method of \citet{Shibayama13}, which assumes that both the star and the flare behave as blackbody radiators in order to estimate the flare luminosity. \edit{The flare energy is calculated from the time integral of the luminosity over the total flare duration, which we discuss in Sect.\,\ref{sec:flare_duration}.} In this method, it is assumed that the flare spectrum can be broadly approximated as a blackbody with a temperature of 9000$\pm$500\,K. This assumption is based on spectroscopic and multi-colour photometric observations \citep{Hawley1992, Kowalski13}. However, while these studies have shown that the optical spectrum of many flares can be broadly approximated with a 9000-10,000\,K blackbody, the flare temperature has been observed to change both between and within individual flare events to values outside of this range \citep[e.g.][]{Howard20temp}. We note that while this limits the use of a single temperature model, we currently cannot predict these changes in the flare temperature, preventing the use of more complex models.  
When calculating the energy we assumed the underlying variation of the stellar flux is approximately linear during the flare. For each flare a baseline of the stellar flux was estimated by linearly interpolating the light curve between just before and just after the flare, as determined by visual inspection. To calculate the uncertainty on the flare energy we performed our flare energy calculation using a Monte Carlo process of 10,000 runs. This enabled us to propagate the uncertainties in the stellar effective temperature, radius and the flare temperature to the final measured flare energy.

\subsection{Flare Occurrence Rate}  \label{sec:flare_occ_rate_123456}
Previous studies \citep[e.g.][]{Lacy76,Hilton11,Hawley14} have shown that flares occur with a power law distribution in energy. This is typically written as
\begin{equation} \label{eq:edit_powlaw}
    dN(E) = k E^{-\alpha}dE ,
\end{equation}
where $N$ is the number of flares which occur in a given duration, $E$ is the flare energy, $k$ is a constant of proportionality and $\alpha$ is the power law index. The flare frequency distribution, the number of flares per unit time with an energy of $E_{f}$ or greater, typically denoted by $\nu$ is obtained by integrating Eq.\,\ref{eq:edit_powlaw} from $E_{f}$ to infinity, resulting in 
\begin{equation} \label{eq:ffd}
    \log{\nu} = C + \beta \log{E_{f}}
\end{equation}
where $C = \log\big(\frac{k}{1-\alpha}\big)$ and $\beta= 1-\alpha$ \citep[][]{Hawley14}. 

The value of $\alpha$ dictates how often high energy flares occur relative to those of lower energy and has been observed to vary across different spectral types. For example, the Sun has $\alpha\approx$1.75 \citep[e.g.][]{Crosby93,Shimizu95,Aschwanden2000}, while ultracool dwarfs have had measured $\alpha$ values from 1.4 to 2 \citep[e.g.][]{Paudel18}. 

We have used the NGTS data to study flare occurrence rates and where possible measure $k$ and $\alpha$ (and in turn $C$ and $\beta$) for subsets of our sample. In particular we are aiming to study how the average occurrence rate changes as a function of spectral type, by combining observations of flaring and non-flaring stars in our sample. We have done this for our main sequence and pre-main sequence samples.

\subsubsection{Fitting the Average Flare Occurrence Rate} \label{sec:flare_freq3}
When measuring the average occurrence rate in each subset it was important to ensure the sample was complete across both flaring and non-flaring stars. Incompleteness in measured occurrence rates typically manifests itself as a flattening at low energies (in log-log space), where the measured distribution no longer matches the expected power law behaviour. At this turnover, the chosen detection method is no longer sensitive to all flares and fewer are detected. Many studies have gotten around this incompleteness by fitting only to data with flare energies above some limit above the turnover \citep[e.g.][]{Hawley14}. A common way of determining this energy limit is to inject flares of known amplitude, duration and energy and to measure the fraction of flares recovered as a function of energy, $R(E)$. The limiting energy can then be the minimum value recovered \citep[e.g.][]{Paudel18}, or the energy at which some percentage of flares are recovered \citep[e.g. 68 per cent;][]{Davenport16,Jackman20}. This has the advantage of obtaining a sample of flares which are mostly free of detection-sensitivity related effects, but will result in smaller samples due to the rarer nature of higher energy flares. 

Other studies have used the measured flare recovery rate to correct the observed flare frequency distribution, to which a power law is then fit \citep[e.g.][]{Ilin19,Kovari20}. This has the benefit of keeping lower energy flares in the fitted distribution while still accounting for the detection efficiency. However, care must be taken with this method, as the measured flare recovery rate from flare injection is often for individual energies and not for all energies above some value. Consequently, the recovery fraction as a function of energy cannot be directly applied to Eq.\,\ref{eq:ffd}. Instead, it must be applied to Eq.\,\ref{eq:edit_powlaw} and the new ``observed'' differential distribution of flare energy should be integrated from some energy $E_{f}$ to infinity. We do this for a generalised manner in Appendix\,\ref{sec:cum_freq_derivation} to obtain the observed flare frequency distribution, accounting for detection sensitivity effects
\begin{equation} \label{eq:obs_ffd}
    N(E>E_{f}) = \frac{k}{\alpha-1} \Bigg(R(E_{f})E_{f}^{-\alpha + 1} +
    \int_{E_{f}}^{E_{max}}R'(E)E^{-\alpha+1}dE\Bigg)
\end{equation}
where $R'(E)$ is the differentiated flare recovery fraction and $E_{max}$ is the energy at which the recovery fraction saturates ($R'(E)$=0). At low energies this will give the observed turnover effect, while at high energies it will be equivalent to Eq.\,\ref{eq:ffd}. \citet{Medina20} recently presented a similar analysis specifically for \tess\ observations, which highlights the necessity of taking the recovery fraction into account in flare studies.

\subsection{Flare Duration} \label{sec:flare_duration}
\edit{In our analysis we calculated both the full flare duration and the $t_{1/2}$ timescale, defined as the time where the flare is above half of its maximum flux \citep[e.g.][]{Davenport2014}. The full flare duration was measured by eye for each flare and was used in our energy calculation. We have used the $t_{1/2}$ timescale when comparing our flare durations with the flare amplitude and energy and with other studies. } 
\edit{We have chosen the $t_{1/2}$ timescale for this analysis over the e-folding time as used in e.g. \citet{Maehara16} or the full flare duration as it uses the brightest, and best resolved, part of the flare to provide a precise measurement of the timescale of both the flare rise and decay. }
We used the 13 second cadence NGTS data when determining the $t_{1/2}$ value. Each flare was visually inspected to make sure the measured $t_{1/2}$ values were sensible (e.g. not thrown off due to outliers). Some flares are known to exhibit multiple peaks, particularly during the flare decay, where the flux may rise back above the threshold value. When this occurs we measure the first and last point where the lightcurve drops below half of the flux maximum. This way we can obtain a minimum and maximum timescale for the more complex flares.

\subsection{Completeness} \label{sec:completeness}
An important factor to consider when analysing the distribution of flare stars is the completeness of our sample. This is particularly vital when comparing our sample of flaring stars to non-flaring stars, such as when probing the fraction of stars to show a flare of a given energy. \edit{For fainter stars we expect our detection threshold to be at higher flare amplitudes than for brighter stars, due to the increased relative variance in their lightcurves. This will result in us only detecting higher energy flare events for a faint star than for a bright star of the same spectral type.}

To parameterise the completeness of our sample, we performed flare injection and retrieval tests. These tests were performed on all K and M stars in our sample, whether or not they are on the main sequence. 
We followed the method outlined by \citet{Jackman20}, which itself was based on the flare injection and retrieval method used by \citet{Davenport16}. We injected 50 artificial flares into each night of every K and M star lightcurve in our sample. For the average lightcurve in our sample, this resulted in roughly 6000 flares being injected and tested.

These flares were created using the \citet{Davenport2014} empirical flare model. Amplitudes for the artificially generated flares were chosen randomly from a uniform distribution between 0.01 and 4 times the lightcurve median. The \thalf\ durations were chosen randomly from a uniform distribution between 30 seconds and 70 minutes. Every flare had its energy calculated using the method given in Sect.\,\ref{sec:flare_energy}, using the TIC v8 effective temperatures, radii and associated uncertainties from Sect.\,\ref{sec:stellar_characterisation}. 

We then used our detection method from Sect.\,\ref{sec:FLARE DETECT} with our lightcurves with injected flares, to see how many flares we could retrieve. If an injected flare was flagged as a candidate by our detection method, it was given a flag of 1. If it was not detected, it was given a flag of 0 for that star. The end result for each star was a catalogue of flares, with known energies, and a corresponding array for whether each flare, if it had occurred on that star, would have been detected in our survey. We calculated how many flares were retrieved as a function of energy for each star. We used 20 logarithmically spaced bins in energy which, following \citet{Davenport16}, were smoothed using a Wiener filter of three bins. The flare energies and recovery fractions were then recorded for later use.
\begin{figure*}
	\includegraphics[width=\textwidth]{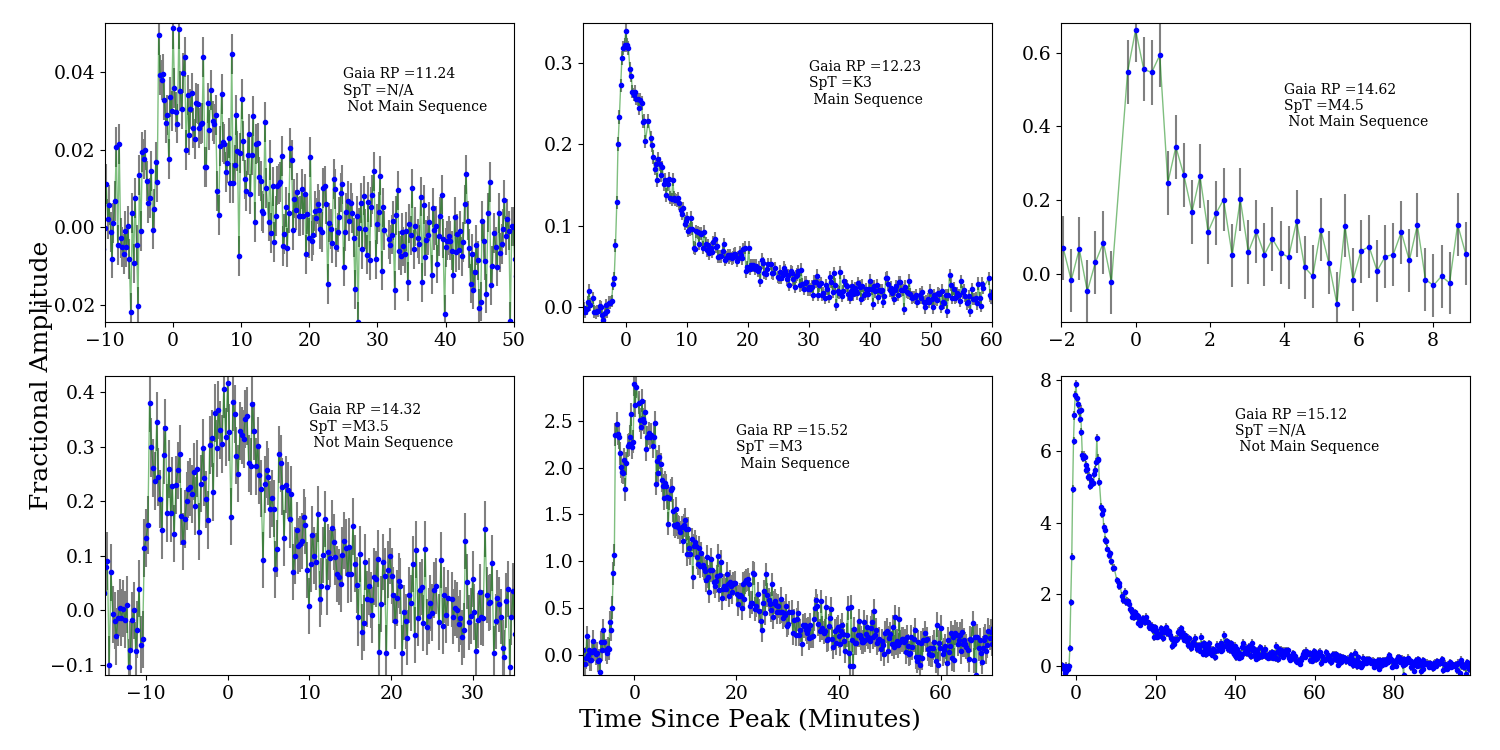}
    \caption{A selection of flares detected in the first data release from NGTS. The blue circles are individual NGTS data points, at the standard 13 second cadence. The green line joins these points to aid the eye. The bottom row shows examples of complex flares. Inset in each panel is the \gaia\ RP magnitude, spectral type used in our analysis and whether the source lies on the main sequence. In this plot, all sources marked as not being on the main sequence are all above the main sequence. Note the short duration of the flare on the top right, which would not have been resolved in either \kepler\ 1 minute or \tess\ 2 minute cadence modes.}
    \label{fig:flare_selection}
\end{figure*}

\subsection{Stellar Rotation}\label{sec:rotation}
Many active stars show periodic modulation in their lightcurves. This is due to the presence of dark starspots on their surface which appear and disappear from view as the star rotates \citep[e.g.][]{Maehara17,Doyle18}. We searched the lightcurves of all flare stars for signs of periodic modulation consistent with rotation. To measure rotation periods we used the \textsc{astropy} package LombScargle \citep{Astropy13}. We initially masked any identified flares in our lightcurves to remove their effect on the measured power spectrum. We then constructed a frequency grid formed of 2000 periods spaced linearly between 13 seconds and 1 day, then 20,000 periods spaced linearly between 1 and 30 days. For each flare star lightcurve we used the full time resolution data to calculate a Generalised Lomb Scargle Periodogram, which returns normalised power values between 0 and 1 (corresponding to a perfect straight line and a perfect sine wave respectively). To determine the best fitting period of our sample we used a similar method as \citet{Oelkers18}. Firstly we remove the effects of aliases, by rejecting periods within five per cent of single day aliases and those due to the lunar month ($29\pm3$ days) and ignoring the associated signals in our analysis. This is done iteratively for each star until the best remaining period is not due to these aliases. We then checked the remaining periods to see whether their harmonics fall into any of our identified alias regions. We checked 0.5P, 1.5P, 2P and 3P for this and rejected any period which fails this test. Again, this is done iteratively until the best remaining period does not fail this test. Once we were satisfied the best period was not due to a systematic alias, we checked whether its normalised power was above 0.1 \citep[][]{Oelkers18}. This was done to remove spurious periods with weak powers in the periodogram and to keep only those where we were confident a true astrophysical signal was present.

\section{Results}
From searching the 196600 stars in \fieldnumber\ fields in NGTS DR1 we have identified \flarenumber\ flares from \starnumber\ stars. A selection of the detected flares is shown in Fig.\,\ref{fig:flare_selection}. Those on the main sequence have spectral types between F8V and M6V. We note that the cut off at spectral type of M6V is due to the limiting magnitude of I=16 for the standard NGTS lightcurves.

\subsection{Distribution} \label{sec:distribution}

To probe the distribution of stars in our sample we use the results of our crossmatching pipeline and TIC v8 from Sect.\,\ref{sec:stellar_characterisation}. For sources with centroid shifts, we have manually corrected the crossmatching to give the correct \gaia\ DR2 source where possible.

\subsubsection{Centroiding}
Through centroiding we found that 11 per cent of flares in our final sample were due to a star not at the centre of the \NGTS\ aperture. This is similar to the 10 per cent of ``false sources'' identified by \citet{Shibayama13} when analysing \kepler\ pixel level data for G dwarf superflares. As \kepler\ and \NGTS\ operate with pixel scales of 4 and 5 arcseconds respectively, the similar level of blending is expected. These percentages highlight the importance of centroiding for accurate flare analysis. This is particularly relevant 
for \tess, which has a pixel scale of 21 arcseconds per pixel \citep[][]{Ricker14}. We therefore expect \tess\ to suffer from a higher fraction of blended flare sources, making careful centroid analysis vital. 

An alternative solution which has been used previously \citep[e.g.][]{Shibayama13,Yang17} is to choose isolated stars, avoiding the need for centroiding of every source. While this has the benefit of providing a cleaner input catalogue, it can limit the number of observed systems and remove potentially interesting events \citep[e.g. the QPP exhibiting flare from][]{JackmanQPP}. Along with this, it will remove serendipitous observations of flares from stars too faint to see in quiescence which may be located close to a brighter source.

\subsubsection{Galactic Distribution} \label{sec:galactic_distribution}
Figure \ref{fig:HR_diagram} shows the position of our flaring sources on a Hertzsprung-Russell (HR) diagram. We calculate that 56 per cent of stars observed with NGTS are consistent with being an isolated main sequence star, according to the criteria from Sect.\,\ref{sec:stellar_characterisation}. We calculate that approximately three per cent of stars observed with NGTS are pre-main sequence. Of the flare sources, 32 per cent of our sample are consistent with being an isolated main sequence star according to the criteria from Sect.\,\ref{sec:stellar_characterisation}. 63 per cent are consistent with being pre-main sequence and the remaining 5 per cent either reside in the gap between main and pre-main sequence sources (making them ``non-main sequence'', a mixture of young and binary systems that cannot be easily distinguished) or are on the subgiant branch in Fig\,\ref{fig:HR_diagram}. 

When investigating the distance distribution of our flare stars, we identified an apparent excess in the number of observed flaring stars between 350 and 400 pc. These additional objects come from a single field (NG0531-0826) and are associated with the Orion Complex. The Orion Complex is vast collection of multiple clusters with many stellar populations \citep[e.g.][]{Kounkel18}. This sample of pre-main sequence stars was analysed separately in \citet{Jackman20} and underwent SED fitting to account for the specific extinction and reddening properties from dust in the Orion region. We use the results of this SED fitting for these stars in the rest of this work. 
For reference these stars are depicted alongside the rest of the flare survey in Fig.\,\ref{fig:HR_diagram}. 

\rewrite{To estimate the age of the remaining pre-main sequence stars, we compared the pre-main sequence cut-off defined in Sect.\,\ref{sec:stellar_characterisation} (-0.753 magnitudes above the fiducial median absolute Gaia G magnitude) to the PARSEC isochronal models \citep[][]{Bressan12}. We found that, from comparison to the PARSEC isochrones, our imposed cut off in absolute \gaia\ G magnitude results in a mass-dependence in the limiting age. This is because the time taken to reach the zero age main sequence is mass dependent, taking longer for lower mass stars \citep[][]{Hayashi61,Bressan12}. We found that late K stars (K6-K8) designated as pre-main sequence are younger than approximately 15 Myr. For M0-M2 and M3-M5 stars this increases to 30 Myr and 40 Myr respectively.}

\begin{figure*}
	\includegraphics[width=\textwidth]{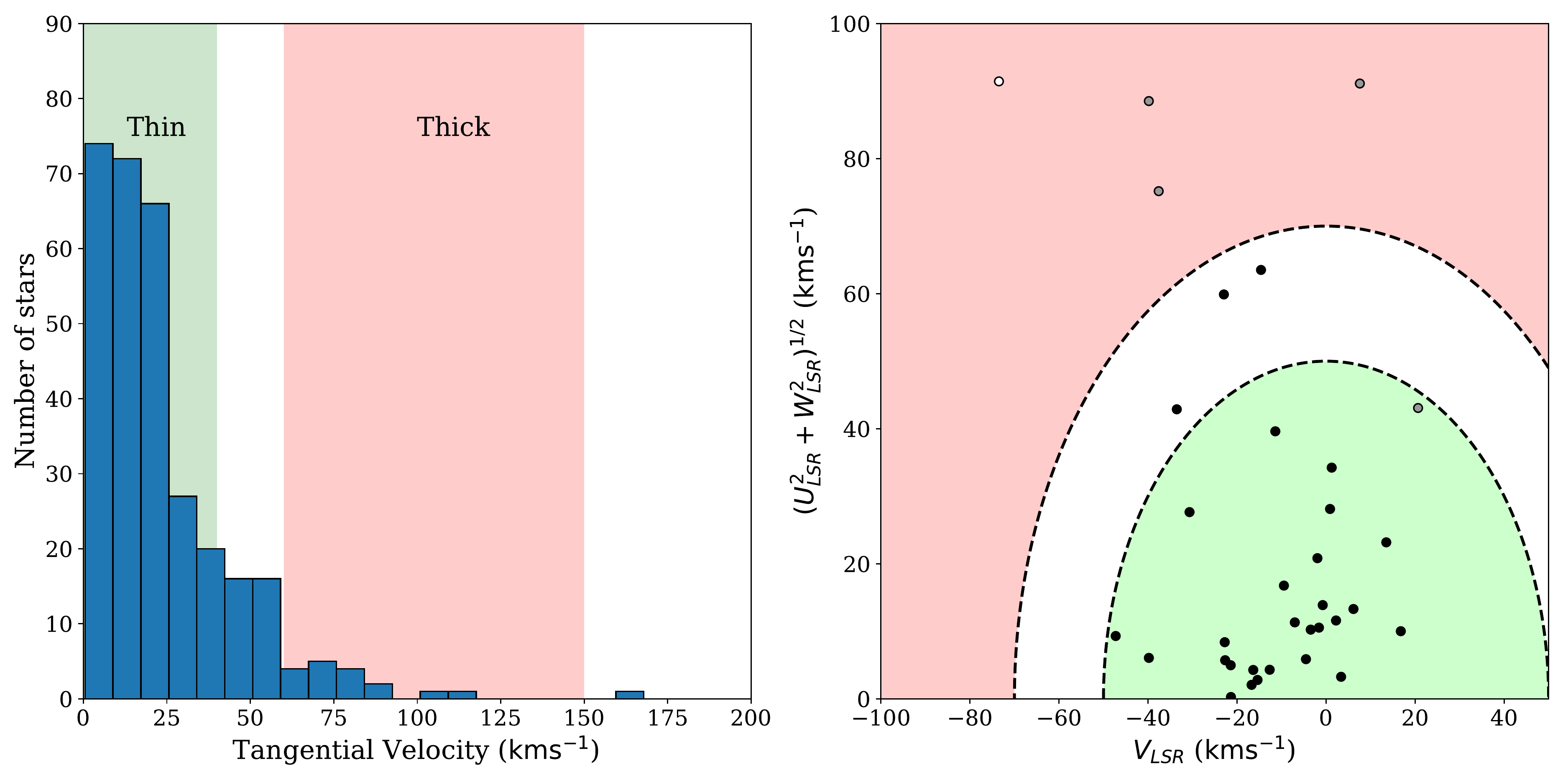}
    \caption{Left: The tangential velocity, $V_{tan}$, distribution for flare stars with \gaia\ DR2 matches. The green and red shaded regions represent approximations for the thin and thick disc respectively \citep[][]{Gaia_HR}. Right: Toomre diagram for flare stars with radial velocity information. Black, grey and white dots indicate thin disc stars, thick disc candidates and thick disc stars respectively, according to the probability classes from \citet{Bensby03}. The green and red regions represent the expected $v_{total}$ distributions for the thin and thick disc from \citep[][]{Bensby14}. We can see that the majority of our flare stars are consistent with the thin disc distribution.}
    \label{fig:galactic_kinematics}
\end{figure*}

To probe the Galactic distribution of our flare stars we have calculated their kinematics. For all stars with distance and proper motion information from a \gaia\ DR2 match we calculate the tangential velocity \vtan. \gaia\ DR2 also provides radial velocity measurements for selected stars brighter than $G_{RVS}=12$ \citep[][]{Katz18}. Consequently, where a measurement of the radial velocity is available from our matching with \gaia\ or RAVE, we also calculate the velocities relative to the LSR \U,\V,\W.

We are able to calculate full 6D coordinates for 34 stars in our sample and tangential velocities for 309 stars. The distribution of the tangential velocities and kinematics for our sample is shown in Fig.\,\ref{fig:galactic_kinematics}. We use the kinematic selection criteria of \cite{Bensby03} to calculate the thick-disc-thin-disc (TD/D) relative probability for each of the stars with a full 6D solution. We find that 29 out of the 34 stars (85 per cent) with 6D kinematics in our sample are most likely to be thin disc members (P(thick) < 0.1x P(thin)). Of the remaining stars, four are thick disc candidates (P(thick)>0.1x P(thin)) and one is a most likely a member of the thick disc (P(thick)> 10x P(thin)). This likely thick disc source is TYC 5181-713-1, a bright star located at the turn off to the subgiant branch in Fig.\,\ref{fig:HR_diagram}. We believe this source is a subgiant, something we discuss further in Sect.\,\ref{sec:subgiant}. 

When taking the \vtan\ sample separately we again find the majority of stars reside within the regions expected for a thin disc. Thin disc members are typically characterised as younger than those of the thick disc \citep[e.g.][]{Bensby04}. Consequently, we might expect a greater number of flare stars from the thin disc, due to their relative youth (and increased activity).

In Fig.\,\ref{fig:galactic_kinematics} we can see that two stars have \vtan\ greater than 150 km/s, the upper velocity limit for a thick disc. Investigating these stars further reveals that one is LHS 2802, previously classified by \citet{Reid2005} as an sdM4.5 subdwarf. This source can also be seen as an outlier from the main sequence in Fig.\,\ref{fig:HR_diagram} and is analysed further in Jackman et al. (in prep). The other is a subgiant candidate NGTS J060634.4-250020. We discuss subgiant candidates in our sample in Sect.\,\ref{sec:subgiant}. While these stars have higher velocities than expected for the thick disc, they are not high enough to be within the realm of possible halo stars \cite[\vtan>200 km/s][]{Gaia_HR}.

\subsection{Flare Amplitude} \label{sec:flare_amplitude}
\begin{figure}
	\includegraphics[width=\columnwidth]{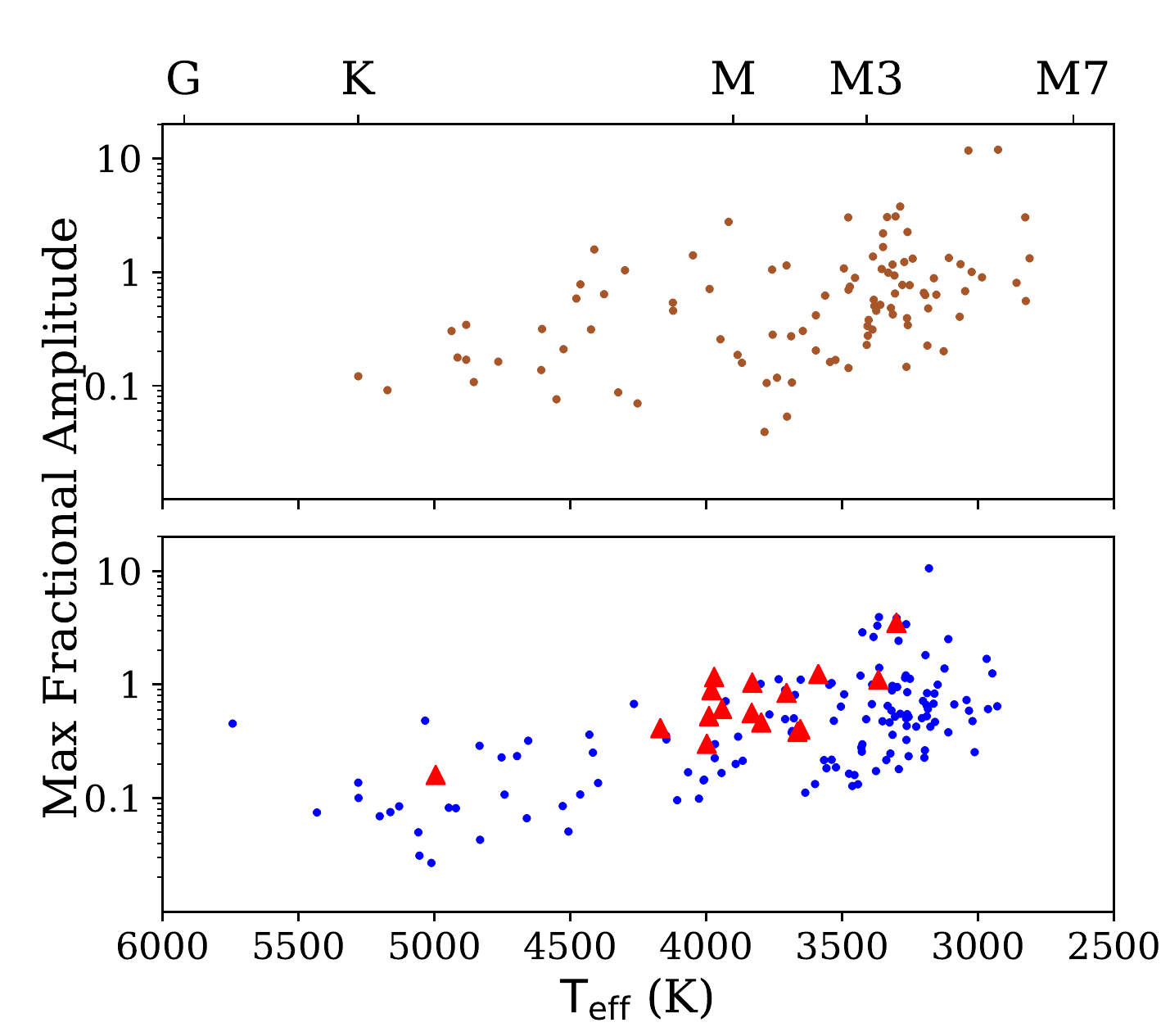}
    \caption{Maximum observed fractional flare amplitude against effective temperature. Top: Main sequence isolated stars. Bottom: Non-main sequence stars. Stars associated with Orion are shown in the bottom panel as red triangles. Note how M stars of all ages can flare with amplitudes equivalent to 10 times the quiescent flux.}
    \label{fig:frac_amp_plot}
\end{figure}

We have calculated the amplitude for all flares in our sample, using the method described in Sect.\,\ref{sec:flare_amplitude_method}. In the case of blended sources this amplitude corresponds to a lower limit. Our measured fractional flare amplitudes range between 0.027 and 12 ($\Delta\,m_{NGTS}=-2.8$), with our highest amplitude flare coming from an M6 star in our main sequence sample. Figure\,\ref{fig:frac_amp_plot} shows the distribution of maximum observed flare amplitudes with effective temperature for the non-blended sources in our sample. As we progress to cooler stars in our sample we observe the expected increase in maximum flare amplitude due to the changing ``contrast ratio'' between the flare blackbody and stellar photosphere within the NGTS filter.

Splitting our sample into main and non-main sequence stars (which includes both binaries and pre-main sequence stars) reveals that some sources in our main sequence sample exhibit higher maximum flare amplitudes than their non-main sequence counterparts of similar temperature. While these sources appear to be consistent with isolated main sequence sources in the HR diagram, they may be binaries where one component has been spun up through tidal interaction and is magnetically active \citep[e.g.][]{JackmanBD}. Alternatively, the stars non-main sequence include binary stars. The extra flux from the second star in these systems will dilute the observed signal and result in a smaller measured flare amplitude. 
Spectroscopic follow up of these sources will help determine whether this is the case. Of the non-main sequence stars, we can see that the Orion-associated sample from \citet{Jackman20} appear to flare with the largest maximum amplitudes, which is to be expected for their very young ages.

\subsection{Flare Energy}
\rewrite{Using the method outlined in Sect.\,\ref{sec:flare_energy} we have measured the bolometric energy for the 428 flares from stars with available effective temperatures and radii from TIC v8. These energies range between \minflareenergy\ and \maxflareenergy. The stars with calculated flare energies have effective temperatures between 6170\,K and 2800\,K, equivalent to between F8 and M6 spectral type.}

\subsubsection{Flare Occurrence Rate} \label{sec:flare_frequency2}

\begin{figure}
	\includegraphics[width=\columnwidth]{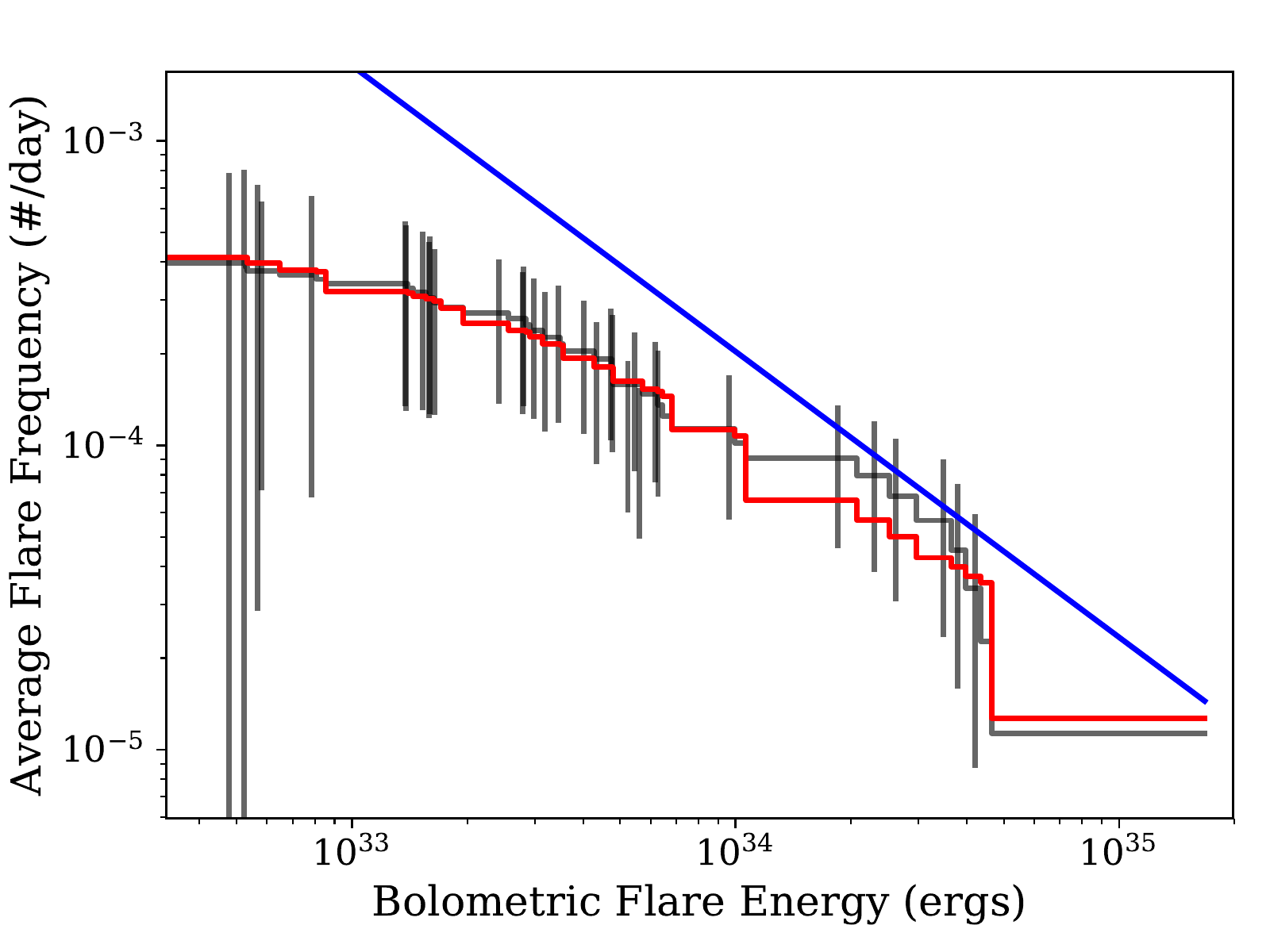}
    \caption{The results of our fitting of the average flare occurrence rate for M0V-M2V stars. The grey line show the observed occurrence rate while the red line is our best fitting model from using Eq.\ref{eq:edit_powlaw}. The blue line is the power law distribution required to give the red line and can be thought of as the intrinsic flare occurrence rate.}
    \label{fig:early_m_example}
\end{figure}

 \begin{figure}
     \centering
     \includegraphics[width=\columnwidth]{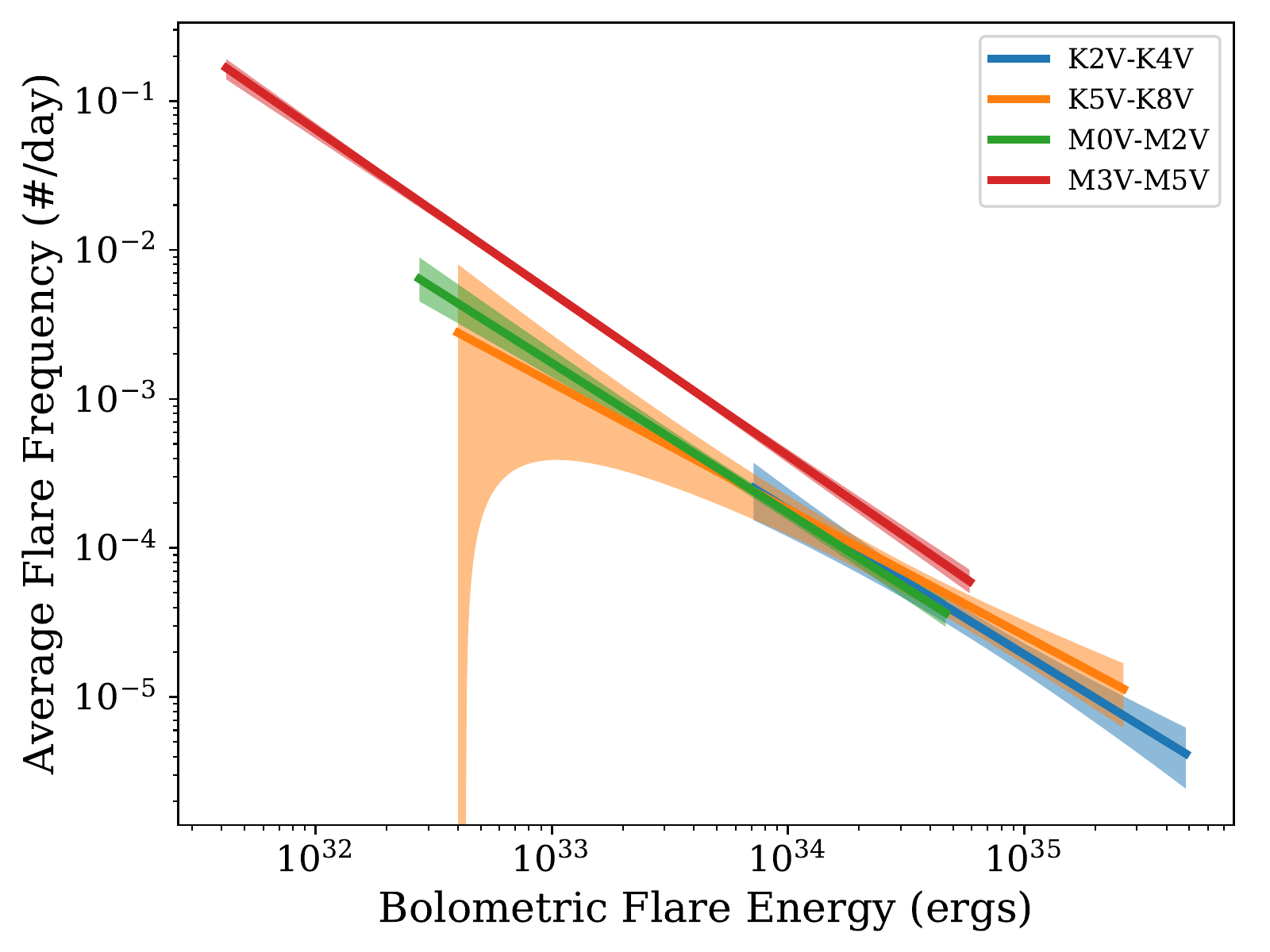}
     \caption{The best fitting average flare occurrence rates for the main sequence K and M subsets. The solid lines indicate the best fitting power laws in the observed energy ranges, while the shaded areas are the 1$\sigma$ regions. The K and early M stars show similar flare rates, while the fully convective mid-M stars flare more frequently at all studied energies.}
     \label{fig:new_occ_rate_example}
 \end{figure}

\begin{figure}
	\includegraphics[width=\columnwidth]{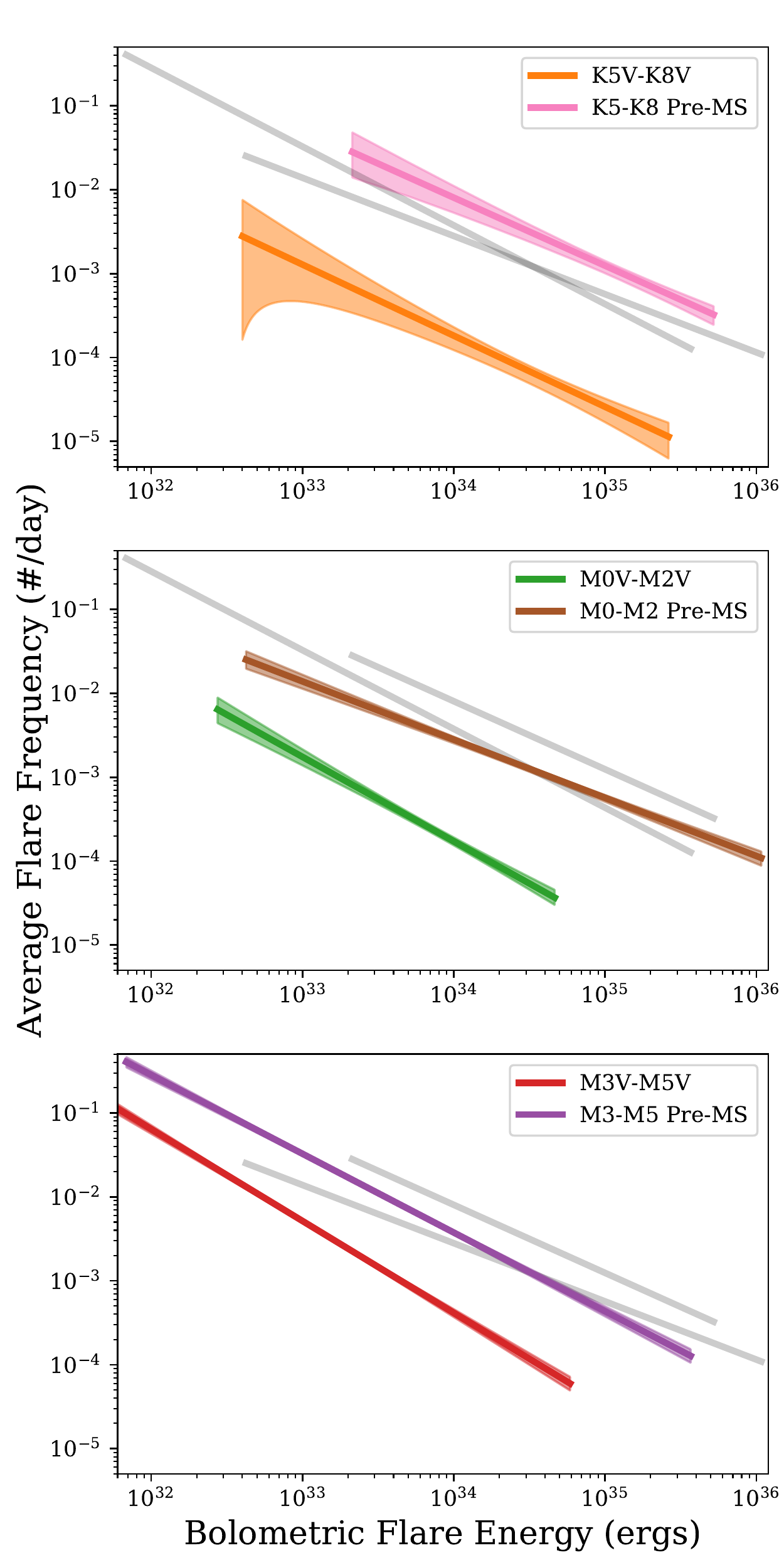}
    \caption{Average flare occurrence rates for the main and pre-main sequence K5-K8, M0-M2 and M3-M5 subsets. The solid line is our best fitting intrinsic flare model and the shaded areas are the corresponding 1$\sigma$ regions. The grey lines show the pre-main sequence occurrence rates from the other panels for reference. Note the decrease in flare rate with age for all spectral types.}
    \label{fig:ms_non_ms_occ_rate}
\end{figure}

We fit Eq.\,\ref{eq:obs_ffd} to the observed average flare occurrence rates for a series of bins in spectral type, shown in Tab.\,\ref{tab:alpha_values}. We did this for both our main and pre-main sequence samples. To calculate an average flare occurrence rate for a specific bin we assumed an ergodic principle that all stars taken together were equivalent to a single, average, star. This average star was observed for the same duration as the sum of the observing times for all stars within that bin (e.g. 10 stars observed for one hour were equal to one average star observed for 10 hours). This method included both the flaring and the non-flaring stars, to avoid biasing our samples to more active stars and to obtain the true average behaviour. \edit{We used Eq.\,\ref{eq:obs_ffd} and the method outlined in Sect.\,\ref{sec:flare_freq3} to account for the decreased detection efficiency of low energy flares from fainter stars relative to brighter stars of the same spectral type. If we did not account for this, then the average flare occurrence rate would be diminished at lower energies, where we can only detect flares from the brighter star. We note that one way around this, as mentioned in Sect.\,\ref{sec:flare_freq3}, would be to select a low energy cut off where we are confident we would detect flares from all the stars in our sample. However, due to the wide range of magnitudes of stars used in our sample (I$\approx$9 to 16), we found that only the highest energy flares would be detected on both the bright and faint stars. As these are the rarest flare events, this resulted in low numbers of flares to use when fitting our occurrence rates. By directly including the recovery fraction in our fitting we were able to include both the high energy flares from faint stars and low energy flares from bright stars, maximising our sample size.}

To calculate the recovery fraction $R(E)$ for our average star we used the results of the injection and recovery tests described in Sect.\,\ref{sec:completeness}. We first generated a grid of flare energies, spaced logarithmically from energies an order of magnitude below the lowest observed flare energy to an order of magnitude above the highest. For each star we multiplied the value of \edit{its own recovery fraction} at each energy in our grid by its observing duration to calculate a series of ``effective durations''. This had the effect of weighting the contribution of each star at different energies, e.g. downweighting the contribution of a faint star at very low flare energies. To calculate $R(E)$ for our average star, we summed the effective durations at each flare energy and then divided it by the total observing duration. We used this average $R(E)$ in our fitting of the observed average flare occurrence rate. 

To fit Eq.\,\ref{eq:obs_ffd} to our data, we used the {\scshape emcee} Python package \citep[][]{Foreman-Mackey13} to generate a Markov Chain Monte Carlo (MCMC) process. We used 32 walkers for 10000 steps and used the final 2000 to sample the posterior distribution. To account for possible uncertainties in our injection and recovery of the smallest flares, we followed \citet{Ilin19} and multiplied the uncertainties in our occurrence rate by $R(E)^{-1/2}$.

The results of our fitting and the marginalised 1D uncertainties are given in Tab.\,\ref{tab:alpha_values}. Fig.\,\ref{fig:early_m_example} shows an example of the recovered average flare occurrence rate compared to the observed. We can see that as expected, the recovered rate is more frequent than the observed behaviour at all but the highest energies, where all flares are recovered. Fig.\,\ref{fig:new_occ_rate_example} shows the recovered rates for our main sequence K and M stars together, while Fig.\,\ref{fig:ms_non_ms_occ_rate} shows the main and pre-main sequence samples together. We can see in Fig.\,\ref{fig:new_occ_rate_example} that the distributions for main sequence K and early M stars overlap, suggesting that at field ages the average flare rate for partially convective stars does not have a strong mass dependence. We can also see in Fig.\,\ref{fig:new_occ_rate_example} that fully convective mid M stars are offset from the other distributions, showing a greater level of activity for field age stars compared to their higher mass counterparts. We believe this behaviour is due to the increased spin-down time and period of activity for fully-convective stars relative to K and early M stars \citep[e.g.][]{Matt15}. We can see in Fig.\,\ref{fig:ms_non_ms_occ_rate} that the average flare occurrence rate for all spectral types has a clear dependence on age, with all pre-main sequence samples flaring more often than their main sequence counterparts. This, along with the behaviour of the main sequence sample, is something we discuss further in Sect.\,\ref{sec:flare_freq_discussion}. 

For comparison with other works we have also provided the values of $C$ and $\beta$. During our analysis we noted that the uncertainties on $k$ and $\alpha$ were correlated and as such, the marginalised 1D uncertainties provided in Tab.\,\ref{tab:alpha_values} will not accurately reproduce our fitted power law distributions and associated uncertainties if used in future works. For this purpose, we have also provided the covariance matrices for each subset in Tab.\,\ref{tab:alpha_cov}.

We have also used our fitted power laws to calculate the average waiting time of flares of different energies for main and pre-main sequence K and M stars. These waiting times are shown in Tab.\,\ref{tab:energy_times}, showing that we expect a $10^{35}$ erg flare to occur on a pre-main sequence M star once every five years.

\begin{table*}
    \centering
    \begin{tabular}{|l|c|c|c|c|c|c|c|c|}
    \hline
    Class & $\log k$ & $\alpha$ & $C$ & $\beta$ & $N_{flares}$ & $N_{stars,flare}$ & $N_{stars,total}$  \tabularnewline \hline
    K2V-K4V & $29.4\pm7.2$ & $1.97\pm0.20$ & $29.4\pm7.1$ & $-0.97\pm0.20$ & 18 & 17 & 15613   \tabularnewline
    K5V-K8V & $25.0\pm7.4$ & $1.85\pm0.21$ & $25.0\pm7.3$ & $-0.85\pm0.21$ & 18 & 14 & 6272   \tabularnewline
    M0V-M2V & $30.4\pm3.4$ & $2.01\pm0.10$ & $30.4\pm3.4$ & $-1.01\pm0.10$ & 34 & 26 & 4495 \tabularnewline
    M3V-M5V & $33.9\pm1.6$ & $2.09\pm0.05$ & $33.8\pm1.5$ & $-1.09\pm0.05$ & 91 & 39 & 626 \tabularnewline \hline
    K5-K8 ($<15$ Myr) & $25.4\pm4.8$ & $1.82\pm0.14$ & $25.5\pm4.8$ & $-0.82\pm0.14$ & 19 & 14 & 405    \tabularnewline
    M0-M2 ($<30$ Myr) & $20.9\pm1.8$ & $1.69\pm0.05$ & $21.0\pm1.7$ & $-0.69\pm0.05$ & 39 & 33 & 705 \tabularnewline
    M3-M5 ($<40$ Myr) & $29.5\pm1.2$ & $1.94\pm0.04$ & $29.5\pm1.2$ & $-0.94\pm0.04$ & 121 & 60 & 311 \tabularnewline
    \hline
    
    \end{tabular}
    \caption{Parameters from the power law fits to main sequence and pre-main sequence subsets of our data. $\mathrm{N_{flares}}$ is the number of flares in each subset.  $\mathrm{N_{stars, flare}}$ is the number of flaring stars in each subset, while $\mathrm{N_{stars,total}}$ is the total number of stars, both flaring and non-flaring. \edit{The uncertainties here are the marginalised 1D values and correspond to the uncertainties on the power law fit for the average flare rate, rather than an explicit range of power laws in the input sample.}}\label{tab:alpha_values} 
\end{table*}

 \renewcommand{\arraystretch}{1.25}

\begin{table}
	\centering
	\begin{tabular}{|l|c|c|c|}
    \hline
    Class & \multicolumn{1}{|p{1.45cm}|}{\centering  E>$10^{33}$ erg \\ (years)} & \multicolumn{1}{|p{1.45cm}|}{\centering  E>$10^{34}$ erg \\ (years)}  & \multicolumn{1}{|p{1.45cm}|}{\centering  E>$10^{35}$ erg \\ (years)}  \tabularnewline	\hline
    \edit{Main Sequence} & & &\tabularnewline	\hline
    K2V-K4V & $1.7^{+2.1}_{-0.9}$ & $15.2^{+7.4}_{-4.3}$ & $144.9^{+43.2}_{-25.5}$ \tabularnewline
    K5V-K8V & $2.2^{+2.6}_{-1.1}$ & $15.9^{+6.5}_{-4.0}$ & $112.7^{+44.4}_{-29.0}$ \tabularnewline
    M0V-M2V & $1.6^{+0.4}_{-0.3}$ &$15.9^{+2.0}_{-1.5}$ & $160.8^{+50.9}_{-37.7}$ \tabularnewline
    M3V-M5V & $0.5\pm0.1$ & $6.6^{+0.8}_{-0.6}$ & $82.4^{+20.8}_{-15.5}$\tabularnewline
    \hline
    \edit{Pre-Main Sequence} & & &\tabularnewline	\hline
    K5-K8 ($<15$ Myr) & $0.1\pm0.1$ & $0.3^{+0.2}_{-0.1}$ & $2.2^{+0.5}_{-0.3}$ \tabularnewline
    M0-M2 ($<30$ Myr) & $0.2\pm0.1$  & $1.0\pm0.1$ & $4.8^{+0.5}_{-0.4}$ \tabularnewline
    M3-M5 ($<40$ Myr) & $0.1\pm0.1$ & $0.7\pm0.1$ & $6.3^{+0.9}_{-0.7}$ \tabularnewline
    \hline

	\end{tabular}
    \caption{Time in years for a flare of a given energy to occur on an average isolated main sequence star in each of our late-type star classes. We can see that the mid M stars flare more often on average in our sample than the early M stars. Note the pre-MS K stars are preferentially younger than their M star counterparts. \edit{We note that the isolated main sequence sample does not include any of the stars in the pre-main sequence sample and can be taken as representative of field age stars. The uncertainties correspond to the 16th and 84th percentiles propagated from our flare frequency power law fits.}}\label{tab:energy_times} 
\end{table}

\subsection{Flare Duration} \label{sec:flare_duration_results}

\begin{figure}
	\includegraphics[width=\columnwidth]{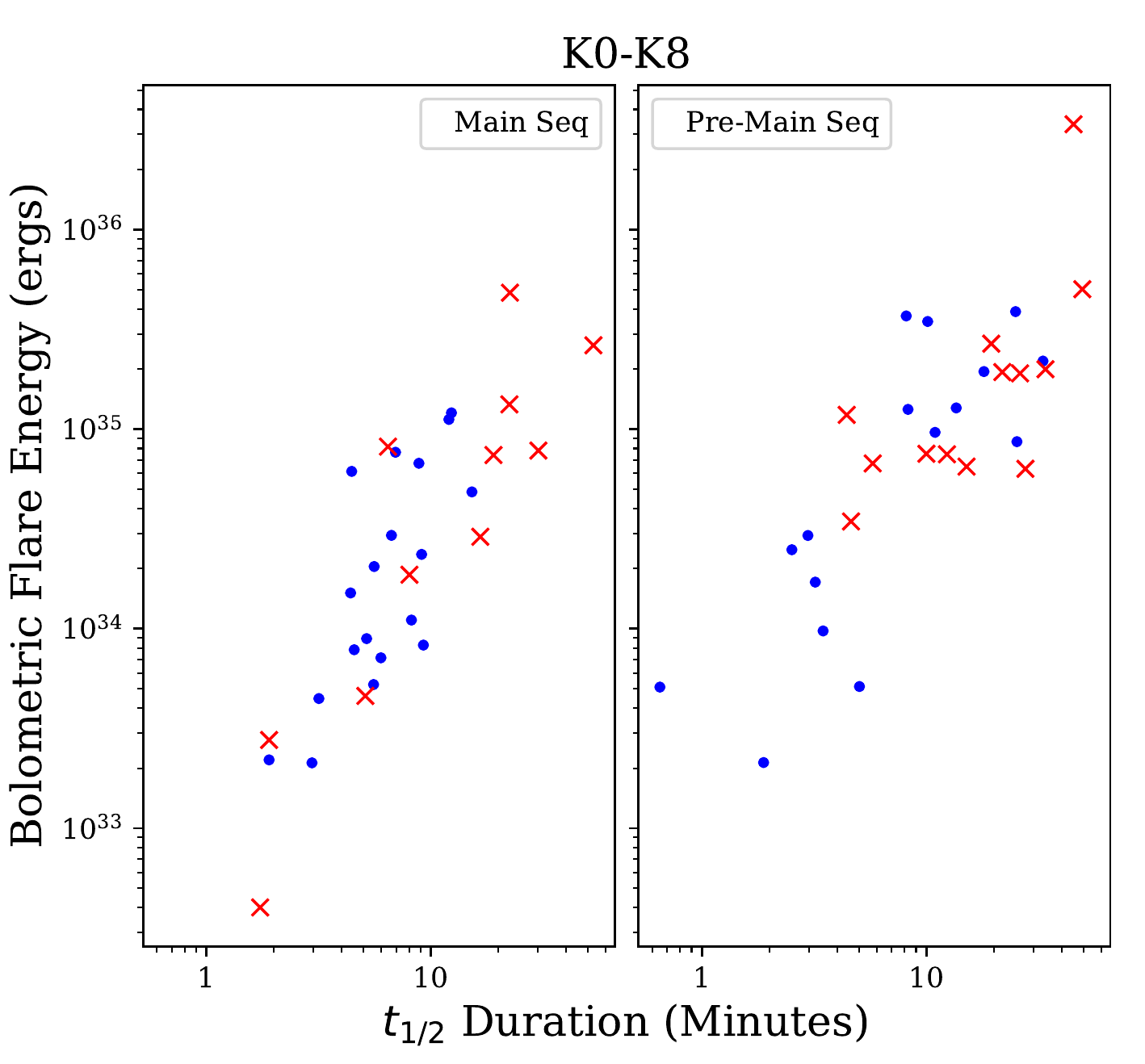}
    \caption{Duration and bolometric energy for flares detected from K stars in NGTS DR1. ``Simple'' flares are shown as blue circles, while ``complex'' events are shown as red crosses. Here we have chosen the ``shortest'' value of \thalf\ as discussed in Sect.\,\ref{sec:flare_duration}. The cut off in detected flares for longer durations in each figure is consistent with a selection effect from our detection method. We can see that complex flares become more prevalent at longer durations and higher energies.}
    \label{fig:k_star_duration_energy}
\end{figure}

\begin{figure}
	\includegraphics[width=\columnwidth]{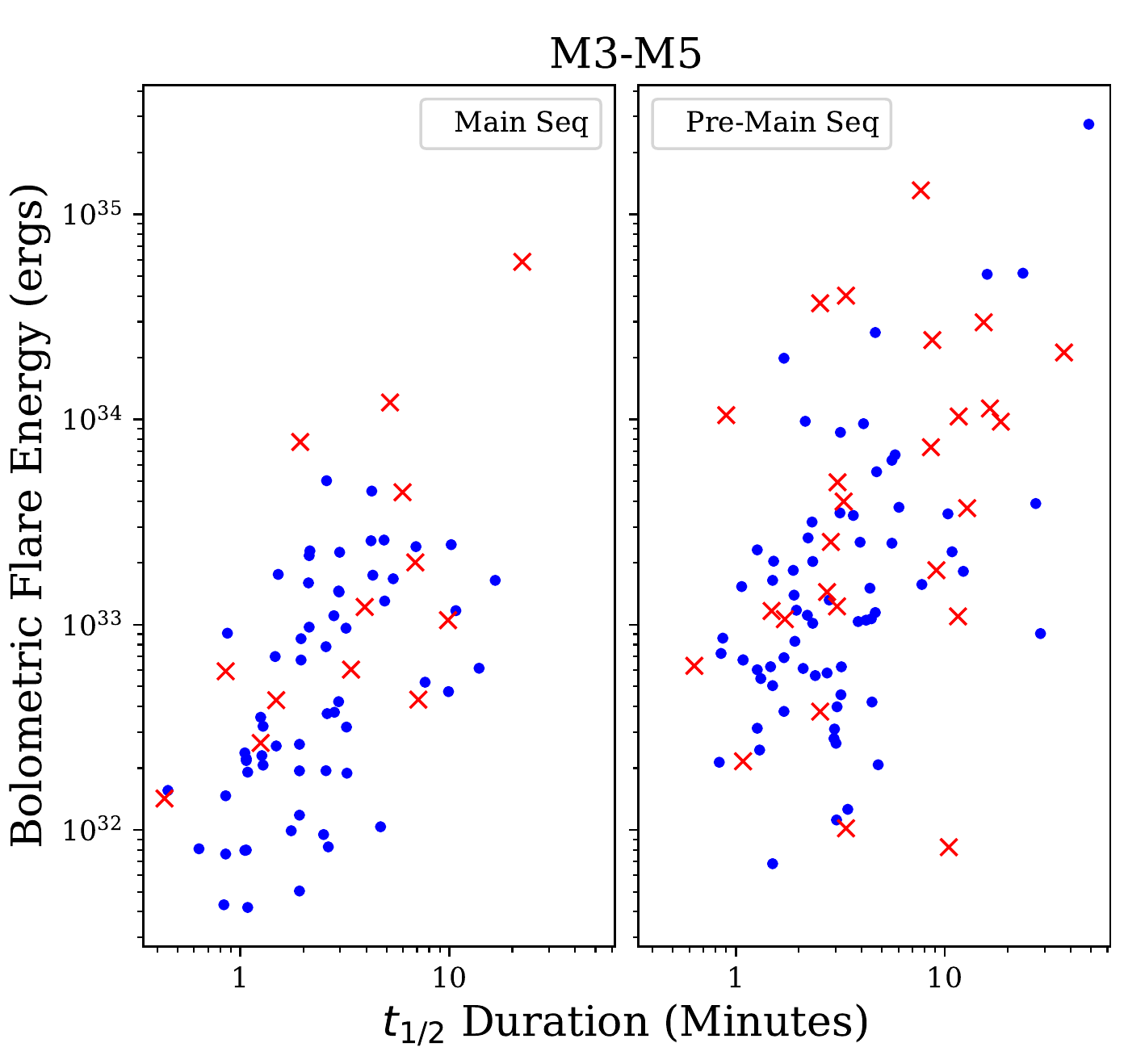}
    \caption{Duration and bolometric energy for flares detected from mid-M stars in NGTS DR1. Symbols are per Fig.\,\ref{fig:k_star_duration_energy}.}
    \label{fig:m_star_duration_energy}
\end{figure}

We have measured the \thalf\ timescale of all observed flare events, using the high time cadence of \NGTS. We measured \thalf\ timescales between 26 seconds and 50 minutes in our full sample. We have also split our sample into K, early-M and mid-M spectral bins to see how the \thalf\ timescale varies with energy in each regime. Fig.\,\ref{fig:k_star_duration_energy} and \ref{fig:m_star_duration_energy} shows this distribution for our K and mid-M star samples. Our early M star sample sits between these distributions. We can see that both our K and mid-M star samples show a positive correlation between the \thalf\ duration and measured flare energy in log-log space, however there is notably more scatter in the mid-M samples. We measured the Pearson rank coefficients of each distribution in Fig.\,\ref{fig:k_star_duration_energy} and Fig.\,\ref{fig:m_star_duration_energy}, measuring 0.74 and 0.80 for the main and pre-main sequence K star samples and 0.55 for both mid-M samples. At first glance this would imply a strong relation between the flare energy and duration for K stars which breaks down towards later spectral types. However, we believe that the reason for this increased scatter at later spectral types is due to a greater number of detected lower amplitude, but longer duration, flares. These flares, due to their lower amplitudes are more easily detected on cooler stars due to the greater contrast between the flare and quiescent spectrum. To test this theory, we compared the observed distributions in Fig.\,\ref{fig:k_star_duration_energy} and \ref{fig:m_star_duration_energy} to the results of our flare injection and recovery tests from Sect.\,\ref{sec:completeness}. We found that the longer duration - lower energy cut off in both the K and M star samples (\edit{and apparent linear relation}) was consistent with a selection effect from our detection method, rather than a \edit{physical} cut off in \edit{\thalf\ duration} for low energy flares.

\subsection{Stellar Rotation}
We measured the periodicity in the lightcurves of our flaring stars in our sample using the method outlined in Sect.\,\ref{sec:rotation}. From this, we detected and measured the periods of 102 stars in our sample. We vetted these stars to remove signals that may more likely be due to pulsation or eclipses than rotation. This left 97 stars, 36 of which were main sequence, 39 pre-main sequence and 22 in between. Overall, these stars had spectral types between K0 and M5.5 and periods between 0.35 and 23.3 days. The main sequence sample had spectral types ranging between K0 and M3.5 and periods between 0.35 and 22.3 days, while the pre-main sequence sample had spectral types ranging between K0 and M5.5 and periods between 0.43 and 23.3 days. The sample in between the isolated main sequence stars and the pre-main sequence stars had periods between 0.2 and 12 days. 

\subsubsection{Flares and Starspot Phase} \label{sec:flare_phase_dis}
Starspots are the result of local concentrations of the magnetic field which inhibit convection and create cooler regions of plasma on the photosphere. Correlations between spot groups and flares are frequently observed on the Sun \citep[e.g.][]{Zinn87,Jiang12,Xan18}, however previous studies \citep[e.g.][]{Hawley14,Doyle18,Feinstein20} have generally found no correlation between starspot modulation and the timing of high amplitude stellar flares.

To search for any relation between flares and starspot phase in our sample of spotted stars, we initially phase folded our lightcurves on the rotation period measured in Sect.\,\ref{sec:rotation}. We then set all stars such that the minimum flux (the dominant starspot or starspot group is facing us) at phase 0.5. During this analysis we assumed the modulation was due to a single large starspot.

In order to avoid selection effects in our analysis we required our flares to have fractional amplitude greater than the amplitude of variation. This is ensured we only used flares which would have been visible at all phases. To test the dependence of these large flares on the starspot phase we perform a single sample KS test. We used a uniform distribution as our reference sample, based on the results of previous studies \citep[e.g.][]{Doyle18}. We tested individual stars in our sample which have more than 3 flares, along with testing the combined sets from Tab.\,\ref{tab:alpha_values}. From our analysis no star, or combination of stars, seems to deviate significantly from a uniform distribution, consistent with previous results for large flares and we discuss the reasons for this in Sect.\,\ref{sec:starspot_discussion}.

\subsubsection{Flares and Starspot Size}
Along with investigating how flares relate to starspot phase, we can also see how flare energy is related to starspot area. To do this we first calculate the starspot area, $A_{spot}$, using the equation \citep[e.g.][]{Notsu13},
\begin{equation}
    \frac{A_{spot}}{A_{star}} = \bigg(\frac{\Delta F_{spot}}{F}\bigg)\bigg[1-\bigg(\frac{T_{spot}}{T_{star}}\bigg)^{4}\bigg]^{-1}
\end{equation}
where $\frac{\Delta F_{spot}}{F}$ is the change in flux due to a spot normalised by the average flux and $A_{star}$ is the apparent area of the star, equal to $\pi R^{2}_{s}$. $T_{star}$ is the temperature of the stellar photosphere, which we have equated to \teff. We calculate the starspot temperature $T_{spot}$ using the relation from \citet{Maehara17},
\begin{equation} \label{eq:spot_temp}
    \begin{aligned}
    \Delta T(T_{star}) = T_{star} - T_{spot} = \\3.58\times10^{-5}T^{2}_{star} + 0.249 T_{star} - 808
    \end{aligned}
\end{equation}
This relation was derived by \citet{Maehara17} using a second order polynomial fit to a compilation of stellar and spot temperatures from \citet{Berdyugina05}, for stars with temperatures between 3300 and 5870K. We also calculate the energy stored in the magnetic field, $E_{mag}$, using \citep[e.g.][]{Notsu13}
\begin{equation}
    E_{mag} \approx \frac{B^{2}}{8\pi}A^{3/2}_{spot}
\end{equation}
where $B$ is the magnetic field strength of the starspot. This represents the maximum flare energy from this starspot, assuming all energy is converted from the magnetic field to the observed flare.

\begin{figure}
	\includegraphics[width=\columnwidth]{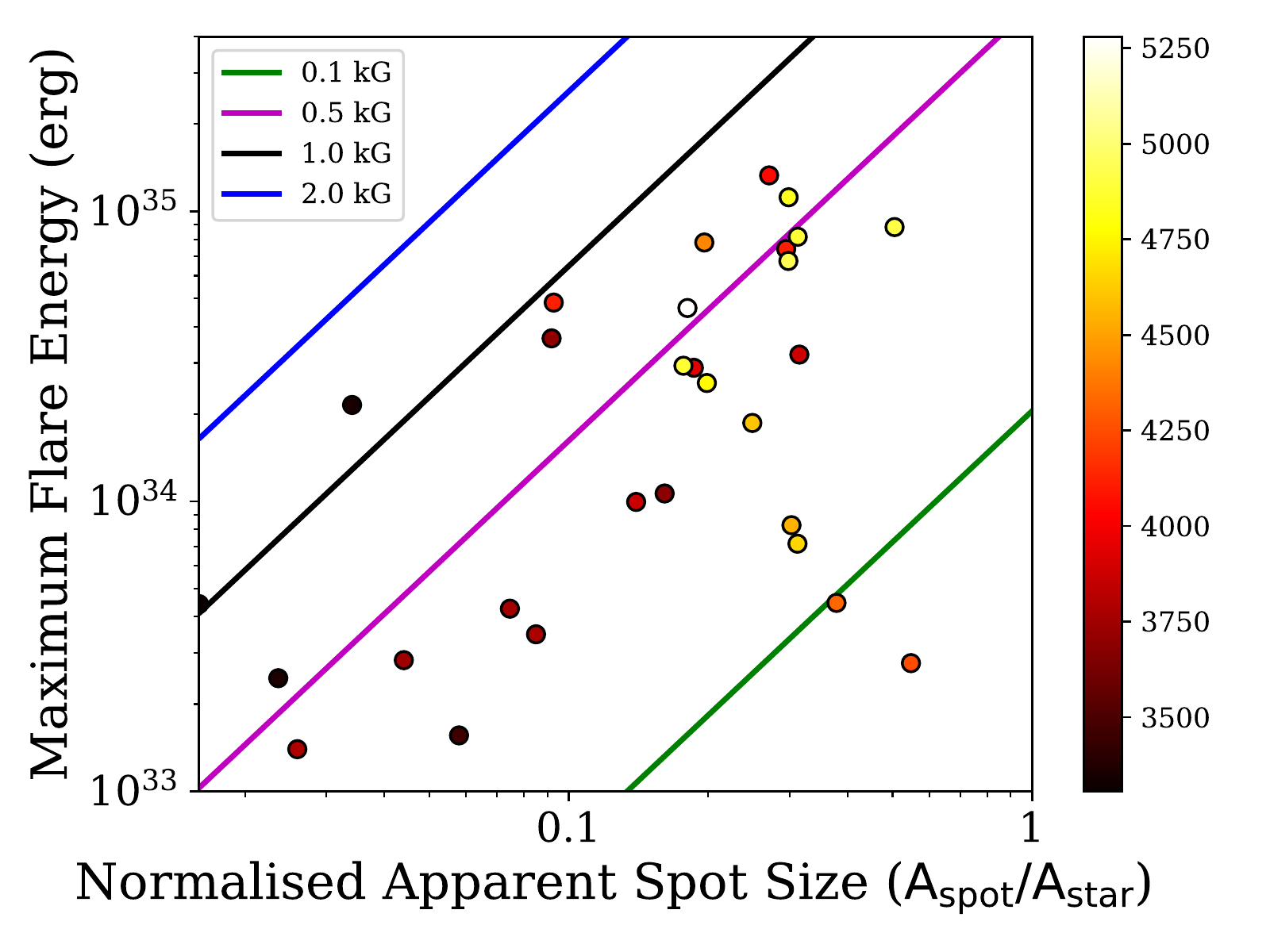}
    \caption{Maximum observed flare energy against the normalised starspot area. The coloured lines represent the maximum flare energy from a spot of a given magnetic field strength. The colour of the spot indicates its effective temperature. Note that all flares in this sample of stars with measured starspot modulation can be explained with a field strength of 2 kG.}
    \label{fig:starspot_flares}
\end{figure}

The results of our analysis are shown in Fig.\,\ref{fig:starspot_flares}. We can see in Fig.\,\ref{fig:starspot_flares} that the majority of stars with detected flares and starspot modulation have maximum observed flare energies which can be explained by a single large spot of 0.5 kG, across K and M stars alike. All flares and spots can be explained with a 2 kG field. We note the presence of some stars with large apparent spot sizes, reaching up to 60 per cent of the visible stellar surface. This can be explained by us overestimating the spot temperature in Eq.\,\ref{eq:spot_temp}. Our magnetic field results agree with those of \citet{Howard19spot}, who performed the same analysis using the EvryScope flare sample for cool stars.
Our results are also consistent with previous observations of kilogauss magnetic field strengths for K and M stars \citep[e.g.][]{Saar86,Shulyak17}. We have assumed in this analysis that the spots are located at the equator. Previous studies \citep[e.g.][]{Notsu19} have noted that the position of the lines in Fig.\,\ref{fig:starspot_flares} will shift upwards for stars with lower inclination angles, or for spots at higher latitudes (e.g. near the pole) that cause smaller brightness variations. Consequently, those flares lying above the 1 and 2 kG lines may in fact be consistent with lower magnetic field strengths if their spots are located at high latitudes or the stars have low inclinations.

\section{Discussion} \label{sec:discussion}
We have presented the results of searching for stellar flares in the first \NGTS\ data release. We have identified \flarenumber\ flares from \starnumber\ stars at a 13 second time cadence. We have been able to find flares from redder stars than the main \kepler\ survey was able to do and are currently probing a similar regime to \tess, but with a higher time resolution for all stars.

\subsection{Flare Occurrence Rates} \label{sec:flare_freq_discussion}

\subsubsection{Main Sequence} 
\label{sec:occ_ms}
\renewcommand{\arraystretch}{1.0}
In Sect.\,\ref{sec:flare_freq3} we measured the average flare occurrence rates for the K and M stars in our sample, the values of which are shown in Tab.\,\ref{tab:alpha_values}. The best fitting power laws of main sequence stars are shown in Fig.\,\ref{fig:new_occ_rate_example}. We can see in Fig.\,\ref{fig:new_occ_rate_example} that the average flare occurrence rates of K and early M stars overlap, while the fully convective mid M stars are offset, showing an increased level of activity. As mentioned in Sect.\,\ref{sec:flare_freq3}, we believe that this offset is due to the increased spin-down time of fully-convective stars relative to K and early M stars \citep[e.g.][]{Matt15}.

By $\approx$ 650 Myr isolated K stars have mostly spun down \citep[][]{Douglas16}, meaning our field age sample will likely be dominated by slow rotating stars. \citet{Newton17} measured the rotation periods of field age M dwarfs with MEarth photometry. The majority of early M stars in their sample have periods longer than 10 days, showing they have mostly spun down. K and early M stars are believed to share a common dynamo. The strength of this dynamo and the relative magnetic activity (e.g. \lxlbol) of field stars is known to be tied to the rotation period, with stars with periods above 10 days showing reduced activity \citep[e.g.][]{Pizzolato03,Wright11}. Consequently, the field age K and early M stars having spun down to the same rotational period (and magnetic activity) distribution would explain our observed common average flare rates.

In contrast to early M stars, mid-M stars show significant scatter in their period distribution, showing that while some stars have spun down, many have not. Studies of the evolution of the quiescent magnetic activity of early and mid M stars using UV emission have found results reflecting this behaviour. 
In a study of the NUV and FUV emission of open cluster and field age M dwarfs \citet{Schneider18} found that the UV emission of early M stars dropped sharply between $\approx$150 Myr and field ages (by a factor of 31 in the FUV), while the emission from mid M stars retained high levels of UV activity into field ages. Therefore, while K and early M stars in our sample have on average spun down and dropped in magnetic activity, the longer spin down of mid M stars helps to sustain their average flaring behaviour into field ages. \edit{This is also in agreement with \citet{West08}, who measured the fraction of active M dwarfs, using \halpha\ luminosity, as a function of spectral type and combined this with dynamical simulations to derive age-activity relations for M dwarfs. They found that the active fractions and activity lifetimes of mid-late M dwarfs were greater than for early M dwarfs, making it likely that our observed increased flare rates of mid-M stars is due to an extended activity lifetime.} 

\citet{Medina20} recently studied the white-light flaring behaviour and rotation periods of nearby field age mid-M dwarfs using \tess\ short-cadence data and MEarth photometry. They found that these field stars fell into two distributions in flaring activity-rotation space --- fast rotating and regularly flaring, and slow rotating ($\approx$100 days) and rarely flaring. This is consistent with some mid-M stars having spun down and decreased in flaring activity, while others have not. We note here that in order to measure the average white-light flare behaviour we will have combined stars from both regimes highlighted by \citet{Medina20}. Therefore, individual mid-M stars may show activity above or below this behaviour, depending on the regime they fall into. We intend to measure the average occurrence rate in each regime in future work.

In Tab.\,\ref{tab:energy_times} we have used our fitted occurrence rates to calculate the average waiting time for flares with energies above $10^{33}$, $10^{34}$ and $10^{35}$ erg. We can see from Tab.\,\ref{tab:energy_times} that on average an isolated main sequence M0-M2 star will flare with an energy of at least $10^{34}$ erg once every \textcolor{black}{16} years and $10^{35}$ erg roughly once every \textcolor{black}{160} years. Mid-M stars show a flare of $10^{34}$ erg once every \textcolor{black}{7} years and $10^{35}$ erg roughly once every \textcolor{black}{82} years. Our K and early M stars subsets have similar waiting times, especially for energies up to $10^{34}$ ergs, namely one flare every 15 years. This illustrates our finding that the K and early M stars have spun down to a common state of activity. \edit{We note that what we have calculated is the average flare rates and waiting time, by combining large samples of active and inactive stars within chosen ranges of age and spectral type. The uncertainties given in Tab.\,\ref{tab:alpha_values} and \ref{tab:energy_times} correspond to the error on the calculated average flare rate for the average star, rather than an explicit range of power laws and waiting times for every star in the sample. Therefore we note that while these values can be used for the flare waiting time of individual stars, they are best suited for predicting the number of flares from large samples of stars and that some stars of interest (e.g. very active M stars) may show activity and values of $C$ outside of our calculated ranges.}

Our measured values of $\alpha$ for main-sequence M and late-K stars are consistent with those measured in previous surveys using ground and space-based observations \citep[e.g.][]{Hilton11,Yang17,Howard19,Lu19}. A value of $\alpha$ greater than 2 implies that, if the distribution given in Eq.\,\ref{eq:ffd} holds at all energies, then the lowest energy flares can dominate the total energy distribution \citep[e.g.][]{Gudel03}, possibly providing a way of heating the Solar and stellar corona through frequent nanoflares \citep[e.g.][]{Doyle85}. Our measured values of $\alpha$ for the K and early M main sequence subsets are all consistent with $\alpha$=2. Interestingly, our best-fitting power law for the fully convective mid M stars has $\alpha>2$, although it is consistent with $\alpha$=2 within 2$\sigma$. Our fitted $\alpha$ values suggest that for K and early M stars large flares dominate the total energy distribution, however our uncertainties do not allow us to confidently rule out the alternative, while our results suggest that small flares dominate the distribution for the average mid-M star.

\subsubsection{Pre-Main Sequence Stars} \label{sec:occ_pre_ms}
We have also measured the average flare occurrence rates for pre-main sequence stars in our sample, the results of which are shown in Tab.\,\ref{tab:alpha_values}. 
We defined our pre-main sequence sample in Sect.\,\ref{sec:stellar_characterisation} using the positions of the stars on the HR diagram in Fig.\,\ref{fig:HR_diagram}. \rewrite{As we discussed in Sect.\,\ref{sec:galactic_distribution} the limit for stars to be designated as pre-main sequence had a mass-dependent age cutoff. Stars with a K6-K8 spectral type are younger than 15 Myr, M0-M2 are younger than 30 Myr and M3-M5 are younger than 40 Myr. Along with this we note that these samples may also be contaminated by young (but older than the pre-main sequence samples) binaries which are shifted upwards on the HR diagram. We initially used the PARSEC isochrones to limit our pre-main sequence M star sample to 15 Myr, however this resulted in too few flares to fit a meaningful power law. Consequently, we combine all pre-main sequence stars in each subset. As these pre-main sequence samples are a mixture of ages we have not used our fitted values here to quantitatively probe how stellar flare activity changes with age as we would for an open cluster \citep[e.g.][]{Ilin19,Jackman20,Ilin20}. Instead we use this sample for comparison with our main-sequence data.}

We have plotted our the average flare occurrence rates for our main and pre-main sequence late-K and M star samples in Fig.\,\ref{fig:ms_non_ms_occ_rate}. We can see that, as expected, the pre-main sequence stars show a greater level of magnetic activity than their main sequence counterparts, flaring more often and with greater energies. As stars age they undergo magnetic braking \citep[e.g][]{Booth17} which carries away angular momentum, reducing differential rotation and their dynamo and magnetic activity. This in turn is expected to reduce flaring rates, as we can see in Fig.\,\ref{fig:ms_non_ms_occ_rate}. This also broadly matches the behaviour of flares in the FUV. \citet{Loyd18}, in their study of the FUV flaring activity of 40 Myr and magnetically inactive field age M stars from \textit{HST} observations, found that for FUV flare energies above $10^{30}$ erg, young M stars flare on average 50 times more often than their field age counterparts. However, we find smaller values for the lowest energy flares in our sample, which would be most comparable to those studied by \citet{Loyd18}, with young M stars flaring 5-10 times as much as their main sequence counterparts. One possibility for this may be because \citet{Loyd18} used known magnetically inactive and isolated field-age M stars from \citet{Loyd18b} to compare to their young sample, whereas our field age sample contains both inactive and active M stars, some of which may not have fully spun down, as discussed in Sect.\,\ref{sec:occ_ms}. This could act to increase the observed gap in flare rates. Further vetting and classification of the quiescent activity of stars in future flare studies will help test whether this is the case, or whether our findings point towards a true discrepancy between the change in flare rates between the optical and the FUV.

We have calculated the average waiting times for flares of various energies from our pre-main sequence samples, shown in Tab.\,\ref{tab:energy_times}. We find that, on average, late K stars with ages below 15 Myr flare with energy of $10^{34}$ erg approximately 50 times more often than for field age stars. For the same flare energy, the difference between field age M3-M5 stars and those with ages below 40 Myr is a factor of nine. While these pre-main sequence M3-M5 stars are older than the K5-K8 sample, the reduced gap in flare rates between young and field age stars highlights how the magnetic activity persists with age for the fully convective sample. As we discussed in Sect.\,\ref{sec:occ_ms}, this is likely due to the longer spin down timescales of fully convective stars.

The best fitting values of $\alpha$ for our pre-main sequence samples are shown in Tab.\,\ref{tab:alpha_values}. Our value of late K stars is consistent with our measured value for the field age sample. However, we can see that both M star samples appear to have lower values of $\alpha$ at young ages, which points towards a steepening of the average flare occurrence rate of M stars with age. The change of $\alpha$ with age has been studied by various groups. From their study of ultracool dwarfs with different kinematic ages, \citet{Paudel18} found evidence for an increase in the average $\alpha$ with age, seemingly in agreement with our finding. However, the authors were unable to conclude any definite relation due to their small sample size. From a study of nearby stars and stars in open clusters, \citet{Shak89} highlighted a dependence of $\alpha$ on age. Their observations showed that $\alpha$ appeared to decrease with age, in contrast with our results. In their study of flares from the Pleaides and Praesepe open clusters \citet{Ilin19} found no strict age dependence for $\alpha$, but did appear to observe a decrease in $\alpha$ between the clusters for early M stars. \citet{Ilin19} discussed the evolution towards a shallower flare rate with age (decreasing $\alpha$), suggesting that the magnetic field topology changes with age, in turn allowing longer build ups of magnetic stress and energy. This then results in fewer small flares and more large ones. Our results would suggest the opposite, that a change in the field topology instead acts in favour of smaller flares. Very young early M stars are expected to be fully convective, forming a radiative core after approximately 100 Myr \citep[e.g.][]{Baraffe15}. Fully convective M stars are believed to predominantly have strong dipolar poloidal fields \citep[e.g.][]{Morin08,Morin10,Shulyak17}, while partially convective stars have multipolar fields with a large scale toroidal element. For Solar flares it is the toroidal fields that supply the energy and the same is typically considered for flares from partially convective M dwarfs \citep[e.g.][]{Mullan18}. However, in a study of the magnetic field and flares from the M8 planet host TRAPPIST-1, \citet{Mullan18Tr1} noted that while a toroidal field would dominate small flares, the polodial field may contribute in part to higher energy flares with a different power law. If this is the case, then young early M dwarfs (which are fully convective) may have a dipolar field which is more preferable for high energy flares. As the star ages and the radiative zone and the interface dynamo are formed, a multipolar field is generated which would reduce the field contribution to high energy flares and results in an apparent preference towards smaller flares. As the radiative zone for early M stars is formed by 100 Myr, this could explain why this behaviour was not observed in the 125 and 650 Myr samples of \citet{Ilin19}. However, this does not explain why the value of $\alpha$ for our mid-M stars, which remain fully convective their entire lives, also appears to increase with age in our sample. This would point towards other elements at work, not considered here. Along with this, it does not explain why the opposite behaviour in $\alpha$ with age was observed by \citet{Shak89}. On a similar note, \citet{Ilin20} recently compiled measurements of $\alpha$ from the literature for young and field age stars, finding a spread of measured values from 1.4 to 2.5. They noted that this spread of measured values, with various ways of determining the uncertainty, currently makes it difficult to assess whether $\alpha$ truly is changing with mass and age. Future studies of flares from open clusters should aim to address the evolution of $\alpha$ in a systematic way to tackle this problem.

\citet{Davenport19} has investigated how the flare occurrence rate varies as a function of age by applying gyrochronology techniques to the \citet{Davenport16} \kepler\ sample. By using rotation period as a proxy for stellar age, they found that flaring rates decrease as stars age and they generated flare evolution models which vary as a function of age and stellar mass. Using this model for a 0.5\Msun\ star flaring at an energy $10^{35}$ erg gives a best fitting age around 2 Gyr for the young samples in Fig.\,\ref{fig:ms_non_ms_occ_rate}. This is not consistent with their position on the HR diagram, suggesting the model from \citet{Davenport19} is overestimating the flare rate for these stars. Both \citet{Jackman20} and \citet{Ilin20} found similar discrepancies of this model with the measured average flare occurrence rates of low mass stars in open clusters. We believe the overestimation of the flare rates is because the \citet{Davenport19} model only used active flare stars as an input, whereas we have used both the flaring and non-flaring stars in our sample. Therefore we note that this model is most applicable to active subsets and should not be used to model the average flaring behaviour of groups of stars.

\subsubsection{Refitting the Orion occurrence rate} \label{sec:orion_analysis}
We have also applied our new method to the average occurrence rate of Orion-associated 3400-3940\,K (approximately M0-M3.5) stars from \citet{Jackman20}. In their analysis, \citet{Jackman20} fit a power law according to Eq.\,\ref{eq:ffd} to flares above an energy of $9\times10^{34}$ erg. Only stars with a 68 per cent completeness limit below this value, determined from injection and recovery tests, were used in the fitting of the average flare occurrence rate. This limited the number of flares and stars that could be used in the analysis, with 11 out of 40 stars being excluded from the analysis. The best fitting power law from \citet{Jackman20} had $\alpha=2.63\pm0.62$ and $C=57.4\pm21.9$. We note that $C$ was measured in units of $\mathrm{year^{-1}}$ and in units of $\mathrm{day^{-1}}$ is $C=54.8\pm21.9$. 
We ran our analysis using the flare injection and recovery results from Sect.\,\ref{sec:completeness}. We measured a best fitting power law with $\alpha=1.94\pm0.16$ and $C=30.7\pm5.7$. These values are consistent with the previously measured values to well within 2$\sigma$, but have smaller measured uncertainties. The inclusion of lower energy flares in our fitting is likely responsible for the decrease in our measured $\alpha$, while the decrease in the measured value of $C$ is due to both the lower value of $\alpha$ and the inclusion of stars which had 68 per cent completeness limits below the specified energy of $9\times10^{34}$ erg. Many of these stars will have contributed non-negligible amounts to our calculated observing duration, bringing the measured rate down.

\subsection{Maximum Flare Energy}

\begin{figure}
	\includegraphics[width=\columnwidth]{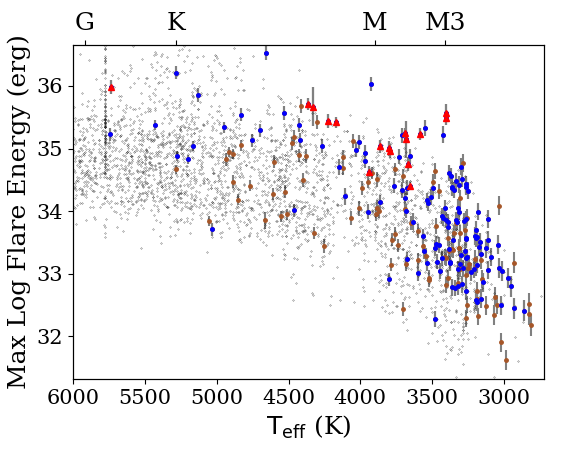}
    \caption{Maximum observed bolometric flare energy against effective temperature for our detected flaring stars. Stars consistent with being isolated and on the main sequence are shown as brown circles. Stars not consistent with being on the main sequence, but with no known association, are shown as blue circles. Stars associated with the Orion complex from \citet{Jackman20} are shown as red triangles. The underlying black points represent stars from \citet{Yang19}, using long cadence \kepler\ observations. We believe the line of stars around 5800\,K is due to an error in the KIC effective temperatures from the \citet{Yang19} sample. Note how the non-main sequence stars appear towards higher energies.}
    \label{fig:energy_colour}
\end{figure}
In Sect.\,\ref{sec:flare_energy} we calculated the energies of the flares in our sample, finding them to range between \minflareenergy\ and \maxflareenergy. One aspect of interest in our study is whether we can use our sample to probe the maximum limit of flare energy. Figure \ref{fig:energy_colour} shows the distribution of maximum measured flare energy with effective temperature for our sample. As we progress to cooler stars we see a decrease in the maximum flare energy, similar to what has been seen in other surveys of maximum flare energy \citep[e.g.][]{Davenport16}. We can see that field age K and early M stars can flare with energies up to approximately $10^{35}$, which is 10,000 times that of the Carrington event on the Sun \citep[][]{Carrington_1859,Carrington_Energy}. In Fig.\,\ref{fig:energy_colour} we have plotted the maximum flare energies from the \kepler\ long cadence sample of \citet{Yang19}. The two samples seem to overlap in the energy space covered, in particular at the upper limit of flare energies for M dwarfs. 

Recent observations of white light flares from M dwarfs with ASAS-SN have recorded V band flare energies up to $10^{36}$ erg and corresponding bolometric energies up to $10^{37}$ erg \citep[][]{Schmidt19,Rodriguez20}, beyond the apparent upper energy limit seen in Fig.\,\ref{fig:energy_colour}. This implies that our observed upper limit is not due to a physical limitation in the amount of energy that can be stored and released through magnetic reconnection, but due to a selection effect from our observing strategy. To test and confirm this, we used the best fitting average flare occurrence rates measured in Sect.\,\ref{sec:flare_frequency2}. For each subset we applied our recovery fraction $R(E)$ to the intrinsic power law distribution using Eq.\,\ref{eq:edit_powlaw} to predict the observed flare distribution. We used this predicted flare distribution to calculate the maximum energy at which we should have seen one flare in our observations. We then compared this value to our measured energies for the subset. If the maximum energy distribution cuts off before this limit, then the cut off would likely be due to a physical limitation. If the limit is below our observed maximum flare energies, then the apparent cut off in our distribution is due to a selection effect, namely not having observed each subset for enough time. Applying this analysis to the main sequence stars in Fig.\,\ref{fig:energy_colour} reveals that the upper energy limit in our sample of K and early to mid M stars is indeed consistent with not having observed each subset for long enough, rather than a physical limit in the energy that can be released from flares. We predict that a main sequence M0V-M2V or M3V-M5V star will flare with an energy of $10^{37}$ erg approximately once every 17000 and 12500 years respectively. This drops to around 80 years for our pre-main sequence samples, for both spectral types. These long waiting times for individual stars highlight how the large sample sizes achieved with wide-field surveys are vital for studying the highest energy flare events.

Splitting our sample of maximum flare energies into those consistent with being isolated main sequence stars and those which are more likely to be binary or pre-main sequence reveals that the non-main sequence sources on average occupy regions of higher energy. This is expected as young stars are known to be more magnetically active than their older counterparts. This activity results in an increased rate of high energy flares, making it more likely we would observe higher energies in our sample. Binary sources in this sample may have undergone tidal interactions which have prolonged this phase of strong magnetic activity, contributing to the increased maximum flare energy.

\subsection{Flare Duration}

In Sect.\,\ref{sec:flare_duration_results} we presented the range of flare \thalf\ durations for our sample. The results for our K and mid-M stars are shown in Fig.\,\ref{fig:k_star_duration_energy} and \ref{fig:m_star_duration_energy}. In Sect.\,\ref{sec:flare_duration_results} we discussed that while there appears to be a strong positive correlation between the flare energy and the \thalf\ duration for K stars that breaks down for later spectral types. However, after testing this distribution using our flare injection and recovery tests, we found that this behaviour was likely due to a selection effect. Something else we can see in Fig.\,\ref{fig:k_star_duration_energy} and \ref{fig:m_star_duration_energy} is that complex flares appear to form the majority of the longest duration events. Long timescale complex flares (particularly those with sharp multiple peaks) have previously been explained as multiple simple flare events \citep[e.g.][]{Davenport2014} occurring in quick succession. In this scenario it is possible the successive flare events are a result of ``sympathetic flaring'', where one large flare can trigger events in the same or nearby active regions, giving the appearance of multiple peaks in a single lightcurve. This trigger is provided through a form of physical link between flare sites. Multiple mechanisms \citep[e.g. a direct link, a travelling disturbance, or the weakening/removal of magnetic fields elsewhere,][]{Jiang09,Wang01,Zuccarello09} have been proposed as the cause for this trigger, with the exact cause still up for debate. Therefore it may be that our high energy complex flares with long durations are in fact formed of several simple flares.

\begin{figure}
	\includegraphics[width=\columnwidth]{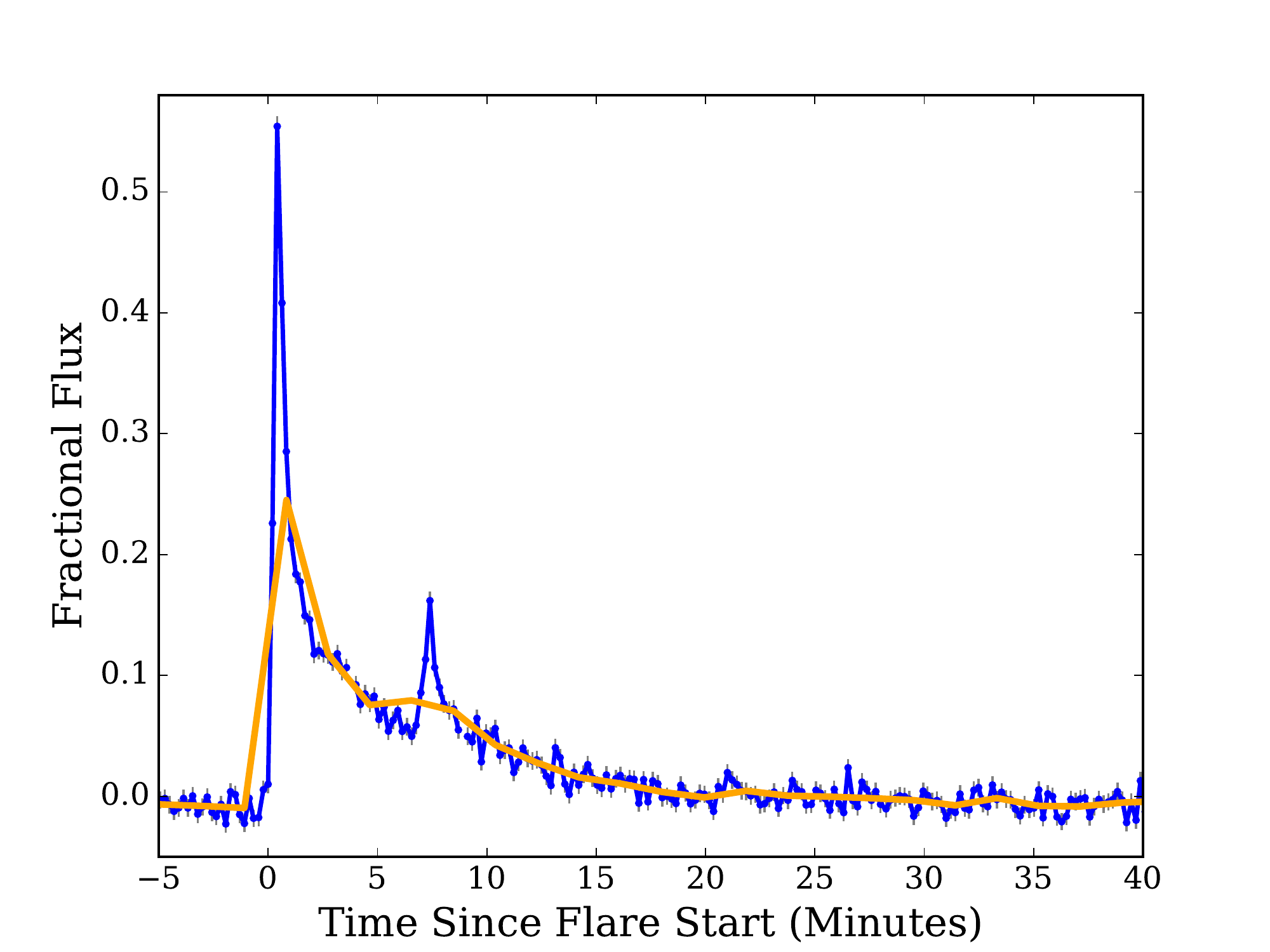}
    \caption{Example of a multi-peaked complex flare with a very short \thalf\ duration ($\approx$30 seconds). NGTS data is shown by the blue line. Overlaid in orange is the simulated \tess\ 2 minute short cadence mode showing the decrease in flare amplitude and resolved substructure.}
    \label{fig:very_impulsive_flare}
\end{figure}

From Fig.\,\ref{fig:k_star_duration_energy} and \ref{fig:m_star_duration_energy} it also appears that, somewhat counter-intuitively, there are complex flares at the shortest \thalf\ values. These flares are impulsive events with high amplitude and short duration peaks, followed by secondary smaller peaks in a long decay which do not reach the \thalf\ limit. Therefore, when using the \thalf\ value (which is primarily a measure of the width of the flare peak), these complex flares can appear very short in length. An example of such a complex flare is shown in Fig.\,\ref{fig:very_impulsive_flare}. We have also presented how this flare would appear in \tess\ two minute cadence observations, showing how longer cadence observations can smear out flare substructure and reduce the measured amplitude \citep[e.g.][]{Yang18}.

\renewcommand{\arraystretch}{1}

\subsection{Flares and Starspot Phase} \label{sec:starspot_discussion}
In Sect.\,\ref{sec:flare_phase_dis} we used our sample of stars with detected flares and measured rotation periods to investigate whether the flare and starspot phases were related. We found that, similar to results from previous studies, flares occurred with a uniform distribution in starspot phase. This uniform distribution of stellar flares with starspot phase has been explained in previous works through various ways. One possibility is that we are observing active regions which are permanently visible, or visible the majority of the time. This can be due to the stellar inclination relative to us, high altitude spots, or a combination of the two. In this scenario flares could still be associated with active regions but appear uniform in phase, while the observed flux modulation would arise from changing visible fractions of the spots.

An alternative explanation is that we are observing multiple active regions, with the observed lightcurve modulation being due to their residual behaviour \citep[e.g. more spots on one side][]{Rackham18}. This would mean a starspot is effectively always in view, providing a relatively uniform distribution of active region associated flares. 

Due to the limits imposed on the flare amplitude, we are also currently only probing the behaviour of the highest energy events. Due to the intrinsic rarity of such events it is possible that more subtle variations are present that we are not sensitive to. This regime was explored by \citet{Roettenbacher18} using long cadence \kepler\ observations of 111 main sequence stars. They found that flares with amplitudes above 1 per cent occurred more often when a single large starspot group was in view, suggesting a spot-flare relation was present for low amplitude flares. Interestingly, they found a similar but less pronounced behaviour for a sample of 425 flares with amplitudes above 5 per cent, which we have not identified. This is likely due to our smaller sample of 82 flares from main sequence K and M stars with measured rotation periods. More recently \citet{Feinstein20} explored both flare amplitude regimes using \tess\ short cadence observations of young stars. They found that for young stars, flares of all amplitudes has no correlation with the rotational modulation phase.
This result is consistent with our observations for our pre-main sequence sample.

\subsection{White-light Flares from Subgiants} \label{sec:subgiant}
In Sect.\,\ref{sec:galactic_distribution} we noted that one of our flaring sources was likely associated with the Galactic thick disc and was located at the turn off to the subgiant branch on the HR diagram in Fig.\,\ref{fig:HR_diagram}. Subgiant stars have begun to evolve off of the main sequence and often show signs of variability. This variability can be due to tidally-induced rotation (e.g. RS CVn) or pulsations (e.g. Delta Scuti stars) and can sometimes be accompanied by flares. Following our analysis of TYC 5181-713-1 we searched for other subgiant candidates within our flare star sample. We did this by selecting stars from our flare star sample based on their position on the HR diagram, specifically around the subgiant turn off from the main sequence, with \teff\ between 4700 and 5600\,K and an absolute \Gaia\ G magnitude less than 4. We identified six more sources in this region near this turn off in Fig.\,\ref{fig:HR_diagram}. These objects are listed in Tab.\,\ref{tab:subgiant_values}, along with their measured periods and modulation amplitudes from our analysis in Sect.\,\ref{sec:rotation}.

\begin{figure}
    \centering
    \includegraphics[width=\columnwidth]{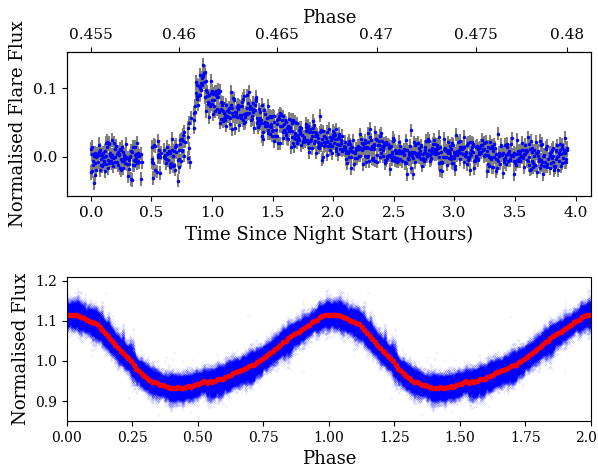}
    \caption{Top panel: The observed flare from NGTS J061739.4-651709 (2MASS J06173939-6517097). The top axis shows the modulation phase. Bottom panel: The phase folded lightcurve of NGTS J061739.4-651709, showing how the modulation is likely due to rotation rather than Cepheid-like variability. The blue points are the phase-folded NGTS lightcurve and the red points are the same data binned to 100 points in phase.}
    \label{fig:cepheid_flare}
\end{figure}

We performed a literature search for each object. Of note was NGTS J061739.4-651709 (2MASS J06173939-6517097), which has previously been classified by \citet{Watson06} and \citet{Chen18} as a Cepheid variable star using AAVSO and \textit{WISE} photometry respectively. The flare and phase-folded lightcurve of NGTS J061739.4-651709 is shown in Fig.\,\ref{fig:cepheid_flare}. Cepheids are pulsating stars for which the period of pulsation is correlated to their luminosities \citep[][]{Leavitt1912}. The presence of a white-light flare on a Cepheid variable would be of interest, due to the current debate about the causes of detected sudden UV and X-ray outbursts near their minimum and maximum radii, and whether this could be due to magnetic reconnection events \citep[e.g.][]{Engle17,Moschou2020}. In particular, we note that the flare from NGTS J061739.4-651709 occurs at the same phase as the X-ray flares observed by \citet{Engle17}, seemingly supporting the idea that these outbursts are due to magnetic reconnection events. However,  \citet{Jayasinghe18} used ASAS-SN photometry to classify the variability of NGTS J061739.4-651709 as due to rotation, rather than Cepheid-like variability. Further inspection of the NGTS lightcurve, shown in Fig.\,\ref{fig:cepheid_flare} shows that while the lightcurve does show asymmetric modulation, it is not consistent with the fast rise, slow decay expected for Cepheids, and is more likely due to rotation.

From our literature search we found that UCAC2 16087743 was identified as a spectroscopic binary candidate from RAVE data by \citet{Birko19,Jack19}. It is possible that this source is an RS CVn. RS CVns are variable tidally-locked close-binary stars in which one component can be a subgiant. RS CVns have previously been observed to show both flaring behaviour and modulation due to starspots, if one of the components is a K or M dwarf \citep[e.g. II Peg;][]{Frasca08}. It is also possible that the remaining subgiant candidates are also RS CVns. However, we note subgiants can have a wide range of periods and amplitudes of variability, making it difficult to classify these sources based off this information alone \citep[see][for a list of variables and associated periods]{Gaia_variable_18}.

\begin{table}
	\centering
	\begin{tabular}{|l|c|c|}
    \hline
    Object & \multicolumn{1}{|p{1cm}|}{\centering Period (days)} & \multicolumn{1}{|p{2cm}|}{\centering Amplitude (per cent)}\tabularnewline\hline
    NGTS J060634.4-250020&1.6&4.8\tabularnewline
    NGTS J061739.4-651709&6.8&18.3\tabularnewline
    NGTS J134437.3-334737 & N/A & N/A \tabularnewline 
    NGTS J141137.1-255427&N/A&N/A\tabularnewline
    NGTS J21440.3-393121& N/A & N/A \tabularnewline
    TYC 5181-713-1 &10.8&7.7\tabularnewline
    UCAC2 16087743&N/A&N/A\tabularnewline
	\hline
	\end{tabular}
    \caption{Parameters of flaring subgiant candidates. N/A indicates no rotation period was measured.}\label{tab:subgiant_values} 
\end{table}

\subsection{Role of flares in exoplanet habitability} \label{sec:exoplanet}
In recent years stellar flares have taken an increasingly important role in the discussions around exoplanet habitability. This is in part due to the discoveries of Earth-sized exoplanets in or near the ``habitable zones'' of M stars \citep[e.g.][]{Dittmann17,Bonfils18}. Due to the cool temperatures of these M stars their habitable zones reside at a close proximity to the host star, exaggerating the effects of flares. Detrimental effects from flares discussed in the literature include atmospheric erosion and ozone depletion due to flare-related XUV irradiation \citep[][]{Segura2010,Tilley17} and from bombardments of charged particles from associated Coronal Mass Ejections (CMEs) \citep[e.g.][]{Lammer07,Kay16}. 

On the other hand, the UV irradiation from flares has been noted as a possible way of facilitating prebiotic chemistry, which may require quiescent NUV flux not available from cool M stars. For active stars the UV irradiation could be provided by regularly flaring activity, with photo-chemistry occurring during flares and halting during stellar quiescence \citep[][]{Ranjan17,Rimmer18}. For tidally locked planets, \citet{Mullan18} have postulated that this activity could provide an analogue to the day/night cycle and a yearly cycle could even be simulated through stellar activity cycles. Through studying the rates of UV chemistry and biomolecular reactions, \citet{Rimmer18} identified minimum flaring limits required to bring a system into the ``abiogenesis zone''. From this work, they found that this mechanism may only be viable for the most active stars. 

To test whether the average flare rates measured in this work would be enough to reach the ``abiogenesis zone'', we compared our best fitting flare occurrence rates from Sect.\,\ref{sec:flare_frequency2} to the requirements from \citet{Rimmer18}, following the method outlined by \citet{Guenther20}. We found that none of our average flare rates, for main or pre-main sequence stars is enough to reach the abiogenesis zone and provide the required NUV flux. We found that a factor of approximately \textcolor{black}{2500} and \textcolor{black}{580} were required to bring the flare rates for main and pre-main sequence mid-M dwarfs into the abiogenesis region. This changed to \textcolor{black}{66000} and \textcolor{black}{3600} for main and pre-main sequence early M stars. Spectroscopic studies have shown that the 9000\,K flare model can often underestimate the flux in the $U$ and NUV regions, due to not including the Balmer lines and the associated jump \citep[][]{Kowalski13,Kowalski19}. However, simultaneous HST NUV and ground based optical spectroscopy observations have only measured the discrepancy to be a factor of two 
\citep{Kowalski19} and cannot provide the required NUV flux. This suggests then, at least for an average star, that flares cannot provide the required NUV flux, as determined by \citet{Rimmer18}, required for surface abiogenesis on habitable zone planets around M stars. However, we cannot rule out the possibility of individual highly-active stars reaching this region \citep[e.g.][]{Guenther20}.

\section{Conclusions}
We have presented the results of a search for stellar flares from the 196600 stars in the first \fieldnumber\ fields observed with the Next Generation Transit Survey. We identified \flarenumber\ flares from \starnumber\ stars. These flaring stars have measured spectral types between F8 and M6. We find the majority of our flares come from stars which are young or in binary systems. We have combined our flare observations with \Gaia\ DR2 to identify that the majority of our sources have kinematics consistent with the Galactic thin disc. We have identified seven likely subgiants which showed flaring activity. 

We have used the NGTS observations of flaring and non-flaring stars to measure the average flare occurrence rates for subsets of K and M stars. We presented a method to measure flare occurrence rates while accounting for how the detection efficiency changes with energy, removing the need to fit only to the highest energy flare events. We found that field age main sequence K and early M stars showed similar average white-light flare occurrence rates, while fully convective field age mid-M stars were more active. We believe this is due to the increased spin down timescale of fully convective M stars, which causes them to remain active into field ages. 
We have compared the average flare occurrence rates for main and pre-main sequence stars. We found that, as expected, for all spectral types pre-main sequence stars flare more often.

We measured the rotation period of 105 stars in our sample which showed visible starspot modulation. We compared the starspot phase to flare occurrence and found no correlation between the two, similar to previous studies, and discussed possible reasons for this. We also used the observed lightcurve modulation to estimate starspot size and combined this with the measured flare energies to estimate the magnetic field strengths for stars in our sample. We found that all flares and spots for these stars could be explained using a 2 kilogauss magnetic field strength.

We also discussed the implications of results for the effect of stellar flares on exoplanet habitability, in particular for exoplanets around M stars. We compared our average flare occurrence rates to the abiogenesis zone from \citet{Rimmer18}. We found that no spectral sample, at any age, in our sample had an average flare occurrence rate which reached the abiogenesis zone at our observed energies. However, we note this does not rule out the possibility of individual active stars flaring at the required rates as found in other surveys.

\section*{Acknowledgements} 
\edit{We are thankful to the referee, Manuel G\"udel, for their helpful comments and suggestions.} This research is based on data collected under the NGTS project at the ESO La Silla Paranal Observatory. The NGTS facility is funded by a consortium of institutes consisting of the University of Warwick,
the University of Leicester,
Queen's University Belfast,
the University of Geneva,
the Deutsches Zentrum f\" ur Luft- und Raumfahrt e.V. (DLR; under the `Gro\ss investition GI-NGTS'),
the University of Cambridge, together with the UK Science and Technology Facilities Council (STFC; project references ST/M001962/1 and ST/S002642/1). 
JAGJ was supported by STFC PhD studentship 1763096 for this work and now acknowledges support from grant HST-GO-15955.004-A from the Space Telescope Science Institute, which is operated by the Association of Universities for Research in Astronomy, Inc., under NASA contract NAS 5-26555. PJW, RGW, SG and BTG acknowledge support by STFC under consolidated grants ST/P000495/1 and ST/T000406/1. BTG also acknowledges support from a Leverhulme Research Fellowship.
CEP acknowledges support from the European
Research Council under the SeismoSun Research Project No. 321141. 
JSJ acknowledges support by FONDECYT grant 1201371, and partial support from CONICYT project Basal AFB-170002.
EG gratefully acknowledges support from the David and Claudia Harding Foundation in the form of a Winton Exoplanet Fellowship.

This publication makes use of data products from the Two Micron All Sky Survey, which is a joint project of the University of Massachusetts and the Infrared Processing and Analysis Center/California Institute of Technology, funded by the National Aeronautics and Space Administration and the National Science Foundation.
This publication makes use of data products from the Wide-field Infrared Survey Explorer, which is a joint project of the University of California, Los Angeles, and the Jet Propulsion Laboratory/California Institute of Technology, funded by the National Aeronautics and Space Administration. 
This work has made use of data from the European Space Agency (ESA) mission {\it Gaia} (\url{https://www.cosmos.esa.int/gaia}), processed by the {\it Gaia}
Data Processing and Analysis Consortium (DPAC,
\url{https://www.cosmos.esa.int/web/gaia/dpac/consortium}). Funding for the DPAC
has been provided by national institutions, in particular the institutions
participating in the {\it Gaia} Multilateral Agreement.

\section*{Data Availability}
Derived stellar and measured flare properties are available upon request to JAGJ.

\bibliographystyle{mnras}
\bibliography{references}

\begin{thebibliography}{}
\makeatletter
\relax
\def\mn@urlcharsother{\let\do\@makeother \do\$\do\&\do\#\do\^\do\_\do\%\do\~}
\def\mn@doi{\begingroup\mn@urlcharsother \@ifnextchar [ {\mn@doi@}
  {\mn@doi@[]}}
\def\mn@doi@[#1]#2{\def\@tempa{#1}\ifx\@tempa\@empty \href
  {http://dx.doi.org/#2} {doi:#2}\else \href {http://dx.doi.org/#2} {#1}\fi
  \endgroup}
\def\mn@eprint#1#2{\mn@eprint@#1:#2::\@nil}
\def\mn@eprint@arXiv#1{\href {http://arxiv.org/abs/#1} {{\tt arXiv:#1}}}
\def\mn@eprint@dblp#1{\href {http://dblp.uni-trier.de/rec/bibtex/#1.xml}
  {dblp:#1}}
\def\mn@eprint@#1:#2:#3:#4\@nil{\def\@tempa {#1}\def\@tempb {#2}\def\@tempc
  {#3}\ifx \@tempc \@empty \let \@tempc \@tempb \let \@tempb \@tempa \fi \ifx
  \@tempb \@empty \def\@tempb {arXiv}\fi \@ifundefined
  {mn@eprint@\@tempb}{\@tempb:\@tempc}{\expandafter \expandafter \csname
  mn@eprint@\@tempb\endcsname \expandafter{\@tempc}}}

\bibitem[\protect\citeauthoryear{{Alekseev} et~al.,}{{Alekseev}
  et~al.}{1994}]{Alekseev94}
{Alekseev} I.~Y.,  et~al., 1994, \aap, \href
  {http://adsabs.harvard.edu/abs/1994A%26A...288..502A} {288, 502}

\bibitem[\protect\citeauthoryear{{Andrews} \& {Panagi}}{{Andrews} \&
  {Panagi}}{1996}]{Andrews96}
{Andrews} A.~D.,  {Panagi} P.~M.,  1996, Irish Astronomical Journal, \href
  {http://adsabs.harvard.edu/abs/1996IrAJ...23..171A} {23}

\bibitem[\protect\citeauthoryear{{Anfinogentov}, {Nakariakov}, {Mathioudakis},
  {Van Doorsselaere}  \& {Kowalski}}{{Anfinogentov}
  et~al.}{2013}]{Anfinogentov13}
{Anfinogentov} S.,  {Nakariakov} V.~M.,  {Mathioudakis} M.,  {Van Doorsselaere}
  T.,   {Kowalski} A.~F.,  2013, \mn@doi [\apj] {10.1088/0004-637X/773/2/156},
  \href {https://ui.adsabs.harvard.edu/#abs/2013ApJ...773..156A} {773, 156}

\bibitem[\protect\citeauthoryear{{Arenou} et~al.,}{{Arenou}
  et~al.}{2018}]{Arenou18}
{Arenou} F.,  et~al., 2018, \mn@doi [\aap] {10.1051/0004-6361/201833234}, \href
  {https://ui.adsabs.harvard.edu/abs/2018A&A...616A..17A} {616, A17}

\bibitem[\protect\citeauthoryear{{Aschwanden}, {Tarbell}, {Nightingale},
  {Schrijver}, {Title}, {Kankelborg}, {Martens}  \& {Warren}}{{Aschwanden}
  et~al.}{2000}]{Aschwanden2000}
{Aschwanden} M.~J.,  {Tarbell} T.~D.,  {Nightingale} R.~W.,  {Schrijver} C.~J.,
   {Title} A.,  {Kankelborg} C.~C.,  {Martens} P.,   {Warren} H.~P.,  2000,
  \mn@doi [\apj] {10.1086/308867}, \href
  {https://ui.adsabs.harvard.edu/\#abs/2000ApJ...535.1047A} {535, 1047}

\bibitem[\protect\citeauthoryear{{Astropy Collaboration} et~al.,}{{Astropy
  Collaboration} et~al.}{2013}]{Astropy13}
{Astropy Collaboration} et~al., 2013, \mn@doi [\aap]
  {10.1051/0004-6361/201322068}, \href
  {http://adsabs.harvard.edu/abs/2013A%26A...558A..33A} {558, A33}

\bibitem[\protect\citeauthoryear{{Bailer-Jones}, {Rybizki}, {Fouesneau},
  {Mantelet}  \& {Andrae}}{{Bailer-Jones} et~al.}{2018}]{BailerJones18}
{Bailer-Jones} C.~A.~L.,  {Rybizki} J.,  {Fouesneau} M.,  {Mantelet} G.,
  {Andrae} R.,  2018, \mn@doi [\aj] {10.3847/1538-3881/aacb21}, \href
  {https://ui.adsabs.harvard.edu/#abs/2018AJ....156...58B} {156, 58}

\bibitem[\protect\citeauthoryear{{Balona}, {Broomhall}, {Kosovichev},
  {Nakariakov}, {Pugh}  \& {Van Doorsselaere}}{{Balona}
  et~al.}{2015}]{Balona15}
{Balona} L.~A.,  {Broomhall} A.-M.,  {Kosovichev} A.,  {Nakariakov} V.~M.,
  {Pugh} C.~E.,   {Van Doorsselaere} T.,  2015, \mn@doi [\mnras]
  {10.1093/mnras/stv661}, \href
  {http://adsabs.harvard.edu/abs/2015MNRAS.450..956B} {450, 956}

\bibitem[\protect\citeauthoryear{{Baraffe}, {Homeier}, {Allard}  \&
  {Chabrier}}{{Baraffe} et~al.}{2015}]{Baraffe15}
{Baraffe} I.,  {Homeier} D.,  {Allard} F.,   {Chabrier} G.,  2015, \mn@doi
  [\aap] {10.1051/0004-6361/201425481}, \href
  {https://ui.adsabs.harvard.edu/abs/2015A&A...577A..42B} {577, A42}

\bibitem[\protect\citeauthoryear{{Bensby}, {Feltzing}  \&
  {Lundstr{\"o}m}}{{Bensby} et~al.}{2003}]{Bensby03}
{Bensby} T.,  {Feltzing} S.,   {Lundstr{\"o}m} I.,  2003, \mn@doi [\aap]
  {10.1051/0004-6361:20031213}, \href
  {https://ui.adsabs.harvard.edu/#abs/2003A&A...410..527B} {410, 527}

\bibitem[\protect\citeauthoryear{{Bensby}, {Feltzing}  \&
  {Lundstr{\"o}m}}{{Bensby} et~al.}{2004}]{Bensby04}
{Bensby} T.,  {Feltzing} S.,   {Lundstr{\"o}m} I.,  2004, \mn@doi [\aap]
  {10.1051/0004-6361:20035957}, \href
  {https://ui.adsabs.harvard.edu/#abs/2004A&A...421..969B} {421, 969}

\bibitem[\protect\citeauthoryear{{Bensby}, {Feltzing}  \& {Oey}}{{Bensby}
  et~al.}{2014}]{Bensby14}
{Bensby} T.,  {Feltzing} S.,   {Oey} M.~S.,  2014, \mn@doi [\aap]
  {10.1051/0004-6361/201322631}, \href
  {https://ui.adsabs.harvard.edu/#abs/2014A&A...562A..71B} {562, A71}

\bibitem[\protect\citeauthoryear{{Benz}}{{Benz}}{2017}]{Benz17}
{Benz} A.~O.,  2017, \mn@doi [Living Reviews in Solar Physics]
  {10.1007/s41116-016-0004-3}, \href
  {http://adsabs.harvard.edu/abs/2017LRSP...14....2B} {14, 2}

\bibitem[\protect\citeauthoryear{{Benz} \& {G{\"u}del}}{{Benz} \&
  {G{\"u}del}}{2010}]{Benz10}
{Benz} A.~O.,  {G{\"u}del} M.,  2010, \mn@doi [\araa]
  {10.1146/annurev-astro-082708-101757}, \href
  {http://adsabs.harvard.edu/abs/2010ARA%26A..48..241B} {48, 241}

\bibitem[\protect\citeauthoryear{{Berdyugina}}{{Berdyugina}}{2005}]{Berdyugina05}
{Berdyugina} S.~V.,  2005, \mn@doi [Living Reviews in Solar Physics]
  {10.12942/lrsp-2005-8}, \href
  {https://ui.adsabs.harvard.edu/abs/2005LRSP....2....8B} {2, 8}

\bibitem[\protect\citeauthoryear{{Birko} et~al.,}{{Birko}
  et~al.}{2019}]{Birko19}
{Birko} D.,  et~al., 2019, \mn@doi [\aj] {10.3847/1538-3881/ab3cc1}, \href
  {https://ui.adsabs.harvard.edu/abs/2019AJ....158..155B} {158, 155}

\bibitem[\protect\citeauthoryear{{Bonfils} et~al.,}{{Bonfils}
  et~al.}{2018}]{Bonfils18}
{Bonfils} X.,  et~al., 2018, \mn@doi [\aap] {10.1051/0004-6361/201731973},
  \href {https://ui.adsabs.harvard.edu/#abs/2018A&A...613A..25B} {613, A25}

\bibitem[\protect\citeauthoryear{{Booth}, {Poppenhaeger}, {Watson}, {Silva
  Aguirre}  \& {Wolk}}{{Booth} et~al.}{2017}]{Booth17}
{Booth} R.~S.,  {Poppenhaeger} K.,  {Watson} C.~A.,  {Silva Aguirre} V.,
  {Wolk} S.~J.,  2017, \mn@doi [\mnras] {10.1093/mnras/stx1630}, \href
  {https://ui.adsabs.harvard.edu/#abs/2017MNRAS.471.1012B} {471, 1012}

\bibitem[\protect\citeauthoryear{{Borucki} et~al.,}{{Borucki}
  et~al.}{2010}]{Borucki2010}
{Borucki} W.~J.,  et~al., 2010, \mn@doi [Science] {10.1126/science.1185402},
  \href {http://adsabs.harvard.edu/abs/2010Sci...327..977B} {327, 977}

\bibitem[\protect\citeauthoryear{{Bressan}, {Marigo}, {Girardi}, {Salasnich},
  {Dal Cero}, {Rubele}  \& {Nanni}}{{Bressan} et~al.}{2012}]{Bressan12}
{Bressan} A.,  {Marigo} P.,  {Girardi} L.,  {Salasnich} B.,  {Dal Cero} C.,
  {Rubele} S.,   {Nanni} A.,  2012, \mn@doi [\mnras]
  {10.1111/j.1365-2966.2012.21948.x}, \href
  {https://ui.adsabs.harvard.edu/abs/2012MNRAS.427..127B} {427, 127}

\bibitem[\protect\citeauthoryear{{Carrington}}{{Carrington}}{1859}]{Carrington_1859}
{Carrington} R.~C.,  1859, \mn@doi [\mnras] {10.1093/mnras/20.1.13}, \href
  {http://adsabs.harvard.edu/abs/1859MNRAS..20...13C} {20, 13}

\bibitem[\protect\citeauthoryear{{Chen}, {Wang}, {Deng}, {de Grijs}  \&
  {Yang}}{{Chen} et~al.}{2018}]{Chen18}
{Chen} X.,  {Wang} S.,  {Deng} L.,  {de Grijs} R.,   {Yang} M.,  2018, \mn@doi
  [\apjs] {10.3847/1538-4365/aad32b}, \href
  {https://ui.adsabs.harvard.edu/abs/2018ApJS..237...28C} {237, 28}

\bibitem[\protect\citeauthoryear{{Collier Cameron} et~al.,}{{Collier Cameron}
  et~al.}{2006}]{Collier06}
{Collier Cameron} A.,  et~al., 2006, \mn@doi [\mnras]
  {10.1111/j.1365-2966.2006.11074.x}, \href
  {https://ui.adsabs.harvard.edu/abs/2006MNRAS.373..799C} {373, 799}

\bibitem[\protect\citeauthoryear{{Crosby}, {Aschwanden}  \& {Dennis}}{{Crosby}
  et~al.}{1993}]{Crosby93}
{Crosby} N.~B.,  {Aschwanden} M.~J.,   {Dennis} B.~R.,  1993, \mn@doi
  [\solphys] {10.1007/BF00646488}, \href
  {https://ui.adsabs.harvard.edu/\#abs/1993SoPh..143..275C} {143, 275}

\bibitem[\protect\citeauthoryear{{Cutri} \& {et al.}}{{Cutri} \& {et
  al.}}{2014}]{ALLWISE2014}
{Cutri} R.~M.,  {et al.} 2014, VizieR Online Data Catalog, \href
  {http://cdsads.u-strasbg.fr/abs/2014yCat.2328....0C} {2328}

\bibitem[\protect\citeauthoryear{{Davenport}}{{Davenport}}{2016}]{Davenport16}
{Davenport} J.~R.~A.,  2016, \mn@doi [\apj] {10.3847/0004-637X/829/1/23}, \href
  {http://adsabs.harvard.edu/abs/2016ApJ...829...23D} {829, 23}

\bibitem[\protect\citeauthoryear{{Davenport} et~al.,}{{Davenport}
  et~al.}{2014}]{Davenport2014}
{Davenport} J.~R.~A.,  et~al., 2014, \mn@doi [\apj]
  {10.1088/0004-637X/797/2/122}, \href
  {http://adsabs.harvard.edu/abs/2014ApJ...797..122D} {797, 122}

\bibitem[\protect\citeauthoryear{{Davenport}, {Covey}, {Clarke}, {Boeck},
  {Cornet}  \& {Hawley}}{{Davenport} et~al.}{2019}]{Davenport19}
{Davenport} J. R.~A.,  {Covey} K.~R.,  {Clarke} R.~W.,  {Boeck} A.~C.,
  {Cornet} J.,   {Hawley} S.~L.,  2019, \mn@doi [\apj]
  {10.3847/1538-4357/aafb76}, \href
  {https://ui.adsabs.harvard.edu/abs/2019ApJ...871..241D} {871, 241}

\bibitem[\protect\citeauthoryear{{Dittmann} et~al.,}{{Dittmann}
  et~al.}{2017}]{Dittmann17}
{Dittmann} J.~A.,  et~al., 2017, \mn@doi [\nat] {10.1038/nature22055}, \href
  {https://ui.adsabs.harvard.edu/#abs/2017Natur.544..333D} {544, 333}

\bibitem[\protect\citeauthoryear{{Douglas}, {Ag{\"u}eros}, {Covey}, {Cargile},
  {Barclay}, {Cody}, {Howell}  \& {Kopytova}}{{Douglas}
  et~al.}{2016}]{Douglas16}
{Douglas} S.~T.,  {Ag{\"u}eros} M.~A.,  {Covey} K.~R.,  {Cargile} P.~A.,
  {Barclay} T.,  {Cody} A.,  {Howell} S.~B.,   {Kopytova} T.,  2016, \mn@doi
  [\apj] {10.3847/0004-637X/822/1/47}, \href
  {https://ui.adsabs.harvard.edu/abs/2016ApJ...822...47D} {822, 47}

\bibitem[\protect\citeauthoryear{{Doyle} \& {Butler}}{{Doyle} \&
  {Butler}}{1985}]{Doyle85}
{Doyle} J.~G.,  {Butler} C.~J.,  1985, \mn@doi [\nat] {10.1038/313378a0}, \href
  {https://ui.adsabs.harvard.edu/abs/1985Natur.313..378D} {313, 378}

\bibitem[\protect\citeauthoryear{{Doyle}, {Ramsay}, {Doyle}, {Wu}  \&
  {Scullion}}{{Doyle} et~al.}{2018}]{Doyle18}
{Doyle} L.,  {Ramsay} G.,  {Doyle} J.~G.,  {Wu} K.,   {Scullion} E.,  2018,
  preprint, \href {http://adsabs.harvard.edu/abs/2018arXiv180708592D} {}
  (\mn@eprint {arXiv} {1807.08592})

\bibitem[\protect\citeauthoryear{{Doyle}, {Ramsay}, {Doyle}  \& {Wu}}{{Doyle}
  et~al.}{2019}]{Doyle19}
{Doyle} L.,  {Ramsay} G.,  {Doyle} J.~G.,   {Wu} K.,  2019, \mn@doi [\mnras]
  {10.1093/mnras/stz2205}, \href
  {https://ui.adsabs.harvard.edu/abs/2019MNRAS.tmp.2115D} {p.~2115}

\bibitem[\protect\citeauthoryear{{Engle}, {Guinan}, {Harper}, {Cuntz}, {Remage
  Evans}, {Neilson}  \& {Fawzy}}{{Engle} et~al.}{2017}]{Engle17}
{Engle} S.~G.,  {Guinan} E.~F.,  {Harper} G.~M.,  {Cuntz} M.,  {Remage Evans}
  N.,  {Neilson} H.~R.,   {Fawzy} D.~E.,  2017, \mn@doi [\apj]
  {10.3847/1538-4357/aa6159}, \href
  {https://ui.adsabs.harvard.edu/abs/2017ApJ...838...67E} {838, 67}

\bibitem[\protect\citeauthoryear{{Feinstein}, {Montet}, {Ansdell}, {Nord},
  {Bean}, {G{\"u}nther}, {Gully-Santiago}  \& {Schlieder}}{{Feinstein}
  et~al.}{2020}]{Feinstein20}
{Feinstein} A.~D.,  {Montet} B.~T.,  {Ansdell} M.,  {Nord} B.,  {Bean} J.~L.,
  {G{\"u}nther} M.~N.,  {Gully-Santiago} M.~A.,   {Schlieder} J.~E.,  2020,
  arXiv e-prints, \href {https://ui.adsabs.harvard.edu/abs/2020arXiv200507710F}
  {p. arXiv:2005.07710}

\bibitem[\protect\citeauthoryear{{Foreman-Mackey}, {Hogg}, {Lang}  \&
  {Goodman}}{{Foreman-Mackey} et~al.}{2013}]{Foreman-Mackey13}
{Foreman-Mackey} D.,  {Hogg} D.~W.,  {Lang} D.,   {Goodman} J.,  2013, \mn@doi
  [\pasp] {10.1086/670067}, \href
  {http://adsabs.harvard.edu/abs/2013PASP..125..306F} {125, 306}

\bibitem[\protect\citeauthoryear{{Frasca}, {Biazzo}, {Ta{\textcommabelow s}},
  {Evren}  \& {Lanzafame}}{{Frasca} et~al.}{2008}]{Frasca08}
{Frasca} A.,  {Biazzo} K.,  {Ta{\textcommabelow s}} G.,  {Evren} S.,
  {Lanzafame} A.~C.,  2008, \mn@doi [\aap] {10.1051/0004-6361:20077915}, \href
  {https://ui.adsabs.harvard.edu/abs/2008A&A...479..557F} {479, 557}

\bibitem[\protect\citeauthoryear{{Gaia Collaboration} et~al.,}{{Gaia
  Collaboration} et~al.}{2016}]{Gaia2016}
{Gaia Collaboration} et~al., 2016, \mn@doi [\aap]
  {10.1051/0004-6361/201629512}, \href
  {http://adsabs.harvard.edu/abs/2016A%26A...595A...2G} {595, A2}

\bibitem[\protect\citeauthoryear{{Gaia Collaboration} et~al.,}{{Gaia
  Collaboration} et~al.}{2018b}]{Gaia_HR}
{Gaia Collaboration} et~al., 2018b, preprint, \href
  {http://adsabs.harvard.edu/abs/2018arXiv180409378G} {} (\mn@eprint {arXiv}
  {1804.09378})

\bibitem[\protect\citeauthoryear{{Gaia Collaboration}, {Brown}, {Vallenari},
  {Prusti}, {de Bruijne}, {Babusiaux}  \& {Bailer-Jones}}{{Gaia Collaboration}
  et~al.}{2018a}]{GaiaDR2}
{Gaia Collaboration} {Brown} A.~G.~A.,  {Vallenari} A.,  {Prusti} T.,  {de
  Bruijne} J.~H.~J.,  {Babusiaux} C.,   {Bailer-Jones} C.~A.~L.,  2018a,
  preprint, \href {http://adsabs.harvard.edu/abs/2018arXiv180409365G} {}
  (\mn@eprint {arXiv} {1804.09365})

\bibitem[\protect\citeauthoryear{{Gaia Collaboration} et~al.,}{{Gaia
  Collaboration} et~al.}{2019}]{Gaia_variable_18}
{Gaia Collaboration} et~al., 2019, \mn@doi [\aap]
  {10.1051/0004-6361/201833304}, \href
  {https://ui.adsabs.harvard.edu/abs/2019A&A...623A.110G} {623, A110}

\bibitem[\protect\citeauthoryear{{Gershberg} \& {Shakhovskaia}}{{Gershberg} \&
  {Shakhovskaia}}{1983}]{Gershberg83}
{Gershberg} R.~E.,  {Shakhovskaia} N.~I.,  1983, \mn@doi [\apss]
  {10.1007/BF00653631}, \href
  {http://adsabs.harvard.edu/abs/1983Ap%26SS..95..235G} {95, 235}

\bibitem[\protect\citeauthoryear{{Gilliland} et~al.,}{{Gilliland}
  et~al.}{2010}]{Gilliland2011}
{Gilliland} R.~L.,  et~al., 2010, \mn@doi [\apjl]
  {10.1088/2041-8205/713/2/L160}, \href
  {http://adsabs.harvard.edu/abs/2010ApJ...713L.160G} {713, L160}

\bibitem[\protect\citeauthoryear{{Gizis}, {Paudel}, {Schmidt}, {Williams}  \&
  {Burgasser}}{{Gizis} et~al.}{2017}]{Gizis17b}
{Gizis} J.~E.,  {Paudel} R.~R.,  {Schmidt} S.~J.,  {Williams} P.~K.~G.,
  {Burgasser} A.~J.,  2017, \mn@doi [\apj] {10.3847/1538-4357/aa6197}, \href
  {http://adsabs.harvard.edu/abs/2017ApJ...838...22G} {838, 22}

\bibitem[\protect\citeauthoryear{{G{\"u}del}, {Audard}, {Kashyap}, {Drake}  \&
  {Guinan}}{{G{\"u}del} et~al.}{2003}]{Gudel03}
{G{\"u}del} M.,  {Audard} M.,  {Kashyap} V.~L.,  {Drake} J.~J.,   {Guinan}
  E.~F.,  2003, \mn@doi [\apj] {10.1086/344614}, \href
  {https://ui.adsabs.harvard.edu/abs/2003ApJ...582..423G} {582, 423}

\bibitem[\protect\citeauthoryear{{G{\"u}nther} et~al.,}{{G{\"u}nther}
  et~al.}{2017}]{Gunther17}
{G{\"u}nther} M.~N.,  et~al., 2017, \mn@doi [\mnras] {10.1093/mnras/stx1920},
  472, 295

\bibitem[\protect\citeauthoryear{{G{\"u}nther} et~al.,}{{G{\"u}nther}
  et~al.}{2020}]{Guenther20}
{G{\"u}nther} M.~N.,  et~al., 2020, \mn@doi [\aj] {10.3847/1538-3881/ab5d3a},
  \href {https://ui.adsabs.harvard.edu/abs/2020AJ....159...60G} {159, 60}

\bibitem[\protect\citeauthoryear{{Hawley} \& {Fisher}}{{Hawley} \&
  {Fisher}}{1992}]{Hawley1992}
{Hawley} S.~L.,  {Fisher} G.~H.,  1992, \mn@doi [\apjs] {10.1086/191640}, \href
  {http://adsabs.harvard.edu/abs/1992ApJS...78..565H} {78, 565}

\bibitem[\protect\citeauthoryear{{Hawley}, {Davenport}, {Kowalski},
  {Wisniewski}, {Hebb}, {Deitrick}  \& {Hilton}}{{Hawley}
  et~al.}{2014}]{Hawley14}
{Hawley} S.~L.,  {Davenport} J.~R.~A.,  {Kowalski} A.~F.,  {Wisniewski} J.~P.,
  {Hebb} L.,  {Deitrick} R.,   {Hilton} E.~J.,  2014, \mn@doi [\apj]
  {10.1088/0004-637X/797/2/121}, \href
  {http://adsabs.harvard.edu/abs/2014ApJ...797..121H} {797, 121}

\bibitem[\protect\citeauthoryear{{Hayashi}}{{Hayashi}}{1961}]{Hayashi61}
{Hayashi} C.,  1961, \pasj, \href
  {https://ui.adsabs.harvard.edu/abs/1961PASJ...13..450H} {13, 450}

\bibitem[\protect\citeauthoryear{{Henden} \& {Munari}}{{Henden} \&
  {Munari}}{2014}]{APASS_14}
{Henden} A.,  {Munari} U.,  2014, Contributions of the Astronomical Observatory
  Skalnate Pleso, \href {http://adsabs.harvard.edu/abs/2014CoSka..43..518H}
  {43, 518}

\bibitem[\protect\citeauthoryear{{Hilton}, {Hawley}, {Kowalski}  \&
  {Holtzman}}{{Hilton} et~al.}{2011}]{Hilton11}
{Hilton} E.~J.,  {Hawley} S.~L.,  {Kowalski} A.~F.,   {Holtzman} J.,  2011, in
  {Johns-Krull} C.,  {Browning} M.~K.,   {West} A.~A.,  eds,  Astronomical
  Society of the Pacific Conference Series Vol. 448, 16th Cambridge Workshop on
  Cool Stars, Stellar Systems, and the Sun. p.~197

\bibitem[\protect\citeauthoryear{{Howard}, {Corbett}, {Law}, {Ratzloff},
  {Glazier}, {Fors}, {del Ser}  \& {Haislip}}{{Howard}
  et~al.}{2019a}]{Howard19spot}
{Howard} W.~S.,  {Corbett} H.,  {Law} N.~M.,  {Ratzloff} J.~K.,  {Glazier} A.,
  {Fors} O.,  {del Ser} D.,   {Haislip} J.,  2019a, arXiv e-prints, \href
  {https://ui.adsabs.harvard.edu/abs/2019arXiv190710735H} {p. arXiv:1907.10735}

\bibitem[\protect\citeauthoryear{{Howard}, {Corbett}, {Law}, {Ratzloff},
  {Glazier}, {Fors}, {del Ser}  \& {Haislip}}{{Howard}
  et~al.}{2019b}]{Howard19}
{Howard} W.~S.,  {Corbett} H.,  {Law} N.~M.,  {Ratzloff} J.~K.,  {Glazier} A.,
  {Fors} O.,  {del Ser} D.,   {Haislip} J.,  2019b, \mn@doi [\apj]
  {10.3847/1538-4357/ab2767}, \href
  {https://ui.adsabs.harvard.edu/abs/2019ApJ...881....9H} {881, 9}

\bibitem[\protect\citeauthoryear{{Howard} et~al.,}{{Howard}
  et~al.}{2020}]{Howard20temp}
{Howard} W.~S.,  et~al., 2020, \mn@doi [\apj] {10.3847/1538-4357/abb5b4}, \href
  {https://ui.adsabs.harvard.edu/abs/2020ApJ...902..115H} {902, 115}

\bibitem[\protect\citeauthoryear{{Ilin}, {Schmidt}, {Davenport}  \&
  {Strassmeier}}{{Ilin} et~al.}{2019}]{Ilin19}
{Ilin} E.,  {Schmidt} S.~J.,  {Davenport} J. R.~A.,   {Strassmeier} K.~G.,
  2019, \mn@doi [\aap] {10.1051/0004-6361/201834400}, \href
  {https://ui.adsabs.harvard.edu/abs/2019A&A...622A.133I} {622, A133}

\bibitem[\protect\citeauthoryear{{Ilin}, {Schmidt}, {Poppenh{\"a}ger},
  {Davenport}, {Kristiansen}  \& {Omohundro}}{{Ilin} et~al.}{2020}]{Ilin20}
{Ilin} E.,  {Schmidt} S.~J.,  {Poppenh{\"a}ger} K.,  {Davenport} J. R.~A.,
  {Kristiansen} M.~H.,   {Omohundro} M.,  2020, arXiv e-prints, \href
  {https://ui.adsabs.harvard.edu/abs/2020arXiv201005576I} {p. arXiv:2010.05576}

\bibitem[\protect\citeauthoryear{{Jack}}{{Jack}}{2019}]{Jack19}
{Jack} D.,  2019, \mn@doi [Astronomische Nachrichten] {10.1002/asna.201913496},
  \href {https://ui.adsabs.harvard.edu/abs/2019AN....340..386J} {340, 386}

\bibitem[\protect\citeauthoryear{{Jackman} et~al.,}{{Jackman}
  et~al.}{2018}]{Jackman18}
{Jackman} J.~A.~G.,  et~al., 2018, \mn@doi [\mnras] {10.1093/mnras/sty897},
  \href {http://adsabs.harvard.edu/abs/2018MNRAS.477.4655J} {477, 4655}

\bibitem[\protect\citeauthoryear{{Jackman} et~al.,}{{Jackman}
  et~al.}{2019a}]{JackmanQPP}
{Jackman} J. A.~G.,  et~al., 2019a, \mn@doi [\mnras] {10.1093/mnras/sty3036},
  \href {https://ui.adsabs.harvard.edu/abs/2019MNRAS.482.5553J} {482, 5553}

\bibitem[\protect\citeauthoryear{{Jackman} et~al.,}{{Jackman}
  et~al.}{2019b}]{Jackman19}
{Jackman} J. A.~G.,  et~al., 2019b, \mn@doi [\mnras] {10.1093/mnrasl/slz039},
  \href {https://ui.adsabs.harvard.edu/abs/2019MNRAS.485L.136J} {485, L136}

\bibitem[\protect\citeauthoryear{{Jackman} et~al.,}{{Jackman}
  et~al.}{2019c}]{JackmanBD}
{Jackman} J. A.~G.,  et~al., 2019c, \mn@doi [\mnras] {10.1093/mnras/stz2496},
  \href {https://ui.adsabs.harvard.edu/abs/2019MNRAS.489.5146J} {489, 5146}

\bibitem[\protect\citeauthoryear{{Jackman} et~al.,}{{Jackman}
  et~al.}{2020}]{Jackman20}
{Jackman} J. A.~G.,  et~al., 2020, \mn@doi [\mnras] {10.1093/mnras/staa1971},
  \href {https://ui.adsabs.harvard.edu/abs/2020MNRAS.tmp.2076J} {}

\bibitem[\protect\citeauthoryear{{Jayasinghe} et~al.,}{{Jayasinghe}
  et~al.}{2018}]{Jayasinghe18}
{Jayasinghe} T.,  et~al., 2018, \mn@doi [\mnras] {10.1093/mnras/sty838}, \href
  {https://ui.adsabs.harvard.edu/abs/2018MNRAS.477.3145J} {477, 3145}

\bibitem[\protect\citeauthoryear{{Jiang}, {Bi}, {Yang}, {Zheng}  \&
  {Wang}}{{Jiang} et~al.}{2009}]{Jiang09}
{Jiang} Y.-C.,  {Bi} Y.,  {Yang} J.-Y.,  {Zheng} R.-S.,   {Wang} J.-X.,  2009,
  \mn@doi [Research in Astronomy and Astrophysics] {10.1088/1674-4527/9/5/011},
  \href {http://adsabs.harvard.edu/abs/2009RAA.....9..603J} {9, 603}

\bibitem[\protect\citeauthoryear{{Jiang}, {Zheng}, {Yang}, {Hong}, {Yi}  \&
  {Yang}}{{Jiang} et~al.}{2012}]{Jiang12}
{Jiang} Y.,  {Zheng} R.,  {Yang} J.,  {Hong} J.,  {Yi} B.,   {Yang} D.,  2012,
  \mn@doi [\apj] {10.1088/0004-637X/744/1/50}, \href
  {https://ui.adsabs.harvard.edu/abs/2012ApJ...744...50J} {744, 50}

\bibitem[\protect\citeauthoryear{{K{\H{o}}v{\'a}ri} et~al.,}{{K{\H{o}}v{\'a}ri}
  et~al.}{2020}]{Kovari20}
{K{\H{o}}v{\'a}ri} Z.,  et~al., 2020, arXiv e-prints, \href
  {https://ui.adsabs.harvard.edu/abs/2020arXiv200505397K} {p. arXiv:2005.05397}

\bibitem[\protect\citeauthoryear{{Katz} et~al.,}{{Katz} et~al.}{2018}]{Katz18}
{Katz} D.,  et~al., 2018, \mn@doi [\aap] {10.1051/0004-6361/201832865}, \href
  {https://ui.adsabs.harvard.edu/#abs/2018A&A...616A..11G} {616, A11}

\bibitem[\protect\citeauthoryear{{Kay}, {Opher}  \& {Kornbleuth}}{{Kay}
  et~al.}{2016}]{Kay16}
{Kay} C.,  {Opher} M.,   {Kornbleuth} M.,  2016, \mn@doi [\apj]
  {10.3847/0004-637X/826/2/195}, \href
  {http://adsabs.harvard.edu/abs/2016ApJ...826..195K} {826, 195}

\bibitem[\protect\citeauthoryear{{Kounkel} et~al.,}{{Kounkel}
  et~al.}{2018}]{Kounkel18}
{Kounkel} M.,  et~al., 2018, preprint, \href
  {http://adsabs.harvard.edu/abs/2018arXiv180504649K} {} (\mn@eprint {arXiv}
  {1805.04649})

\bibitem[\protect\citeauthoryear{{Kowalski}, {Hawley}, {Wisniewski}, {Osten},
  {Hilton}, {Holtzman}, {Schmidt}  \& {Davenport}}{{Kowalski}
  et~al.}{2013}]{Kowalski13}
{Kowalski} A.~F.,  {Hawley} S.~L.,  {Wisniewski} J.~P.,  {Osten} R.~A.,
  {Hilton} E.~J.,  {Holtzman} J.~A.,  {Schmidt} S.~J.,   {Davenport} J.~R.~A.,
  2013, \mn@doi [\apjs] {10.1088/0067-0049/207/1/15}, \href
  {http://adsabs.harvard.edu/abs/2013ApJS..207...15K} {207, 15}

\bibitem[\protect\citeauthoryear{{Kowalski} et~al.,}{{Kowalski}
  et~al.}{2019}]{Kowalski19}
{Kowalski} A.~F.,  et~al., 2019, \mn@doi [\apj] {10.3847/1538-4357/aaf058},
  \href {https://ui.adsabs.harvard.edu/abs/2019ApJ...871..167K} {871, 167}

\bibitem[\protect\citeauthoryear{{Kunder} et~al.,}{{Kunder}
  et~al.}{2017}]{RAVE17}
{Kunder} A.,  et~al., 2017, \mn@doi [\aj] {10.3847/1538-3881/153/2/75}, \href
  {http://adsabs.harvard.edu/abs/2017AJ....153...75K} {153, 75}

\bibitem[\protect\citeauthoryear{{Lacy}, {Moffett}  \& {Evans}}{{Lacy}
  et~al.}{1976}]{Lacy76}
{Lacy} C.~H.,  {Moffett} T.~J.,   {Evans} D.~S.,  1976, \mn@doi [\apjs]
  {10.1086/190358}, \href {http://adsabs.harvard.edu/abs/1976ApJS...30...85L}
  {30, 85}

\bibitem[\protect\citeauthoryear{{Lammer} et~al.,}{{Lammer}
  et~al.}{2007}]{Lammer07}
{Lammer} H.,  et~al., 2007, \mn@doi [Astrobiology] {10.1089/ast.2006.0128},
  \href {http://adsabs.harvard.edu/abs/2007AsBio...7..185L} {7, 185}

\bibitem[\protect\citeauthoryear{{Leavitt} \& {Pickering}}{{Leavitt} \&
  {Pickering}}{1912}]{Leavitt1912}
{Leavitt} H.~S.,  {Pickering} E.~C.,  1912, Harvard College Observatory
  Circular, \href {https://ui.adsabs.harvard.edu/abs/1912HarCi.173....1L} {173,
  1}

\bibitem[\protect\citeauthoryear{{Loyd}, {Shkolnik}, {Schneider}, {Barman},
  {Meadows}, {Pagano}  \& {Peacock}}{{Loyd} et~al.}{2018a}]{Loyd18}
{Loyd} R.~O.~P.,  {Shkolnik} E.~L.,  {Schneider} A.~C.,  {Barman} T.~S.,
  {Meadows} V.~S.,  {Pagano} I.,   {Peacock} S.,  2018a, \mn@doi [\apj]
  {10.3847/1538-4357/aae2ae}, \href
  {https://ui.adsabs.harvard.edu/abs/2018ApJ...867...70L} {867, 70}

\bibitem[\protect\citeauthoryear{{Loyd} et~al.,}{{Loyd}
  et~al.}{2018b}]{Loyd18b}
{Loyd} R.~O.~P.,  et~al., 2018b, \mn@doi [\apj] {10.3847/1538-4357/aae2bd},
  \href {https://ui.adsabs.harvard.edu/abs/2018ApJ...867...71L} {867, 71}

\bibitem[\protect\citeauthoryear{{Lu}, {Zhang}, {Shi}, {Han}, {Fan}, {Long}  \&
  {Pi}}{{Lu} et~al.}{2019}]{Lu19}
{Lu} H.-p.,  {Zhang} L.-y.,  {Shi} J.,  {Han} X.~L.,  {Fan} D.,  {Long} L.,
  {Pi} Q.,  2019, \mn@doi [\apjs] {10.3847/1538-4365/ab2f8f}, \href
  {https://ui.adsabs.harvard.edu/abs/2019ApJS..243...28L} {243, 28}

\bibitem[\protect\citeauthoryear{{Maehara} et~al.,}{{Maehara}
  et~al.}{2012}]{Maehara2012}
{Maehara} H.,  et~al., 2012, \mn@doi [\nat] {10.1038/nature11063}, \href
  {http://adsabs.harvard.edu/abs/2012Natur.485..478M} {485, 478}

\bibitem[\protect\citeauthoryear{{Maehara}, {Shibayama}, {Notsu}, {Notsu},
  {Honda}, {Nogami}  \& {Shibata}}{{Maehara} et~al.}{2016}]{Maehara16}
{Maehara} H.,  {Shibayama} T.,  {Notsu} Y.,  {Notsu} S.,  {Honda} S.,  {Nogami}
  D.,   {Shibata} K.,  2016, in {Kosovichev} A.~G.,  {Hawley} S.~L.,
  {Heinzel} P.,  eds,  IAU Symposium Vol. 320, Solar and Stellar Flares and
  their Effects on Planets. pp 144--149 (\mn@eprint {arXiv} {1601.01132}),
  \mn@doi{10.1017/S1743921316000065}

\bibitem[\protect\citeauthoryear{{Maehara}, {Notsu}, {Notsu}, {Namekata},
  {Honda}, {Ishii}, {Nogami}  \& {Shibata}}{{Maehara} et~al.}{2017}]{Maehara17}
{Maehara} H.,  {Notsu} Y.,  {Notsu} S.,  {Namekata} K.,  {Honda} S.,  {Ishii}
  T.~T.,  {Nogami} D.,   {Shibata} K.,  2017, \mn@doi [\pasj]
  {10.1093/pasj/psx013}, \href
  {https://ui.adsabs.harvard.edu/abs/2017PASJ...69...41M} {69, 41}

\bibitem[\protect\citeauthoryear{{Martin} et~al.,}{{Martin}
  et~al.}{2005}]{GALEX_05}
{Martin} D.~C.,  et~al., 2005, \mn@doi [\apjl] {10.1086/426387}, \href
  {http://adsabs.harvard.edu/abs/2005ApJ...619L...1M} {619, L1}

\bibitem[\protect\citeauthoryear{{Matt}, {Brun}, {Baraffe}, {Bouvier}  \&
  {Chabrier}}{{Matt} et~al.}{2015}]{Matt15}
{Matt} S.~P.,  {Brun} A.~S.,  {Baraffe} I.,  {Bouvier} J.,   {Chabrier} G.,
  2015, \mn@doi [\apjl] {10.1088/2041-8205/799/2/L23}, \href
  {https://ui.adsabs.harvard.edu/abs/2015ApJ...799L..23M} {799, L23}

\bibitem[\protect\citeauthoryear{{Medina}, {Winters}, {Irwin}  \&
  {Charbonneau}}{{Medina} et~al.}{2020}]{Medina20}
{Medina} A.~A.,  {Winters} J.~G.,  {Irwin} J.~M.,   {Charbonneau} D.,  2020,
  arXiv e-prints, \href {https://ui.adsabs.harvard.edu/abs/2020arXiv201015635M}
  {p. arXiv:2010.15635}

\bibitem[\protect\citeauthoryear{{Moffett}}{{Moffett}}{1974}]{Moffett74}
{Moffett} T.~J.,  1974, \mn@doi [The Astrophysical Journal Supplement Series]
  {10.1086/190330}, \href
  {https://ui.adsabs.harvard.edu/abs/1974ApJS...29....1M} {29, 1}

\bibitem[\protect\citeauthoryear{{Morin} et~al.,}{{Morin}
  et~al.}{2008}]{Morin08}
{Morin} J.,  et~al., 2008, \mn@doi [\mnras] {10.1111/j.1365-2966.2008.13809.x},
  \href {https://ui.adsabs.harvard.edu/abs/2008MNRAS.390..567M} {390, 567}

\bibitem[\protect\citeauthoryear{{Morin}, {Donati}, {Petit}, {Delfosse},
  {Forveille}  \& {Jardine}}{{Morin} et~al.}{2010}]{Morin10}
{Morin} J.,  {Donati} J.~F.,  {Petit} P.,  {Delfosse} X.,  {Forveille} T.,
  {Jardine} M.~M.,  2010, \mn@doi [\mnras] {10.1111/j.1365-2966.2010.17101.x},
  \href {https://ui.adsabs.harvard.edu/abs/2010MNRAS.407.2269M} {407, 2269}

\bibitem[\protect\citeauthoryear{{Moschou}, {Vlahakis}, {Drake}, {Evans},
  {Neilson}, {Guzik}  \& {ZuHone}}{{Moschou} et~al.}{2020}]{Moschou2020}
{Moschou} S.-P.,  {Vlahakis} N.,  {Drake} J.~J.,  {Evans} N.~R.,  {Neilson}
  H.~R.,  {Guzik} J.~A.,   {ZuHone} J.,  2020, \mn@doi [\apj]
  {10.3847/1538-4357/aba8fa}, \href
  {https://ui.adsabs.harvard.edu/abs/2020ApJ...900..157M} {900, 157}

\bibitem[\protect\citeauthoryear{{Mullan} \& {Bais}}{{Mullan} \&
  {Bais}}{2018}]{Mullan18}
{Mullan} D.~J.,  {Bais} H.~P.,  2018, preprint, \href
  {https://ui.adsabs.harvard.edu/#abs/2018arXiv180705267M} {p.
  arXiv:1807.05267} (\mn@eprint {arXiv} {1807.05267})

\bibitem[\protect\citeauthoryear{{Mullan}, {MacDonald}, {Dieterich}  \&
  {Fausey}}{{Mullan} et~al.}{2018}]{Mullan18Tr1}
{Mullan} D.~J.,  {MacDonald} J.,  {Dieterich} S.,   {Fausey} H.,  2018, \mn@doi
  [\apj] {10.3847/1538-4357/aaee7c}, \href
  {https://ui.adsabs.harvard.edu/abs/2018ApJ...869..149M} {869, 149}

\bibitem[\protect\citeauthoryear{{Newton}, {Irwin}, {Charbonneau}, {Berlind},
  {Calkins}  \& {Mink}}{{Newton} et~al.}{2017}]{Newton17}
{Newton} E.~R.,  {Irwin} J.,  {Charbonneau} D.,  {Berlind} P.,  {Calkins}
  M.~L.,   {Mink} J.,  2017, \mn@doi [\apj] {10.3847/1538-4357/834/1/85}, \href
  {https://ui.adsabs.harvard.edu/abs/2017ApJ...834...85N} {834, 85}

\bibitem[\protect\citeauthoryear{{Notsu} et~al.,}{{Notsu}
  et~al.}{2013}]{Notsu13}
{Notsu} Y.,  et~al., 2013, \mn@doi [\apj] {10.1088/0004-637X/771/2/127}, \href
  {https://ui.adsabs.harvard.edu/abs/2013ApJ...771..127N} {771, 127}

\bibitem[\protect\citeauthoryear{{Notsu} et~al.,}{{Notsu}
  et~al.}{2019}]{Notsu19}
{Notsu} Y.,  et~al., 2019, \mn@doi [\apj] {10.3847/1538-4357/ab14e6}, \href
  {https://ui.adsabs.harvard.edu/abs/2019ApJ...876...58N} {876, 58}

\bibitem[\protect\citeauthoryear{{Oelkers} et~al.,}{{Oelkers}
  et~al.}{2018}]{Oelkers18}
{Oelkers} R.~J.,  et~al., 2018, \mn@doi [\aj] {10.3847/1538-3881/aa9bf4}, \href
  {http://adsabs.harvard.edu/abs/2018AJ....155...39O} {155, 39}

\bibitem[\protect\citeauthoryear{{Oskanian}}{{Oskanian}}{1969}]{Oskanian69}
{Oskanian} V.,  1969, Information Bulletin on Variable Stars, \href
  {http://adsabs.harvard.edu/abs/1969IBVS..345....1O} {345}

\bibitem[\protect\citeauthoryear{{Paudel}, {Gizis}, {Mullan}, {Schmidt},
  {Burgasser}, {Williams}  \& {Berger}}{{Paudel} et~al.}{2018}]{Paudel18}
{Paudel} R.~R.,  {Gizis} J.~E.,  {Mullan} D.~J.,  {Schmidt} S.~J.,  {Burgasser}
  A.~J.,  {Williams} P. K.~G.,   {Berger} E.,  2018, \mn@doi [\apj]
  {10.3847/1538-4357/aab8fe}, \href
  {https://ui.adsabs.harvard.edu/\#abs/2018ApJ...858...55P} {858, 55}

\bibitem[\protect\citeauthoryear{{Pecaut} \& {Mamajek}}{{Pecaut} \&
  {Mamajek}}{2013}]{Pecaut13}
{Pecaut} M.~J.,  {Mamajek} E.~E.,  2013, \mn@doi [\apjs]
  {10.1088/0067-0049/208/1/9}, \href
  {https://ui.adsabs.harvard.edu/abs/2013ApJS..208....9P} {208, 9}

\bibitem[\protect\citeauthoryear{{Pettersen}, {Coleman}  \&
  {Evans}}{{Pettersen} et~al.}{1984}]{Pettersen84}
{Pettersen} B.~R.,  {Coleman} L.~A.,   {Evans} D.~S.,  1984, \mn@doi [\apjs]
  {10.1086/190934}, \href {http://adsabs.harvard.edu/abs/1984ApJS...54..375P}
  {54, 375}

\bibitem[\protect\citeauthoryear{{Pizzolato}, {Maggio}, {Micela}, {Sciortino}
  \& {Ventura}}{{Pizzolato} et~al.}{2003}]{Pizzolato03}
{Pizzolato} N.,  {Maggio} A.,  {Micela} G.,  {Sciortino} S.,   {Ventura} P.,
  2003, \mn@doi [\aap] {10.1051/0004-6361:20021560}, \href
  {https://ui.adsabs.harvard.edu/abs/2003A&A...397..147P} {397, 147}

\bibitem[\protect\citeauthoryear{{Pugh}, {Armstrong}, {Nakariakov}  \&
  {Broomhall}}{{Pugh} et~al.}{2016}]{Pugh2016}
{Pugh} C.~E.,  {Armstrong} D.~J.,  {Nakariakov} V.~M.,   {Broomhall} A.-M.,
  2016, \mn@doi [\mnras] {10.1093/mnras/stw850}, \href
  {http://adsabs.harvard.edu/abs/2016MNRAS.459.3659P} {459, 3659}

\bibitem[\protect\citeauthoryear{{Rackham}, {Apai}  \& {Giampapa}}{{Rackham}
  et~al.}{2018}]{Rackham18}
{Rackham} B.~V.,  {Apai} D.,   {Giampapa} M.~S.,  2018, \mn@doi [\apj]
  {10.3847/1538-4357/aaa08c}, \href
  {https://ui.adsabs.harvard.edu/abs/2018ApJ...853..122R} {853, 122}

\bibitem[\protect\citeauthoryear{{Ranjan}, {Wordsworth}  \&
  {Sasselov}}{{Ranjan} et~al.}{2017}]{Ranjan17}
{Ranjan} S.,  {Wordsworth} R.,   {Sasselov} D.~D.,  2017, \mn@doi [\apj]
  {10.3847/1538-4357/aa773e}, \href
  {https://ui.adsabs.harvard.edu/#abs/2017ApJ...843..110R} {843, 110}

\bibitem[\protect\citeauthoryear{{Reid} \& {Gizis}}{{Reid} \&
  {Gizis}}{2005}]{Reid2005}
{Reid} I.~N.,  {Gizis} J.~E.,  2005, \mn@doi [Publications of the Astronomical
  Society of the Pacific] {10.1086/430462}, \href
  {https://ui.adsabs.harvard.edu/#abs/2005PASP..117..676R} {117, 676}

\bibitem[\protect\citeauthoryear{{Ricker} et~al.,}{{Ricker}
  et~al.}{2014}]{Ricker14}
{Ricker} G.~R.,  et~al., 2014, in Space Telescopes and Instrumentation 2014:
  Optical, Infrared, and Millimeter Wave. p. 914320 (\mn@eprint {arXiv}
  {1406.0151}), \mn@doi{10.1117/12.2063489}

\bibitem[\protect\citeauthoryear{{Ricker} et~al.,}{{Ricker}
  et~al.}{2015}]{Ricker15}
{Ricker} G.~R.,  et~al., 2015, \mn@doi [Journal of Astronomical Telescopes,
  Instruments, and Systems] {10.1117/1.JATIS.1.1.014003}, \href
  {https://ui.adsabs.harvard.edu/\#abs/2015JATIS...1a4003R} {1, 014003}

\bibitem[\protect\citeauthoryear{{Rimmer}, {Xu}, {Thompson}, {Gillen},
  {Sutherland}  \& {Queloz}}{{Rimmer} et~al.}{2018}]{Rimmer18}
{Rimmer} P.~B.,  {Xu} J.,  {Thompson} S.~J.,  {Gillen} E.,  {Sutherland} J.~D.,
    {Queloz} D.,  2018, \mn@doi [Science Advances] {10.1126/sciadv.aar3302},
  \href {https://ui.adsabs.harvard.edu/#abs/2018SciA....4R3302R} {4, eaar3302}

\bibitem[\protect\citeauthoryear{{Rodono}}{{Rodono}}{1974}]{Rodono74}
{Rodono} M.,  1974, \aap, \href
  {http://adsabs.harvard.edu/abs/1974A%26A....32..337R} {32, 337}

\bibitem[\protect\citeauthoryear{{Rodr{\'\i}guez Mart{\'\i}nez}, {Lopez},
  {Shappee}, {Schmidt}, {Jayasinghe}, {Kochanek}, {Auchettl}  \&
  {Holoien}}{{Rodr{\'\i}guez Mart{\'\i}nez} et~al.}{2020}]{Rodriguez20}
{Rodr{\'\i}guez Mart{\'\i}nez} R.,  {Lopez} L.~A.,  {Shappee} B.~J.,  {Schmidt}
  S.~J.,  {Jayasinghe} T.,  {Kochanek} C.~S.,  {Auchettl} K.,   {Holoien} T.
  W.~S.,  2020, \mn@doi [\apj] {10.3847/1538-4357/ab793a}, \href
  {https://ui.adsabs.harvard.edu/abs/2020ApJ...892..144R} {892, 144}

\bibitem[\protect\citeauthoryear{{Roettenbacher} \& {Vida}}{{Roettenbacher} \&
  {Vida}}{2018}]{Roettenbacher18}
{Roettenbacher} R.~M.,  {Vida} K.,  2018, preprint, \href
  {https://ui.adsabs.harvard.edu/#abs/2018arXiv181004762R} {p.
  arXiv:1810.04762} (\mn@eprint {arXiv} {1810.04762})

\bibitem[\protect\citeauthoryear{{Saar}, {Linsky}  \& {Beckers}}{{Saar}
  et~al.}{1986}]{Saar86}
{Saar} S.~H.,  {Linsky} J.~L.,   {Beckers} J.~M.,  1986, \mn@doi [\apj]
  {10.1086/164040}, \href
  {https://ui.adsabs.harvard.edu/abs/1986ApJ...302..777S} {302, 777}

\bibitem[\protect\citeauthoryear{{Schaefer}}{{Schaefer}}{1989}]{Schaefer_1989}
{Schaefer} B.~E.,  1989, \mn@doi [\apj] {10.1086/167162}, \href
  {http://adsabs.harvard.edu/abs/1989ApJ...337..927S} {337, 927}

\bibitem[\protect\citeauthoryear{{Schaefer}, {King}  \&
  {Deliyannis}}{{Schaefer} et~al.}{2000}]{Schaefer_2000}
{Schaefer} B.~E.,  {King} J.~R.,   {Deliyannis} C.~P.,  2000, \mn@doi [\apj]
  {10.1086/308325}, \href {http://adsabs.harvard.edu/abs/2000ApJ...529.1026S}
  {529, 1026}

\bibitem[\protect\citeauthoryear{{Schmidt}, {Cruz}, {Bongiorno}, {Liebert}  \&
  {Reid}}{{Schmidt} et~al.}{2007}]{Schmidt07}
{Schmidt} S.~J.,  {Cruz} K.~L.,  {Bongiorno} B.~J.,  {Liebert} J.,   {Reid}
  I.~N.,  2007, \mn@doi [\aj] {10.1086/512158}, \href
  {https://ui.adsabs.harvard.edu/abs/2007AJ....133.2258S} {133, 2258}

\bibitem[\protect\citeauthoryear{{Schmidt} et~al.,}{{Schmidt}
  et~al.}{2016}]{Schmidt16}
{Schmidt} S.~J.,  et~al., 2016, \mn@doi [\apjl] {10.3847/2041-8205/828/2/L22},
  \href {http://adsabs.harvard.edu/abs/2016ApJ...828L..22S} {828, L22}

\bibitem[\protect\citeauthoryear{{Schmidt} et~al.,}{{Schmidt}
  et~al.}{2019}]{Schmidt19}
{Schmidt} S.~J.,  et~al., 2019, \mn@doi [\apj] {10.3847/1538-4357/ab148d},
  \href {https://ui.adsabs.harvard.edu/abs/2019ApJ...876..115S} {876, 115}

\bibitem[\protect\citeauthoryear{{Schneider} \& {Shkolnik}}{{Schneider} \&
  {Shkolnik}}{2018}]{Schneider18}
{Schneider} A.~C.,  {Shkolnik} E.~L.,  2018, \mn@doi [\aj]
  {10.3847/1538-3881/aaaa24}, \href
  {https://ui.adsabs.harvard.edu/abs/2018AJ....155..122S} {155, 122}

\bibitem[\protect\citeauthoryear{{Segura}, {Walkowicz}, {Meadows}, {Kasting}
  \& {Hawley}}{{Segura} et~al.}{2010}]{Segura2010}
{Segura} A.,  {Walkowicz} L.~M.,  {Meadows} V.,  {Kasting} J.,   {Hawley} S.,
  2010, \mn@doi [Astrobiology] {10.1089/ast.2009.0376}, \href
  {http://adsabs.harvard.edu/abs/2010AsBio..10..751S} {10, 751}

\bibitem[\protect\citeauthoryear{{Shakhovskaia}}{{Shakhovskaia}}{1989}]{Shak89}
{Shakhovskaia} N.~I.,  1989, \mn@doi [\solphys] {10.1007/BF00161707}, \href
  {https://ui.adsabs.harvard.edu/abs/1989SoPh..121..375S} {121, 375}

\bibitem[\protect\citeauthoryear{{Shibata}}{{Shibata}}{1999}]{Shibata1999}
{Shibata} K.,  1999, \mn@doi [\apss] {10.1023/A:1002413214356}, \href
  {http://adsabs.harvard.edu/abs/1999Ap%26SS.264..129S} {264, 129}

\bibitem[\protect\citeauthoryear{{Shibayama} et~al.,}{{Shibayama}
  et~al.}{2013}]{Shibayama13}
{Shibayama} T.,  et~al., 2013, \mn@doi [\apjs] {10.1088/0067-0049/209/1/5},
  \href {http://adsabs.harvard.edu/abs/2013ApJS..209....5S} {209, 5}

\bibitem[\protect\citeauthoryear{{Shimizu}}{{Shimizu}}{1995}]{Shimizu95}
{Shimizu} T.,  1995, Publications of the Astronomical Society of Japan, \href
  {https://ui.adsabs.harvard.edu/\#abs/1995PASJ...47..251S} {47, 251}

\bibitem[\protect\citeauthoryear{{Shulyak}, {Reiners}, {Engeln}, {Malo},
  {Yadav}, {Morin}  \& {Kochukhov}}{{Shulyak} et~al.}{2017}]{Shulyak17}
{Shulyak} D.,  {Reiners} A.,  {Engeln} A.,  {Malo} L.,  {Yadav} R.,  {Morin}
  J.,   {Kochukhov} O.,  2017, \mn@doi [Nature Astronomy]
  {10.1038/s41550-017-0184}, \href
  {https://ui.adsabs.harvard.edu/abs/2017NatAs...1E.184S} {1, 0184}

\bibitem[\protect\citeauthoryear{{Skrutskie} et~al.,}{{Skrutskie}
  et~al.}{2006}]{2MASS_2006}
{Skrutskie} M.~F.,  et~al., 2006, \mn@doi [\aj] {10.1086/498708}, \href
  {http://adsabs.harvard.edu/abs/2006AJ....131.1163S} {131, 1163}

\bibitem[\protect\citeauthoryear{{Stassun} et~al.,}{{Stassun}
  et~al.}{2019}]{Stassun19}
{Stassun} K.~G.,  et~al., 2019, arXiv e-prints, \href
  {https://ui.adsabs.harvard.edu/abs/2019arXiv190510694S} {p. arXiv:1905.10694}

\bibitem[\protect\citeauthoryear{{Tilley}, {Segura}, {Meadows}, {Hawley}  \&
  {Davenport}}{{Tilley} et~al.}{2017}]{Tilley17}
{Tilley} M.~A.,  {Segura} A.,  {Meadows} V.~S.,  {Hawley} S.,   {Davenport} J.,
   2017, arXiv e-prints, \href
  {https://ui.adsabs.harvard.edu/abs/2017arXiv171108484T} {p. arXiv:1711.08484}

\bibitem[\protect\citeauthoryear{Tsurutani, Gonzalez, Lakhina  \&
  Alex}{Tsurutani et~al.}{2003}]{Carrington_Energy}
Tsurutani B.~T.,  Gonzalez W.~D.,  Lakhina G.~S.,   Alex S.,  2003, \mn@doi
  [Journal of Geophysical Research: Space Physics] {10.1029/2002JA009504}, 108,
  n/a

\bibitem[\protect\citeauthoryear{{Wang}, {Chae}, {Yurchyshyn}, {Yang},
  {Steinegger}  \& {Goode}}{{Wang} et~al.}{2001}]{Wang01}
{Wang} H.,  {Chae} J.,  {Yurchyshyn} V.,  {Yang} G.,  {Steinegger} M.,
  {Goode} P.,  2001, \mn@doi [\apj] {10.1086/322377}, \href
  {http://adsabs.harvard.edu/abs/2001ApJ...559.1171W} {559, 1171}

\bibitem[\protect\citeauthoryear{{Watson}, {Henden}  \& {Price}}{{Watson}
  et~al.}{2006}]{Watson06}
{Watson} C.~L.,  {Henden} A.~A.,   {Price} A.,  2006, Society for Astronomical
  Sciences Annual Symposium, \href
  {https://ui.adsabs.harvard.edu/abs/2006SASS...25...47W} {25, 47}

\bibitem[\protect\citeauthoryear{{West}, {Hawley}, {Bochanski}, {Covey},
  {Reid}, {Dhital}, {Hilton}  \& {Masuda}}{{West} et~al.}{2008}]{West08}
{West} A.~A.,  {Hawley} S.~L.,  {Bochanski} J.~J.,  {Covey} K.~R.,  {Reid}
  I.~N.,  {Dhital} S.,  {Hilton} E.~J.,   {Masuda} M.,  2008, \mn@doi [\aj]
  {10.1088/0004-6256/135/3/785}, \href
  {https://ui.adsabs.harvard.edu/abs/2008AJ....135..785W} {135, 785}

\bibitem[\protect\citeauthoryear{{Wheatley} et~al.,}{{Wheatley}
  et~al.}{2018}]{Wheatley18}
{Wheatley} P.~J.,  et~al., 2018, \mn@doi [\mnras] {10.1093/mnras/stx2836},
  \href {http://adsabs.harvard.edu/abs/2018MNRAS.475.4476W} {475, 4476}

\bibitem[\protect\citeauthoryear{{Winters} et~al.,}{{Winters}
  et~al.}{2019}]{Winters19}
{Winters} J.~G.,  et~al., 2019, \mn@doi [\aj] {10.3847/1538-3881/ab05dc}, \href
  {https://ui.adsabs.harvard.edu/abs/2019AJ....157..216W} {157, 216}

\bibitem[\protect\citeauthoryear{{Wright}, {Drake}, {Mamajek}  \&
  {Henry}}{{Wright} et~al.}{2011}]{Wright11}
{Wright} N.~J.,  {Drake} J.~J.,  {Mamajek} E.~E.,   {Henry} G.~W.,  2011,
  \mn@doi [\apj] {10.1088/0004-637X/743/1/48}, \href
  {http://adsabs.harvard.edu/abs/2011ApJ...743...48W} {743, 48}

\bibitem[\protect\citeauthoryear{{Yan}, {Wang}, {Pan}, {Kong}, {Xue}, {Yang},
  {Li}  \& {Feng}}{{Yan} et~al.}{2018}]{Xan18}
{Yan} X.~L.,  {Wang} J.~C.,  {Pan} G.~M.,  {Kong} D.~F.,  {Xue} Z.~K.,  {Yang}
  L.~H.,  {Li} Q.~L.,   {Feng} X.~S.,  2018, \mn@doi [\apj]
  {10.3847/1538-4357/aab153}, \href
  {https://ui.adsabs.harvard.edu/abs/2018ApJ...856...79Y} {856, 79}

\bibitem[\protect\citeauthoryear{{Yang} \& {Liu}}{{Yang} \&
  {Liu}}{2019}]{Yang19}
{Yang} H.,  {Liu} J.,  2019, \mn@doi [The Astrophysical Journal Supplement
  Series] {10.3847/1538-4365/ab0d28}, \href
  {https://ui.adsabs.harvard.edu/abs/2019ApJS..241...29Y} {241, 29}

\bibitem[\protect\citeauthoryear{{Yang} et~al.,}{{Yang} et~al.}{2017}]{Yang17}
{Yang} H.,  et~al., 2017, \mn@doi [\apj] {10.3847/1538-4357/aa8ea2}, \href
  {https://ui.adsabs.harvard.edu/#abs/2017ApJ...849...36Y} {849, 36}

\bibitem[\protect\citeauthoryear{{Yang}, {Liu}, {Qiao}, {Zhang}, {Gao}, {Cui}
  \& {Han}}{{Yang} et~al.}{2018}]{Yang18}
{Yang} H.,  {Liu} J.,  {Qiao} E.,  {Zhang} H.,  {Gao} Q.,  {Cui} K.,   {Han}
  H.,  2018, \mn@doi [\apj] {10.3847/1538-4357/aabd31}, \href
  {https://ui.adsabs.harvard.edu/#abs/2018ApJ...859...87Y} {859, 87}

\bibitem[\protect\citeauthoryear{{Zacharias}, {Finch}, {Girard}, {Henden},
  {Bartlett}, {Monet}  \& {Zacharias}}{{Zacharias} et~al.}{2013}]{UCAC4_13}
{Zacharias} N.,  {Finch} C.~T.,  {Girard} T.~M.,  {Henden} A.,  {Bartlett}
  J.~L.,  {Monet} D.~G.,   {Zacharias} M.~I.,  2013, \mn@doi [\aj]
  {10.1088/0004-6256/145/2/44}, \href
  {http://adsabs.harvard.edu/abs/2013AJ....145...44Z} {145, 44}

\bibitem[\protect\citeauthoryear{{Zirin} \& {Liggett}}{{Zirin} \&
  {Liggett}}{1987}]{Zinn87}
{Zirin} H.,  {Liggett} M.~A.,  1987, \mn@doi [\solphys] {10.1007/BF00147707},
  \href {https://ui.adsabs.harvard.edu/abs/1987SoPh..113..267Z} {113, 267}

\bibitem[\protect\citeauthoryear{{Zuccarello} et~al.,}{{Zuccarello}
  et~al.}{2009}]{Zuccarello09}
{Zuccarello} F.,  et~al., 2009, \mn@doi [\aap] {10.1051/0004-6361:200809887},
  \href {https://ui.adsabs.harvard.edu/#abs/2009A&A...493..629Z} {493, 629}

\makeatother
\end{thebibliography}

\appendix

\section{Derivation of the observed occurrence rate} \label{sec:cum_freq_derivation}
Here we will outline how we derived the observed flare occurrence rate used in Sect.\,\ref{sec:flare_freq3}. As outlined in Sect.\,\ref{sec:flare_occ_rate_123456}, stellar flares are expected to occur with a power-law, or power-law like, distribution in energy. The differential distribution of the number of flares in a given duration, $N$, with energy $E$ is written as
\begin{equation}
\frac{dN(E)}{dE} = k E^{-\alpha}
\end{equation}
where $k$ is a normalisation constant and $\alpha$ is the power law index. When integrated from a flare energy $E_{f}$ to infinity, we obtain the standard cumulative flare frequency distribution. The observed distribution of flares with energy also depends on the efficiency of the instrument being used to sample the flares. This efficiency, also called the recovery rate $R(E)$, is a function of energy. At high energies, we expect $R(E)$ to be near or equal to 1, meaning we observe all flare events. At low energies, $R(E)$ tends to zero, resulting in fewer flare detections. Therefore we can write the observed number of flares with energy as
\begin{equation}
 dN(E) = k R(E) E^{-\alpha} dE.
\end{equation}
$R(E)$ can be sampled through flare injection and recovery tests. The cumulative flare frequency distribution is then
\begin{equation}
    N(E>E_{flare}) = k\int_{E_{flare}}^{\infty}R(E)E^{-\alpha}dE
\end{equation}
Integrating by parts then gives
\begin{equation}
    N(E>E_{f}) = k \Bigg(\bigg[\frac{R(E)E^{-\alpha + 1}}{1-\alpha}\bigg]^{\infty}_{E_{flare}} -
    \int_{E_{f}}^{\infty}\frac{R'(E)E^{-\alpha+1}}{1-\alpha}dE\Bigg)
\end{equation}
Where $R'(E)$ is the differentiated $R(E)$. Assuming $\alpha$ is always greater than 1, this results in
\begin{equation}
    N(E>E_{f}) = k \Bigg(-\frac{R(E_{f})E_{f}^{-\alpha + 1}}{1-\alpha} -
    \int_{E_{f}}^{\infty}\frac{R'(E)E^{-\alpha+1}}{1-\alpha}dE\Bigg)
\end{equation}
and then
\begin{equation}
    N(E>E_{f}) = \frac{k}{\alpha-1} \Bigg(R(E_{f})E_{f}^{-\alpha + 1} +
    \int_{E_{f}}^{\infty}R'(E)E^{-\alpha+1}dE\Bigg)
\end{equation}
Above some energy $E_{max}$, the recovery fraction will not change, i.e. it has reached it's maximum efficiency, so $R'(E)$ will equal zero. If this known (e.g from flare injection and recovery tests), then we can write the observed cumulative flare frequency as
\begin{equation}
    N(E>E_{f}) = \frac{k}{\alpha-1} \Bigg(R(E_{f})E_{f}^{-\alpha + 1} +
    \int_{E_{f}}^{E_{max}}R'(E)E^{-\alpha+1}dE\Bigg)
\end{equation}
We can see at high flare energies which can all be detected, then $R'(E)$=0 and R(E)=1, giving the standard cumulative flare distribution. At low energies, where R(E) equals zero, then the distribution plateaus to the value of the integral.

\begin{table*}
	\centering
	\begin{tabular}{|l|c|c|c||}
    \hline
    Class & $log k$ & $\alpha$ & Covariance Matrix \tabularnewline \hline
    K2V-K4V & $29.4\pm7.2$ & $1.97\pm0.20$ & $\begin{bmatrix} 51.3 & 1.46 \\ 1.46 & 0.04\end{bmatrix}$ \tabularnewline
    K5V-K8V  & $25.0\pm7.4$ & $1.85\pm0.21$ & $\begin{bmatrix} 55.46 & 1.56 \\ 1.56 & 0.04\end{bmatrix}$ \tabularnewline
    M0V-M2V  & $30.4\pm3.4$ & $2.01\pm0.10$ & $\begin{bmatrix} 11.7 & 0.34 \\ 0.34 & 0.009 \end{bmatrix}$ \tabularnewline
    M3V-M5V  & $33.9\pm1.55$ & $2.09\pm0.05$ & $\begin{bmatrix} 2.42 & 0.07 \\ 0.07 & 0.002 \end{bmatrix}$ \tabularnewline
    \hline
    K5-K8 ($<15$ Myr)  & $25.4\pm4.8$ & $1.82\pm0.14$ & $\begin{bmatrix} 23.4 & 0.66 \\ 0.66 & 0.002\end{bmatrix}$ \tabularnewline
    M0-M2 ($<30$ Myr) & $20.9\pm1.8$ & $1.69\pm0.05$ & $\begin{bmatrix} 3.17 & 0.09 \\ 0.09 & 0.003 \end{bmatrix}$ \tabularnewline
    M3-M5 ($<40$ Myr) & $29.5\pm1.2$ & $1.94\pm0.04$ & $\begin{bmatrix} 1.47 & 0.04 \\ 0.04 & 0.001 \end{bmatrix}$ \tabularnewline
    \hline
    Orion 3400-3940\,K & $30.7\pm5.7$ & $1.94\pm0.16$ & $\begin{bmatrix} 31.2 & 0.88 \\ 0.88 & 0.02 \end{bmatrix}$ \tabularnewline \hline
 	\end{tabular}
    \caption{Parameters and corresponding covariance matrices from the power law fits to main sequence and pre-main sequence subsets of our data.}\label{tab:alpha_cov} 
\end{table*}


\bsp	
\label{lastpage}
\end{document}